\documentclass[acmsmall,screen,nonacm]{acmart}

\usepackage{threeparttable}
\usepackage{threeparttablex}
\usepackage{quantikz}
\usetikzlibrary{fit,backgrounds}
\usepackage{amsmath}
\usepackage{longtable}
\usepackage[nointegrals]{wasysym}
\usepackage{xspace}
\usepackage{relsize,tipx}
\usepackage{array,xspace,listings}
\usepackage{longtable}
\usepackage{enumitem}
\usepackage{tabularx}
\usepackage{subcaption}
\usepackage{graphicx}
\usepackage{color}
\usepackage{tikz}
\usepackage{colortbl}
\usepackage{multirow,makecell}
\usepackage{dsfont}
\usepackage{proof}
\usepackage{mathpartir}
\usepackage{wrapfig}
\usepackage{setspace}
\usepackage{stmaryrd}
\usepackage[linesnumbered,ruled,vlined]{algorithm2e}
\SetCommentSty{textit}
\SetAlgorithmName{Alg.}{algorithm}{List of Algorithms}
\SetAlFnt{\small}\SetAlCapFnt{\small}\SetAlCapNameFnt{\small}
\SetVlineSkip{0pt}
\SetAlgoSkip{}
\makeatletter
\patchcmd{\@algocf@start}{\addtolength{\hsize}{-1.5em}}{}{}{\PackageWarning{MBU}{algorithm2e right-gap patch failed}}

\patchcmd{\algocf@Vline}{\algocf@push{\skiptext}}{\advance\skiptotal by 0.4pt\algocf@push{\skiptext}}{}{\PackageWarning{MBU}{algorithm2e Vline patch failed}}
\patchcmd{\algocf@Vsline}{\algocf@push{\skiptext}}{\advance\skiptotal by 0.4pt\algocf@push{\skiptext}}{}{\PackageWarning{MBU}{algorithm2e Vsline patch failed}}
\newcommand{\AlgFrame}[2][MBUPayment]{%
  \par\nointerlineskip
  \setbox\@tempboxa=\vbox{\advance\hsize by -2\fboxsep\advance\hsize by -2\fboxrule #2\par}%
  \hbox{\kern-\fboxsep\kern-\fboxrule\fcolorbox{#1}{white}{\box\@tempboxa}}%
  \nointerlineskip}
\newcommand{\AlgShade}[2][MBUPayment!10]{%
  \par\nointerlineskip
  \setbox\@tempboxa=\vbox{\advance\hsize by -2\fboxsep #2\par}%
  \hbox{\kern-\fboxsep\colorbox{#1}{\box\@tempboxa}}%
  \nointerlineskip}
\makeatother
\usepackage{booktabs,longtable}
\usepackage[dvipsnames,table]{xcolor}
\usepackage{placeins}
\usepackage{calc}
\usepackage{mdframed}
\usepackage{needspace}
\usepackage{cleveref}
\usepackage{pifont}
\usetikzlibrary{quantikz2,arrows.meta,calc,positioning,backgrounds,fit,decorations.pathreplacing}

\usepackage{etoolbox}
\tikzcdset{bundle gap/.initial=0.045cm}
\patchcmd{\fullwire}
  {\arrow[arrows,start anchor=#2startone,end anchor=#2endone,#4] {#1}\arrow[arrows,start anchor=#2starttwo,end anchor=#2endtwo,#4] {#1}\arrow[#1,arrows,#4] {}}
  {\arrow[arrows,shift left=\pgfkeysvalueof{/tikz/commutative diagrams/bundle gap},#4] {#1}\arrow[arrows,shift right=\pgfkeysvalueof{/tikz/commutative diagrams/bundle gap},#4] {#1}\arrow[#1,arrows,#4] {}}
  {}{\PackageWarning{includes}{quantikz bundle gap patch failed; bundle wires fall back to classical gap}}

\DeclareMathAlphabet{\pazocal}{OMS}{zplm}{m}{n}

\newcommand{\patom}[1]{\mathsf{#1}}
\newcommand{\prodatom}[1]{\mathbf{p}_{#1}}
\newcommand{\Val}{\mathit{Val}}
\newcommand{\Atom}{\mathit{Atom}}
\newcommand{\Sym}{\mathit{Sym}}
\newcommand{\one}{\mathds{1}}
\newcommand{\symdiff}{\mathbin{\triangle}}
\newcommand{\sem}[1]{\llbracket #1\rrbracket}
\newcommand{\fresh}{\mathit{fresh}()}
\newcommand{\Product}{\textsc{Product}}
\newcommand{\PairProduct}{\textsc{Product}_2}
\newcommand{\Step}{\textsc{Step}}

\newcommand{\XORg}{\mathsf{XOR}}

\newcommand{\qubits}{\mathit{qubits}}
\newcommand{\written}{\mathit{written}}

\newcommand{\rulename}[1]{{\color{MBUCleanup}[\textsc{#1}]}}

\definecolor{mygreen}{rgb}{0.913,0.9535,0.8925}

\newtheorem{theorem}{Theorem}[section]

\AddToHook{env/assumption/begin}{\crefalias{theorem}{assumption}}

\makeatletter
\theoremstyle{acmdefinition}

\undef\definition
\undef\enddefinition
\newtheorem{definition}[theorem]{Definition}

\undef\strategy
\undef\endstrategy
\newtheorem{strategy}[theorem]{Strategy}

\theoremstyle{acmplain}
\makeatother

\makeatletter
\def\@acmplainindent{0pt}
\def\@proofindent{}
\makeatother

\definecolor{MBUCompute}{HTML}{000000}
\definecolor{MBUCleanup}{HTML}{6A458B}
\definecolor{MBUPayment}{HTML}{B66512}
\definecolor{MBUShared}{HTML}{927323}
\definecolor{MBULifeOne}{HTML}{327CAA}
\definecolor{MBULifeTwo}{HTML}{26846A}
\definecolor{MBUWrong}{HTML}{9B1B1B}
\definecolor{MBULastUse}{HTML}{C00000}
\definecolor{MBUInterval}{HTML}{DA7A19}
\makeatletter
\newcommand{\MBUTableStrip}[1]{\textcolor{#1}{\rule[-\dp\@arstrutbox]{3pt}{\dimexpr\ht\@arstrutbox+\dp\@arstrutbox\relax}}}
\newcommand{\MBUTableStripMid}{(\ht\@arstrutbox-\dp\@arstrutbox)/2}
\newcommand{\MBUTableStripUpper}[2][0pt]{\textcolor{#2}{\rule[\dimexpr\MBUTableStripMid+#1\relax]{3pt}{\dimexpr\ht\@arstrutbox-\MBUTableStripMid-#1\relax}}}
\newcommand{\MBUTableStripLower}[2][0pt]{\textcolor{#2}{\rule[-\dp\@arstrutbox]{3pt}{\dimexpr\dp\@arstrutbox+\MBUTableStripMid-#1\relax}}}
\makeatother

\newcommand{\MBUHMExtraGap}{0.1cm}
\newcommand{\MBUResetExtraGap}{0cm}

\newlength{\MBUGateWidth}
\newlength{\MBUGateHeight}
\newcommand{\MBUGateShrink}{-1.2pt}
\tikzset{mbu gate font/.style={font=\footnotesize}}
\NewExpandableDocumentCommand{\MBUGate}{O{MBUCompute}m}{\gate[disable auto height,label style={mbu gate font},style={draw=#1,fill=none,text=#1,mbu gate font,rounded corners=0pt,minimum width=\MBUGateWidth,minimum height=\MBUGateHeight,inner sep=\MBUGateShrink,outer sep=0pt}][\MBUGateWidth][\MBUGateHeight]{#2}}
\NewExpandableDocumentCommand{\MBUWideGate}{O{MBUCompute}mm}{\gate[disable auto height,label style={mbu gate font},style={draw=#1,fill=white,text=#1,mbu gate font,rounded corners=0pt,minimum width=#2,minimum height=\MBUGateHeight,inner sep=\MBUGateShrink,outer sep=0pt}][#2][\MBUGateHeight]{#3}}
\NewExpandableDocumentCommand{\MBUCtrl}{O{MBUCompute}m}{\ctrl[style={draw=#1,fill=#1},wire style={draw=#1}]{#2}}
\NewExpandableDocumentCommand{\MBUTarget}{O{MBUCompute}m}{\targ[style={draw=#1,fill=none,internal/.append style={draw=#1}}]{#2}}
\NewExpandableDocumentCommand{\MBUOCtrl}{O{MBUCompute}m}{\octrl[style={draw=#1,fill=white},wire style={draw=#1}]{#2}}
\newcommand{\MBUDot}{\control[style={draw=MBUCompute,fill=MBUCompute}]{}}
\tikzset{mbu bundle dot/.style={rectangle,rounded corners=0.65mm,minimum width=1.3mm,minimum height=3mm,inner sep=0pt}}

\newcommand{\MBUCleanupTarget}{\MBUTarget[MBUCleanup]{}}
\newcommand{\MBUCleanupCtrl}[1]{\MBUCtrl[MBUCleanup]{#1}}
\newcommand{\MBUPaymentCtrl}[1]{\MBUCtrl[MBUPayment]{#1}}
\newcommand{\MBUPaymentDot}{\control[style={draw=MBUPayment,fill=MBUPayment}]{} }
\newcommand{\MBUH}{\MBUGate{H}}
\newsavebox{\mbufitbox}
\newcommand{\MBUFitWidth}[1]{\sbox{\mbufitbox}{#1}\ifdim\wd\mbufitbox>\textwidth\resizebox{\textwidth}{!}{\usebox{\mbufitbox}}\else\usebox{\mbufitbox}\fi}
\newcommand{\MBUPaymentGate}[1]{\gate[disable auto height,label style={mbu gate font},style={draw=MBUPayment,fill=white,text=MBUPayment,mbu gate font,rounded corners=0pt,minimum width=\MBUGateWidth,minimum height=\MBUGateHeight,inner sep=\MBUGateShrink,outer sep=0pt}][\MBUGateWidth][\MBUGateHeight]{#1}}
\newcommand{\MBUMeasure}[1]{\meter[style={draw=black,fill=none,color=black,minimum width=\MBUGateWidth,minimum height=\MBUGateHeight,inner sep=\MBUGateShrink,outer sep=0pt}][\MBUGateWidth][\MBUGateHeight]{}\wire[d][#1]{c}}
\newcommand{\MBUReset}{\push{\ket{0}}\wireoverride{n}}
\newcommand{\MBUMeasureInline}{\meter[style={draw=black,fill=none,color=black,minimum width=\MBUGateWidth,minimum height=\MBUGateHeight,inner sep=\MBUGateShrink,outer sep=0pt}][\MBUGateWidth][\MBUGateHeight]{}\wireoverride{q}\setwiretype{c}}

\newcommand{\MBUClassicalBegin}{\setwiretype{c}}
\newcommand{\MBUClassicalEnd}{\setwiretype{n}}
\newcommand{\MBUClassicalCtrl}[1]{\ctrl[vertical wire=c,style={draw=MBUPayment,fill=MBUPayment},wire style={draw=black}]{#1}}
\newcommand{\MBUOutcomeLabelAbove}[5]{\node[above=2pt,text=black,font=\scriptsize,inner sep=0pt] at ($(#1-#2-#3.center)!0.5!(#1-#2-#4.center)$) {#5};}
\newcommand{\MBUOutcomeLabel}[5]{\node[below=3pt,text=black,font=\scriptsize,inner sep=0pt] at ($(#1-#2-#3.center)!0.5!(#1-#2-#4.center)$) {#5};}
\newcommand{\MBUOutcomeLabelLeft}[5]{\node[left=3pt,text=black,font=\scriptsize,inner sep=0pt] at (#1-#2-#3.center) {#5};}
\tikzset{mbu number distance/.initial=8pt}
\newcommand{\MBUNumberSteps}[2]{%
  \foreach \mbuStep in {1,...,#2}{%
    \pgfmathtruncatemacro{\mbuStepColumn}{\mbuStep+1}%
    \node[above=\pgfkeysvalueof{/tikz/mbu number distance},text=black,font=\scriptsize,inner sep=0pt] at (#1-1-\mbuStepColumn.center |- #1-1-1.center) {\mbuStep};
  }%
}

\newcommand{\MBUNumberAt}[4][black]{\node[above=\pgfkeysvalueof{/tikz/mbu number distance},text=#1,font=\scriptsize,inner sep=0pt] at (#2-1-#3.center |- #2-1-1.center) {#4};}
\newcommand{\MBUNumberCorrection}[3]{\MBUNumberAt[MBUPayment]{#1}{#2}{#3$'$}}
\tikzset{mbu lifetime label distance/.initial=11pt}
\tikzset{mbu write/.pic={\draw[semithick] (0,0) circle (0.11cm) (-0.11cm,0) -- (0.11cm,0) (0,-0.11cm) -- (0,0.2cm);},
         mbu use/.pic={\fill (0,0) circle (0.07cm); \draw[semithick] (0,0) -- (0,0.2cm);}}

\tikzset{mbu value label distance/.initial=7pt}
\tikzset{mbu value label font/.initial=\scriptsize}
\newcommand{\MBUValueLabel}[5]{\node[below=\pgfkeysvalueof{/tikz/mbu value label distance},text=black,font=\pgfkeysvalueof{/tikz/mbu value label font},inner sep=0pt] at ($(#1-#2-#3.center)!0.5!(#1-#2-#4.center)$) {#5};}
\newcommand{\MBULifetimeX}[7][below]{%
  \begin{pgfonlayer}{background}
    \coordinate (mbu-life-l) at #4;
    \coordinate (mbu-life-r) at #5;
    \coordinate (mbu-life-left) at (mbu-life-l |- #2-#3-1.center);
    \coordinate (mbu-life-right) at (mbu-life-r |- #2-#3-1.center);
    \path[fill=#6!16,draw=none,rounded corners=1pt] ($(mbu-life-left)+(0,-4.5pt)$) rectangle ($(mbu-life-right)+(0,4.5pt)$);
    \node[#1=\pgfkeysvalueof{/tikz/mbu lifetime label distance},text=#6!80!black,font=\scriptsize,inner sep=0pt] at ($(mbu-life-left)!0.5!(mbu-life-right)$) {#7};
  \end{pgfonlayer}%
}
\newcommand{\MBULifetime}[7][below]{%
  \begin{pgfonlayer}{background}
    \coordinate (mbu-life-left) at (#2-col#4.west |- #2-#3-#4.center);
    \coordinate (mbu-life-right) at (#2-col#5.east |- #2-#3-#5.center);
    \path[fill=#6!16,draw=none,rounded corners=1pt] ($(mbu-life-left)+(0,-4.5pt)$) rectangle ($(mbu-life-right)+(0,4.5pt)$);
    \node[#1=\pgfkeysvalueof{/tikz/mbu lifetime label distance},text=#6!80!black,font=\scriptsize,inner sep=0pt] at ($(mbu-life-left)!0.5!(mbu-life-right)$) {#7};
  \end{pgfonlayer}%
}

\begin{document}

\title{MBUndo: Sound, Automatic, and Efficient Measurement-Based Uncomputation for Quantum Circuits}

\author{Shengyuan Yang}
\orcid{0009-0006-3298-7999}
\email{syang686@wisc.edu}
\author{Charles Yuan}
\orcid{0000-0002-4918-4467}
\affiliation{%
  \institution{University of Wisconsin--Madison}
  \country{USA}
}
\email{charlesyuan@cs.wisc.edu}

\renewcommand{\shortauthors}{Yang and Yuan}

\settopmatter{}
\authorsaddresses{}

\begin{abstract}
Quantum programs that use temporary values, known as \emph{ancilla} qubits, must explicitly \emph{uncompute} each ancilla back to zero to ensure that the program produces a correct output.
Conventional uncomputation uses unitary quantum gates, whereas leading quantum algorithms now leverage \emph{measurement-based uncomputation} (MBU), which replaces costly unitary $T$ gates with cheaper measurements and phase corrections on hardware.

Current quantum programming frameworks can use measurement to uncompute specific gates such as Toffoli.
No automatic transformation, however, accepts an arbitrary circuit whose ancillas are computed and uncomputed via the standard approach and converts the circuit to use MBU\@.
Researchers instead apply MBU to individual circuit designs by hand, which limits algorithm development and compiler optimization.

Such a transformation has not existed due to two fundamental unknowns: where MBU applies and which gates realize MBU for an arbitrary computation.
This work answers both questions.
First, we present a static analysis that identifies, for each ancilla qubit, each circuit position where measurement is sound --- subsuming basic assumptions that MBU only applies at the end of a circuit.
Next, we present circuit transformations that soundly turn unitary uncomputation gates into efficient phase corrections after measurement --- generalizing rewrite rules from prior literature that are either limited to a single gate or not applicable to a general circuit.

We present \textsf{MBUndo}, an integrated compiler pass that automatically converts circuits to measurement-based uncomputation.
It requires no user annotations other than which qubits are ancillas, the same as needed by existing uncomputation synthesizers, and can enhance their output circuits.
We evaluate \textsf{MBUndo} on 22 benchmarks to reveal \emph{T} gate reductions of ${>}36.5\%$ over uncomputation synthesizers.
\end{abstract}

\maketitle

\section{Introduction}
\label{sec:intro}

Like classical programs, quantum programs that implement quantum algorithms generate temporary values as part of execution.
These values are stored in \emph{ancilla} qubits alongside the data qubits that store the ultimate output of the computation.
Classically, any temporary value can simply be discarded and destroyed once used.
But in a quantum program, ancilla qubits are \emph{entangled} with the data qubits they are computed from, meaning that destroying a temporary value can also corrupt the program state and jeopardize correctness of the program~\cite[\S 4]{QCQI}.

The remedy for this problem is \emph{uncomputation}~\cite{reversibility}, the process of undoing the computation of a temporary value to explicitly return it to zero and thereby eliminate all entanglement with the data qubits.
Writing uncomputation by hand is common yet tedious and error-prone for programmers. Thus, researchers have developed techniques to \emph{automatically uncompute} variables in programming languages~\cite{silq,Qurts,modular-synthesis} or qubits in the quantum circuits to which languages compile~\cite{unqomp,reqomp,dirty-and-clean-uncomputation}.

\subsection{Unitary Uncomputation and Measurement-Based Uncomputation}

Whether manual or automatic, uncomputation is typically~\cite{silq,Qurts,modular-synthesis,unqomp,reqomp,tower} implemented as a \emph{unitary operator} --- a series of quantum logic gates that do not perform measurement.
This approach, which we will call \emph{unitary uncomputation}, characteristically executes the inverse of every logic gate used to compute each ancilla qubit.
Unitary uncomputation executes as many gates as the computation being undone, and thus accounts for up to half the runtime cost of a typical quantum circuit.

\begin{figure}[t!]
\centering%
\begin{subfigure}[b]{0.57\textwidth}%
\centering%
\parbox[c][1.95cm][c]{\linewidth}{\centering\mbox{\small%
\begin{quantikz}[wire types={b,q},row sep={0.65cm,between origins},column sep=0.225cm,align equals at=1.5,classical gap=0.025cm,bundle gap=0.06cm,/tikz/commutative diagrams/every matrix/.append style={/tikz/name=mbuintro},/tikz/execute at end picture={\node at (mbuintro-2-2.center) {$\oplus f(x)$};}]
\lstick{$\ket{x}$} & \gate[2]{\vphantom{\oplus f(x)}\hspace{0.70cm}} & \rstick{$\ket{x}$} \\
\lstick{\footnotesize $\ket{f(x)}_a$} & & \rstick{$\ket{0}_a$}
\end{quantikz}%
\hspace{0.05cm}$=$\hspace{0.05cm}
\begin{quantikz}[wire types={b,q},row sep={0.65cm,between origins},column sep=0.2cm,align equals at=1.5,classical gap=0.025cm,bundle gap=0.06cm]
\lstick{} & & &[-0.5cm] \gate{\footnotesize (-1)^{f(x)}} & \rstick{} \\
\lstick{} & \gate{H} & \meter{}\wireoverride{q}\setwiretype{c} & \ctrl[vertical wire=c]{-1} & \setwiretype{n}
\end{quantikz}%
}}
\vspace{-1.75ex}
\caption{MBU of an ancilla holding $f(x)$, as in \citet{gidney-spooky-pebble}.}
\label{fig:mbu-intro}
\end{subfigure}\hfill%
\begin{subfigure}[b]{0.42\textwidth}
\centering
\parbox[c][1.95cm][c]{\linewidth}{\centering\mbox{\small%
\begin{quantikz}[row sep={0.5cm,between origins},column sep=0.225cm,align equals at=2,classical gap=0.025cm]
\lstick{$\ket{x_1}$} & \ctrl{1} & \rstick{$\ket{x_1}$} \\
\lstick{$\ket{x_2}$} & \ctrl{1} & \rstick{$\ket{x_2}$} \\
\lstick{$\ket{x_1x_2}_a$} & \targ{} & \rstick{$\ket{0}_a$}
\end{quantikz}%
\hspace{0.1cm}$=$\hspace{0.1cm}
\begin{quantikz}[row sep={0.5cm,between origins},column sep=0.2cm,align equals at=2,classical gap=0.025cm]
& & & \ctrl{1} & \\
& & & \gate{Z} & \\
& \gate{H} & \meter{}\wireoverride{q}\setwiretype{c} & \ctrl[vertical wire=c]{-1} & \setwiretype{n}
\end{quantikz}%
}}
\vspace{-1.75ex}
\caption{MBU of a Toffoli, as in \citet{gidney-toffli-mbu}.}
\label{fig:mbu-toffoli}
\end{subfigure}%
\setlength{\abovecaptionskip}{1.25ex}%
\setlength{\belowcaptionskip}{-0.5ex}%
\caption{Equivalence specifications of unitary uncomputation and measurement-based uncomputation (MBU).}
\label{fig:mbu}
\end{figure}
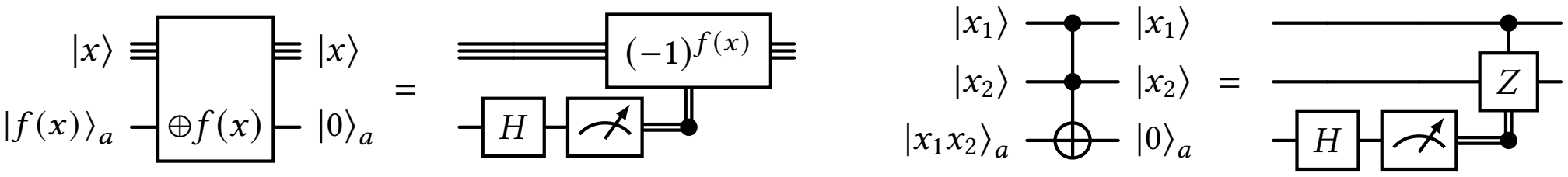

\emph{Measurement-based uncomputation} (MBU) is an alternative to unitary uncomputation introduced by \citet{gidney-toffli-mbu,gidney-spooky-pebble} that can be far cheaper.
Figure~\ref{fig:mbu-intro} depicts a unitary uncomputation circuit and the MBU it is equivalent to.
Both circuits start with an ancilla $a$ holding a temporary $f(x)$, computed
from data qubits $x$, that is no longer needed.
The starting program state is $\sum_x \alpha_x\ket{x}\ket{f(x)}_a$, which uncomputation must transform to $\sum_x\alpha_x\ket{x}\ket{0}_a$,
zeroing the ancilla while leaving the data qubits unaffected.
To do so, unitary uncomputation runs the inverse of the circuit that computed $f(x)$. This operation, denoted $\oplus f(x)$, repeats the cost of the original computation of $f(x)$.

By contrast, MBU cleans up the ancilla $a$ by performing a destructive \emph{X-basis measurement} --- it applies a Hadamard gate and then measures $a$ in the computational basis.
This $H$ gate transforms the state into \smash{$\textstyle\frac{1}{\sqrt2}\big(\sum_x\alpha_x\ket{x}\ket{0}_a
+\sum_x\alpha_x(-1){}^{f(x)}\ket{x}\ket{1}_a\big)$}, so measuring $a$
yields $0$ or $1$ with equal probability. Upon outcome $0$, the state collapses to $\sum_x\alpha_x\ket{x}\ket{0}_a$,
which is already the desired state. Upon outcome $1$, the state collapses to $\sum_x\alpha_x(-1){}^{f(x)}\ket{x}\ket{1}_a$, where the ancilla can be safely reset to zero, but the data qubits retain an unwanted phase $(-1){}^{f(x)}$ that depends on~$x$. MBU thus concludes with a \emph{phase correction}, controlled on $x$ and outcome $1$, to cancel this phase.

To see why MBU is often cheaper than unitary uncomputation, consider its simplest and best-known instantiation, which is for a Toffoli gate --- the quantum analogue of AND\@.
Figure~\ref{fig:mbu-toffoli} shows how when an ancilla holds $x_1x_2 = x_1 \land x_2$, unitary uncomputation executes a Toffoli gate controlled on $x_1$ and $x_2$ ($\oplus$ and Toffoli are self-inverse).
MBU instead performs an $X$-basis measurement and then a single controlled-$Z$ (CZ) gate, when the outcome is $1$, to cancel the phase $(-1){}^{x_1x_2}$.

MBU is cheaper here because on a fault-tolerant quantum computer, a Toffoli gate costs much more than the measurement, $H$, and CZ gates that replace it.
Toffoli is not a \emph{Clifford gate}, meaning that realizing it on a quantum error-correcting code requires decomposing Toffoli into elementary gates including the $T$ gate, which is a major architectural bottleneck on hardware~\citep{beverland-scaling}.
Even after recent advances that substantially reduce the cost of $T$ gates, each $T$ gate is still expected to have cost equivalent to multiple lattice surgery operations on the surface code~\citep{gidney2024}, whereas Clifford gates and single-qubit measurements frequently require none~\citep{litinski-gosc}.
One Toffoli gate costs between four to seven $T$ gates, depending on the decomposition~\cite{Jones_2013,Selinger2013}.
Hadamard and CZ gates are Clifford gates, which cost zero $T$ gates.

Moreover, the CZ gate is needed only half of the time, since with probability \textonehalf{} the measurement yields $0$ and the uncomputation is complete.
By applying MBU to an adder circuit, each of whose ancillas is computed by a Toffoli, \citet{gidney-toffli-mbu} halves the $T$ count of quantum addition.

\subsection{Automating Measurement-Based Uncomputation Soundly and Efficiently}
To date, researchers have applied MBU to individual circuit designs~\cite{windowed-quantum-arithmetic,MBU-QLZOC,MBU-GF2-multiplication}, including by generalizing Gidney's rule (Figure~\ref{fig:mbu-toffoli}) to multiply-controlled-$X$ gates~\citep{amy-ross-rule}.
Quantum programming frameworks such as Q\#~\cite{qsharp}, Qualtran~\cite{qualtran}, and Qrisp~\citep{qrisp} automate MBU for Toffoli gates, via an \texttt{AND} primitive whose uncomputation is optimized as in Figure~\ref{fig:mbu-toffoli} for this three-qubit case.

But to date, no automatic transformation can take an arbitrary circuit, whose ancillas have been computed and uncomputed via the standard approach, and converts its unitary uncomputation into MBU\@.
Such a transformation generalizes the benefit of MBU to circuits beyond those specifically designed and constructed to use it.
In particular, it further enhances the value of existing synthesis methods for unitary uncomputation~\cite{unqomp,reqomp,modular-synthesis,dirty-and-clean-uncomputation}.
Making this transformation possible requires answering two fundamental questions:

\paragraph{Where does MBU apply?}
In complex circuits, either designed by hand or produced by uncomputation synthesizers, a physical ancilla \emph{wire} is typically reused for several logical temporaries, each computed and uncomputed in turn, corresponding to multiple logical ancilla \emph{lifetimes}.
For example, given a 12-bit multiplier, the uncomputation synthesizer Unqomp~\cite{unqomp} maps 144 ancilla lifetimes onto 23 physical ancilla wires.
Each lifetime ends with the uncomputation of its temporary, so a wire offers several MBU opportunities rather than one.
Correctly exploiting these opportunities requires reconstructing the lifetimes of each physical wire from an input circuit.

\begin{wrapfigure}{r}{0.47\textwidth}%
\vspace{-3ex}%
\centering%
\mbox{\small%
\begin{quantikz}[wire types={b,q,n},row sep={0.6cm,between origins},column sep=0.13cm,classical gap=0.025cm,bundle gap=0.06cm,/tikz/commutative diagrams/every matrix/.append style={/tikz/name=mono},/tikz/execute at end picture={\node at (mono-1-5.center) {};\node at (mono-2-5.center) {$\oplus f(x)$};\draw[double,double distance=0.025cm,line width=0.7pt] (mono-3-5.center) -- ($(mono-2-5.center)+(0,-0.3cm)$);\fill[black] (mono-3-5.center) circle (2.2pt);}]
\lstick{$\ket{x}$} & & & & \gate[2]{\vphantom{\oplus f(x)}\hspace{0.70cm}} & & & \rstick{$\ket{x}$} \\
\lstick{\footnotesize $\ket{f(x)}_a$} & \gate{H} & \meter{}\wire[d][1]{c} & \gate{H} & & \gate{H} & \gate{X} & \rstick{$\ket{0}_a$} \\
 & & & \ctrl[vertical wire=c]{-1}\MBUClassicalBegin & & \ctrl[vertical wire=c]{-1} & \ctrl[vertical wire=c]{-1} & \MBUClassicalEnd
\end{quantikz}%
}
\vspace{-2ex}
\caption{Monolithic MBU of an ancilla holding $f(x)$.}
\label{fig:monolithic}
\vspace{-2.5ex}
\end{wrapfigure}
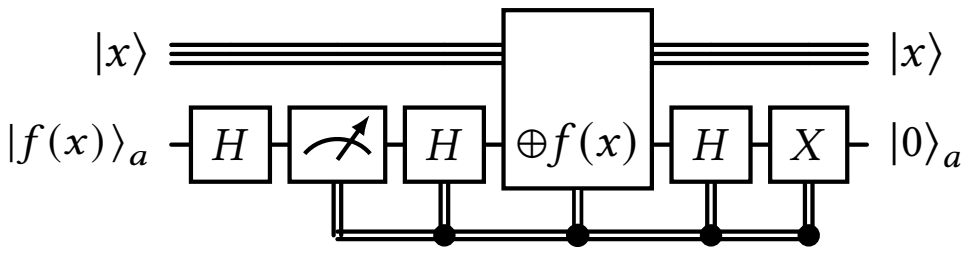
\paragraph{How is MBU realized in gates?}
Fig.~\ref{fig:mbu-intro} specifies that MBU needs to realize $(-1){}^{f(x)}$, but not how it can do so.
For circuits~\cite{gidney-toffli-mbu,MBU-QLZOC} where $f(x)$ is computed by one gate, one can apply Gidney's rule in Fig.~\ref{fig:mbu-toffoli}.
When $f(x)$ needs many gates, \citet[Fig.~24]{MBU-for-modular-arithmetic} propose a different rule, shown in Figure~\ref{fig:monolithic}.
This rule, which we will call \emph{monolithic MBU}, treats the unitary uncomputation circuit for $f(x)$ as an opaque \emph{block}.
In Fig.~\ref{fig:monolithic}, the block corresponds to the $\oplus f(x)$ component from Figure~\ref{fig:mbu-intro}.
Instead of unconditionally executing this block, the rule performs an $X$-basis measurement and then conditionally executes the block.
When the $X$-basis measurement of the ancilla yields $1$, an $H$ gate brings the ancilla to $\ket{-}$, the block kicks back the phase correction $(-1){}^{f(x)}$ onto the data qubits, and subsequent $H$ and $X$ gates restore the ancilla to $\ket{0}$.

Monolithic MBU has numerous shortcomings.
First, only half the time does Figure~\ref{fig:monolithic} reduce the cost of the circuit at all.
Unlike Gidney's rule, it cannot exploit the structure of $f(x)$ to synthesize a phase correction that always has lower $T$ cost.
Second, Figure~\ref{fig:monolithic} is only sound if the $\oplus f(x)$ block exists as a contiguous unit within the circuit, not interspersed with gates that affect other qubits.
If the block included gates that modify the data~$x$, then Fig.~\ref{fig:monolithic} would corrupt the data.
And if $f(x)$ is stored in multiple ancillas, Fig.~\ref{fig:monolithic} fails to uncompute all but one.
Both cases arise in practice.

\paragraph{This Work}
In this paper, we answer the two fundamental questions above and establish the first automatic transformation from an arbitrary quantum circuit containing unitary uncomputation into one that uses measurement-based uncomputation and derives its performance benefits.

We answer the question of \emph{where} to perform MBU in a circuit by presenting a static analysis that recovers the lifetimes of every ancilla via a symbolic execution tracking the basis value on each qubit.
Using the analysis results, we answer \emph{how} to realize MBU by incrementally constructing two circuit transformations to replace unitary uncomputation with MBU for each ancilla lifetime.

The first, \emph{endpoint MBU}, directly automates the rule of \citet{gidney-toffli-mbu} as generalized by \citet{amy-ross-rule}.
It converts the \emph{last} uncomputation gate within each ancilla lifetime into a measurement and an appropriate phase correction gate.
The second transformation, \emph{interval MBU}, further turns a multi-gate unitary uncomputation for each ancilla into an efficient \emph{sequence} of phase correction gates.
Unlike the monolithic MBU of \citet{MBU-for-modular-arithmetic}, it replaces unitary uncomputation while guaranteeing soundness and minimizing $T$-gate cost.
In Section~\ref{sec:overview}, we walk through examples to illustrate the challenges and insights behind the static analysis and both transformations.

We integrate our analyses and transformations into a tool, \textsf{MBUndo}, that programmers can use to automatically convert circuits to correct and efficient measurement-based uncomputation.
\textsf{MBUndo} requires no user annotations on input circuits other than which qubits are ancillas -- the same as required by existing uncomputation synthesizers such as Unqomp~\citep{unqomp} and Reqomp~\citep{reqomp} -- and can operate as a compiler pass on their output circuits.

We evaluate \textsf{MBUndo} using 22 benchmarks drawn from uncomputation synthesizers and quantum algorithms.
Endpoint MBU reduces $T$ gates over Reqomp by 36.5\% in geomean.
Interval MBU further reduces $T$ gates by another 2.3 to 18.4 percentage points for benchmarks containing multi-gate uncomputation.
Moreover, \textsf{MBUndo} automatically matches 3 of 4 evaluated specialized MBU circuits from prior literature, and processes each benchmark within 43 ms on a laptop.

\paragraph{Contributions}
In this work, we present the following contributions:
\begin{itemize}
\item[\S\ref{sec:symbolic}:] A static analysis to identify the multiple logical lifetimes that exist on each physical ancilla in a circuit, maximizing opportunities to introduce measurement-based uncomputation.
\item[\S\ref{sec:transform-endpoint}:] Endpoint MBU, a simple transformation that soundly automates the rule of \citet{gidney-toffli-mbu} by converting the last uncomputation gate in each ancilla lifetime into a measurement.
\item[\S\ref{sec:transform-interval}:] Interval MBU, a more advanced transformation that further converts a multi-gate unitary uncomputation for each ancilla into an efficient sequence of phase correction gates.
\item[\S\ref{sec:impl}:] \textsf{MBUndo}, a tool that integrates the analyses and transformations into a compiler pass that automatically enhances the output circuits of existing uncomputation synthesizers.
\item[\S\ref{sec:eval}:] An empirical evaluation of \textsf{MBUndo}'s performance on 22 benchmarks against uncomputation synthesizers, MBU circuits from prior work, and state-of-the-art circuit optimizers.
\end{itemize}

This work introduces into quantum compilers a powerful technique previously accessible only to expert circuit designers. As quantum programs become more complex and fault-tolerant devices come online, our contributions lower the barrier to practically executing them on hardware.

\section{Overview and Examples}
\label{sec:overview}

In this section, we explain the problem of transforming unitary uncomputation into measurement-based uncomputation.
Using several example circuits, we illustrate the static analysis to determine ancilla \emph{lifetimes} (Sec.~\ref{sec:informal-analysis}) and the \emph{endpoint MBU} transformation that soundly converts the final uncomputation gate in every ancilla lifetime to measurement (Sec.~\ref{sec:informal-endpoint}).
We present the limitations of this method and associated challenges.
Then, we present the \emph{interval MBU} transformation that converts complex gate sequences to MBU (Sec.~\ref{sec:informal-interval}).
We deliberately keep all examples simple, but they preview the structure of the practical applications that we evaluate in Section~\ref{sec:eval}.

\subsection{Static Analysis for Ancilla Lifetimes: Discovering MBU Opportunities}
\label{sec:informal-analysis}

We consider the circuit in Figure~\ref{fig:basic-example}, which implements one round of a Feistel cipher~\cite{feistel-round-example}, the structure underlying block ciphers such as DES\@.
Given a function $F$, a round mixes half of its state into the other, mapping $(y,t)$ to $(t\oplus F,\,y)$.
In this circuit, $F=y\oplus xy\oplus xyz$, and the implementation uses an ancilla wire $a$ as temporary scratch space to compute $F$.

\begin{figure}[b]
\centering
\begin{tabular}[c]{@{}c@{}}
\begin{quantikz}[row sep={0.435cm,between origins},column sep={0.8cm,between origins},/tikz/commutative diagrams/every matrix/.append style={/tikz/name=coherent},/tikz/execute at end picture={\MBULifetimeX{coherent}{5}{(coherent-col2.west)}{($(coherent-col4.center)!0.5!(coherent-col5.center)+(-1pt,0)$)}{MBULifeOne}{}\MBULifetimeX{coherent}{5}{($(coherent-col4.center)!0.5!(coherent-col5.center)+(1pt,0)$)}{(coherent-col9.east)}{MBULifeTwo}{}\MBUNumberSteps{coherent}{8}\MBUValueLabel{coherent}{5}{2}{3}{$\patom{q}$}\MBUValueLabel{coherent}{5}{3}{4}{$\patom{q}$}\MBUValueLabel{coherent}{5}{4}{5}{$0$}\MBUValueLabel{coherent}{5}{5}{6}{$\patom{p}$}\MBUValueLabel{coherent}{5}{6}{7}{$\patom{p}$}\MBUValueLabel{coherent}{5}{7}{8}{$\patom{p}$}\MBUValueLabel{coherent}{5}{8}{9}{$\patom{p}$}\draw[-{Stealth[length=4pt,width=3.5pt]},black!60,semithick] ([yshift=-15pt]coherent-5-9.center) -- ([yshift=-15pt]coherent-5-2.center) node[midway,below=2pt,font=\scriptsize,text=black!70,inner sep=0pt] {backward symbolic execution};}]
\lstick{$x$} &[-0.45cm,between origins] \MBUCtrl{1} & & \MBUCtrl{1} & \MBUCtrl{1} & & & & \MBUCtrl{2} &[-0.45cm,between origins] \rstick{$x'$} \\
\lstick{$y$} & \MBUCtrl{3} & & \MBUCtrl{3} & \MBUCtrl{1} & & \MBUTarget{} & \MBUCtrl{2} & & \rstick{$y'$} \\
\lstick{$z$} & & & & \MBUCtrl{2} & & & & \MBUCtrl{1} & \rstick{$z'$} \\
\lstick{$t$} & & \MBUTarget{} & & & \MBUTarget{} & \MBUCtrl{-2} & \MBUTarget{} & \MBUCtrl{1} & \rstick{$t'$} \\
\lstick{$a$} & \MBUTarget{} & \MBUCtrl{-1} & \MBUTarget{} & \MBUTarget{} & \MBUCtrl{-1} & & & \MBUTarget{} & \rstick{$\ket{0}$}
\end{quantikz}
\end{tabular}%
\hspace{0.5cm}%
{\scriptsize\setlength{\tabcolsep}{3pt}\renewcommand{\arraystretch}{1.11}%
\begin{tabular}[c]{@{}c|ccc@{\hspace{4pt}}c@{}}
\multicolumn{1}{c}{position} & $y$ & $t$ & $a$ & \\
\cline{1-4}
post-circuit & $y'$ & $t'$ & \textbf{0} & \MBUTableStripLower{MBULifeTwo} \\
gate 7--8 & \textcolor{gray}{$y'$} & \textcolor{gray}{$t'$} & $\patom{p}$ & \MBUTableStrip{MBULifeTwo} \\
gate 6--7 & \textcolor{gray}{$y'$} & $\textcolor{gray}{t'}\oplus y'$ & \textcolor{gray}{$\patom{p}$} & \MBUTableStrip{MBULifeTwo} \\
gate 5--6 & $t'$ & \textcolor{gray}{$t'\oplus y'$} & \textcolor{gray}{$\patom{p}$} & \MBUTableStrip{MBULifeTwo} \\
gate 4--5 & \textcolor{gray}{$t'$} & $\textcolor{gray}{t'\oplus y'}\oplus \patom{p}$ & \textcolor{gray}{$\patom{p}$} & \MBUTableStrip{MBULifeTwo} \\
gate 3--4 & \textcolor{gray}{$t'$} & \textcolor{gray}{$t'\oplus y'\oplus \patom{p}$} & \textbf{0} & \rlap{\MBUTableStripUpper[1pt]{MBULifeTwo}}\MBUTableStripLower[1pt]{MBULifeOne} \\
gate 2--3 & \textcolor{gray}{$t'$} & \textcolor{gray}{$t'\oplus y'\oplus \patom{p}$} & $\patom{q}$ & \MBUTableStrip{MBULifeOne} \\
gate 1--2 & \textcolor{gray}{$t'$} & $\textcolor{gray}{t'\oplus y'\oplus \patom{p}}\oplus \patom{q}$ & \textcolor{gray}{$\patom{q}$} & \MBUTableStrip{MBULifeOne} \\
pre-circuit & \textcolor{gray}{$t'$} & \textcolor{gray}{$t'\oplus y'\oplus \patom{p}\oplus \patom{q}$} & \textbf{0} & \MBUTableStripUpper{MBULifeOne} \\
\end{tabular}}

\vspace{2mm}
{\small
\textcolor{MBULifeOne}{\rule{10pt}{4pt}}\ \ First lifetime of $a$, holding $\patom{q} = x't'$.\qquad
\textcolor{MBULifeTwo}{\rule{10pt}{4pt}}\ \ Second lifetime of $a$, holding $\patom{p} = x'z't'$.\par\smallskip
}
\vspace{-2ex}
\caption{Example circuit demonstrating lifetime analysis. Left: the input circuit,
with $a$'s symbolic values and lifetimes inferred by our
backward symbolic execution. Right: how symbolic values of $y$, $t$, and
$a$ are updated.}
\label{fig:basic-example}
\end{figure}

Converting this circuit to MBU requires replacing with measurements the \emph{multiply-controlled $X$} ($\mathrm{C}^k \mathrm{X}$;\ generalized Toffoli) gates that uncompute $a$ back to zero.
The rule of \citet{gidney-toffli-mbu} applies to the final gate~8: its output leaves $a$ as zero, matching the rule in Figure~\ref{fig:mbu-toffoli} so that gate~8 can be soundly replaced by a measurement followed by an appropriate controlled-phase gate.

\paragraph{Lifetimes}
Given the circuit as written, Gidney's rule replaces gate~8 only.
A second opportunity for MBU hides in the fact that Figure~\ref{fig:basic-example} uses $a$ to store two distinct temporaries in turn.
Specifically, gate~1 computes $\textsf{q} = x \land y = xy$ into $a$, gate~3 uncomputes $\textsf{q} = xy$, gate~4 computes $\textsf{p} = x y z$, and gate~8 uncomputes $\textsf{p} = x' z' t'$ (a different formula because $y$ is modified in the output).

Identifying that gate~3 also leaves $a$ as zero yields a second candidate for MBU, but this fact is invisible in Figure~\ref{fig:basic-example}.
Thus, in Section~\ref{sec:symbolic} we present a static analysis that, given a circuit, soundly identifies MBU candidates for each distinct \emph{lifetime} of a temporary value on each ancilla wire.
This analysis requires no programmer input aside from which wires are zero when the circuit ends, i.e., which wires are ancillas.
From this information, it performs a backward symbolic execution to find a conservative set of positions where ancillas are cleaned up to zero and are thus eligible for MBU\@.

\paragraph{Efficient Symbolic Values}
To avoid the exponential blowup of explicit quantum states, our static analysis tracks the value of each qubit as a \emph{symbolic value}, a Boolean function of data qubits.
We define a symbolic value as a set of \emph{atoms} joined by XOR ($\oplus$).
The empty set represents $0$, and a cheap symmetric difference of sets joins two values.
Each atom is either a \emph{symbol} for an unknown Boolean or a \emph{product} of multiple symbolic values via $\land$.
Our design is more compact than algebraic normal form (ANF)~\cite{zhegalkin1927} as used by prior symbolic executors for quantum circuits~\cite{CaretteANF}.
It exploits usual algebraic laws of $\oplus$ and $\land$, such as $s \,{\oplus}\, s \,{=}\, 0$ and $s \,{\land}\, s \,{=}\, ss \,{=}\, s$, but not the general distributivity of $\land$ over $\oplus$.
Its number of atoms is linear in the number of gates, whereas ANF expands $(x \,{\oplus}\, y)(z \,{\oplus}\, w) \mapsto xz \,{\oplus}\, xw \,{\oplus}\, yz \,{\oplus}\, yw$ and is exponentially large in general.

\paragraph{Symbolic Execution}
Our analysis starts from the end of the circuit, where the symbolic value of every ancilla is $0$ and every data qubit has a fresh symbol for its unknown output (e.g., $y'$ and $t'$ for $y$ and $t$).
It processes each gate from right to left.
For a $\mathrm{C}^k \mathrm{X}$ gate, it XORs the product of symbolic values of the controls into the target; for a phase gate ($Z$, $S$, $T$), it leaves symbolic values unchanged; and for a basis-changing gate such as $H$, it conservatively replaces the target with a fresh symbol.
Finally, the analysis delimits each lifetime of an ancilla by where its symbolic value is $0$.

For the example, Fig.~\ref{fig:basic-example} lists the symbolic values after each analysis step.
Gate~8 XORs its controls $x$, $z$, and $t$ into $a$, which then stores the atom $\patom{p} = x'z't'$.
The CNOT gates~7 and~6 on $y$ and $t$ yield $t=t'\oplus y'$ and then $y=t'$.
Gate~5 XORs $a = \patom{p}$ into $t$.
Gate~4 XORs its controls $x$, $y$, and $z$ into $a$; since $y = t'$, $xyz = x't'z' = \patom{p}$ and the state of $a$ becomes $a=\patom{p}\oplus \patom{p}=0$.
The analysis thus identifies a lifetime spanning gates 4--8, and continues on to identify a second lifetime from gates 1--3.

\begin{figure}[b]
\vspace{-0.7ex}
\centering
\resizebox{\textwidth}{!}{\input{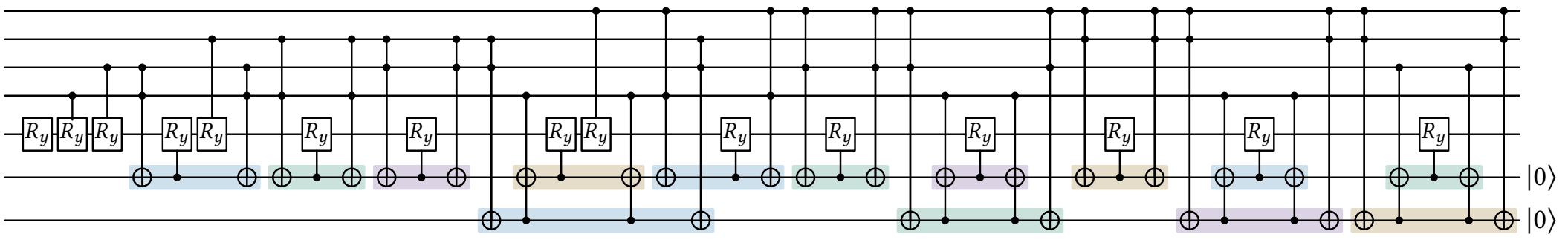}}
\vspace{-4.85ex}
\caption{The 14 ancilla lifetimes identified by our static analysis on the polynomial Pauli rotation circuit on five data qubits and two ancilla wires with uncomputation synthesized by Unqomp~\cite{unqomp}.}
\label{fig:lifetime-overview}
\end{figure}

\paragraph{Result}
Our static analysis soundly identifies, given an arbitrary circuit annotated only by which qubits are ancillas, a conservative set of lifetimes of distinct temporary values on each ancilla that each expose an opportunity for MBU to the transformations we present in the next section.
Beyond the small example in Figure~\ref{fig:basic-example}, our analysis can scalably handle more complex circuits.
For instance, Figure~\ref{fig:lifetime-overview} shows the 14 distinct lifetimes identified by our analysis on the {polynomial Pauli rotation} benchmark circuit constructed by \citet{unqomp} in order to evaluate Unqomp.

\subsection{Endpoint MBU:\ Soundly Applying Gidney's Rule to Final Gates}
\label{sec:informal-endpoint}

\begin{figure}[t]
\centering
\MBUFitWidth{%
\begin{quantikz}[row sep={0.435cm,between origins},column sep={0.58cm,between origins},/tikz/commutative diagrams/every matrix/.append style={/tikz/name=coherent},baseline={(coherent-1-1.center)},/tikz/execute at end picture={\MBULifetimeX{coherent}{5}{(coherent-col2.west)}{($(coherent-col4.center)!0.5!(coherent-col5.center)+(-1pt,0)$)}{MBULifeOne}{}\MBULifetimeX{coherent}{5}{($(coherent-col4.center)!0.5!(coherent-col5.center)+(1pt,0)$)}{(coherent-col9.east)}{MBULifeTwo}{}\MBUNumberSteps{coherent}{8}}]
\lstick{$x$} &[-0.30cm,between origins] \MBUCtrl{1} & & \MBUCleanupCtrl{1} & \MBUCtrl{1} & & & & \MBUCleanupCtrl{2} &[-0.30cm,between origins] \\
\lstick{$y$} & \MBUCtrl{3} & & \MBUCleanupCtrl{3} & \MBUCtrl{1} & & \MBUTarget{} & \MBUCtrl{2} & & \\
\lstick{$z$} & & & & \MBUCtrl{2} & & & & \MBUCleanupCtrl{1} & \\
\lstick{$t$} & & \MBUTarget{} & & & \MBUTarget{} & \MBUCtrl{-2} & \MBUTarget{} & \MBUCleanupCtrl{1} & \\
\lstick{$a$} & \MBUTarget{} & \MBUCtrl{-1} & \MBUCleanupTarget & \MBUTarget{} & \MBUCtrl{-1} & & & \MBUCleanupTarget & \rstick{$\ket{0}$}
\end{quantikz}%
\hspace{-0.3cm}%
\raisebox{\dimexpr-0.5\height-0.87cm\relax}{\includegraphics[width=1.25cm]{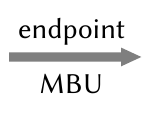}}%
\hspace{0.05cm}%
\begin{quantikz}[row sep={0.435cm,between origins},column sep={0.48cm,between origins},classical gap=0.025cm,wire types={q,q,q,q,q,n},/tikz/commutative diagrams/every matrix/.append style={/tikz/name=measured},baseline={(measured-1-1.center)},/tikz/execute at end picture={\MBULifetime{measured}{5}{2}{5}{MBULifeOne}{}\MBULifetime{measured}{5}{8}{13}{MBULifeTwo}{}\MBUOutcomeLabelLeft{measured}{6}{5}{7}{$m_1$}\MBUOutcomeLabelLeft{measured}{6}{13}{15}{$m_2$}\foreach \mbuCol/\mbuNum in {2/1,3/2,8/4,9/5,10/6,11/7}{\MBUNumberAt{measured}{\mbuCol}{\mbuNum}}\MBUNumberCorrection{measured}{7}{3}\MBUNumberCorrection{measured}{15}{8}}]
\lstick{$x$} &[-0.30cm,between origins] \MBUCtrl{1} & & &[\MBUHMExtraGap,between origins] &[\MBUResetExtraGap,between origins] &[-0.12cm,between origins] \MBUPaymentCtrl{1} & \MBUCtrl{1} & & & & &[\MBUHMExtraGap,between origins] &[\MBUResetExtraGap,between origins] &[-0.12cm,between origins] \MBUPaymentCtrl{2} &[-0.30cm,between origins] \\
\lstick{$y$} & \MBUCtrl{3} & & & & & \MBUPaymentDot & \MBUCtrl{1} & & \MBUTarget{} & \MBUCtrl{2} & & & & & \\
\lstick{$z$} & & & & & & & \MBUCtrl{2} & & & & & & & \MBUPaymentCtrl{1} & \\
\lstick{$t$} & & \MBUTarget{} & & & & & & \MBUTarget{} & \MBUCtrl{-2} & \MBUTarget{} & & & & \MBUPaymentDot & \\
\lstick{$a$} & \MBUTarget{} & \MBUCtrl{-1} & \MBUH & \MBUMeasure{1} & \MBUReset & & \MBUTarget{} & \MBUCtrl{-1} & & & \MBUH & \MBUMeasure{1} & \MBUReset\setwiretype{n} & & \\
 & & & & & \MBUClassicalBegin & \MBUClassicalCtrl{-4} & \MBUClassicalEnd & & & & & & \MBUClassicalBegin & \MBUClassicalCtrl{-2} & \MBUClassicalEnd
\end{quantikz}%
}
{\small
\textcolor{MBULifeOne}{\rule{10pt}{4pt}} First lifetime of $a$.\quad
\textcolor{MBULifeTwo}{\rule{10pt}{4pt}} Second lifetime of $a$.\quad
\textcolor{MBUCleanup}{Purple}: unitary uncomputation.\quad
\textcolor{MBUPayment}{Brown}: phase correction.\par\smallskip
}
\vspace{-2.2ex}
\caption{Endpoint MBU applied to the circuit of Fig.~\ref{fig:basic-example}.
Left: the input circuit. Right: the transformed circuit.}
\vspace*{-1.25ex}
\label{fig:endpoint-example}
\end{figure}

Given lifetimes of each ancilla, one direct next step is to use Gidney's rule at the end of each lifetime, as done manually by~\citet{gidney-toffli-mbu,MBU-QLZOC}.
In Section~\ref{sec:transform-endpoint}, we describe \emph{endpoint MBU}, a circuit transformation that automates this rule in a simple and sound way.
Figure~\ref{fig:endpoint-example} shows its result on the Feistel example from Figure~\ref{fig:basic-example}, whose two unitary uncomputation gates are in \textcolor{MBUCleanup}{purple}.

Endpoint MBU converts the last gate of each lifetime into a measurement and phase correction.
\citet{amy-ross-rule} define the appropriate correction for any $\mathrm{C}^k \mathrm{X}$ gate, which generalizes Fig.~\ref{fig:mbu-toffoli} into the correction $\mathrm{C}^{k-1}\mathrm{Z}$ gate over the $k$ controls.
The $k=2$ case corresponds to Figure~\ref{fig:mbu-toffoli}, whereas in general, MBU uses one control
fewer than the unitary uncomputation gate being replaced.

This transformation reduces the circuit's $T$ count when it replaces $\mathrm{C}^k \mathrm{X}$ with $\mathrm{C}^{k-1} \mathrm{Z}$ gates.
In Figure~\ref{fig:endpoint-example}, the \textcolor{MBUCleanup}{purple} uncomputation gates CCX (gate~3) and CCCX (gate~8) are replaced by $X$-basis measurements followed by CZ and CCZ corrections in \textcolor{MBUPayment}{brown}.
CZ is Clifford and thus costs no $T$ gates, while CCZ differs from CCX only by Hadamard conjugation on its target and can therefore use the same decomposition as CCX, with equal $T$ count~\cite{Selinger2013}.
Even in the worst case when $m_1=m_2=1$, both corrections together cost no more than the original CCX uncomputation gate alone, thereby eliminating the entire $T$ cost of the CCCX unitary uncomputation gate.

\paragraph{Limitations}
Endpoint MBU is most effective when each lifetime is uncomputed by a single gate, like the single Toffoli in Figure~\ref{fig:mbu-toffoli}.
But in more complex circuits, an ancilla is often uncomputed by a sequence of gates, and uncomputing one ancilla can require using another as a control.

\begin{figure}[t]
\centering
\includegraphics[width=0.91\textwidth]{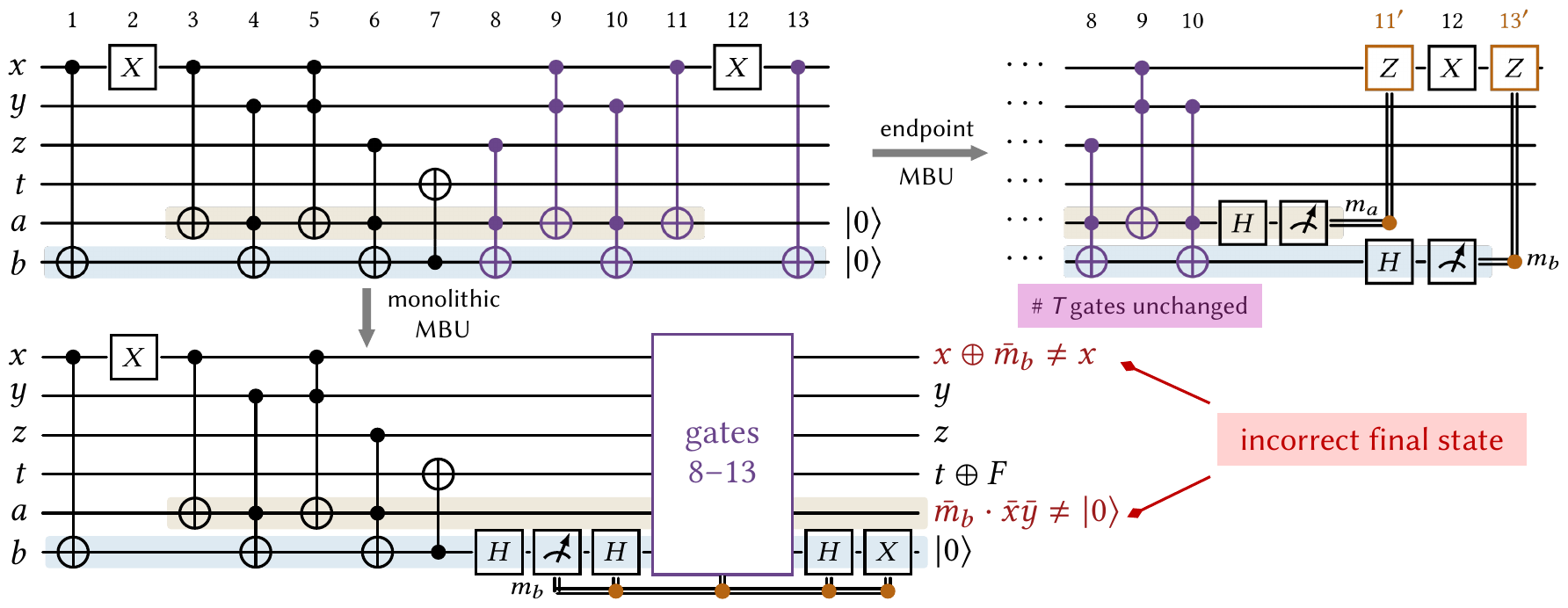}
\vspace{-2.25ex}%
\setlength{\belowcaptionskip}{-1.75ex}%
\caption{A circuit that computes three-bit OR, for which endpoint MBU (top right) yields zero \emph{T}-gate savings, while monolithic MBU~\cite{MBU-for-modular-arithmetic} (bottom) yields unsound and overall inefficient results.}
\label{fig:challenge}
\label{fig:challenge-example}
\label{fig:monolithic-attempt}
\end{figure}

Figure~\ref{fig:challenge-example} depicts a circuit for $t\mapsto t\oplus F$ where $F=x\vee y\vee z$, accumulated from three sub-cases $F=x\oplus\bar{x}y\oplus\bar{x}\bar{y}z$.
Gates~1--6 compute $F$ into the ancilla $b$, where the ancilla $a$ holds $\bar{x}$ (gate~3) and then $\bar{x}\bar{y}$ (gate~5).
Gate~7 adds $F$ to $t$, and gates~8--13 uncompute $b$ and $a$.
The top right of Figure~\ref{fig:challenge-example} shows the result of endpoint MBU, which replaces one CNOT per lifetime: gate~13 by measuring $b$ and correcting $Z(x){}^{m_b}$, and gate~11 by measuring $a$ and correcting $Z(x){}^{m_a}$.
But since gates~13 and~11 are Clifford, the result saves zero $T$ gates, while the Toffoli gates~8, 9, and~10 remain.

\paragraph{Challenges to Generalization}
Efficiently handling ancillas uncomputed by multiple gates or gates that depend on other ancillas requires a more generalized rule for MBU\@.
The candidate in literature is the rule of \citet{MBU-for-modular-arithmetic}, which we presented in Fig.~\ref{fig:monolithic} under the name \emph{monolithic MBU}.
The bottom of Fig.~\ref{fig:challenge-example} applies this rule to the three-bit OR circuit, revealing two fundamental issues:

\begin{itemize}[itemsep=1pt, topsep=1pt,leftmargin=1.5em]
\item \emph{Unsoundness}. Monolithic MBU conditionally applies the gates of unitary uncomputation, which in Figure~\ref{fig:challenge-example} are gates~8--13 in \textcolor{MBUCleanup}{purple}. This is unsound when the circuit intersperses modification of the data (gate~2) with computation of the ancillas. Indeed, $x$ is left flipped with \textonehalf{} probability.
Moreover, monolithic MBU measures only one ancilla but conditionally corrects all ancillas. In Figure~\ref{fig:challenge-example}, the result is that the unmeasured ancilla $a$ is left in a garbled nonzero state. And if it measured $a$, then the phase correction would be wrong for $b$ as gates 8 and 10 depend on $a$.
\item \emph{Inefficiency}. Monolithic MBU never decreases the $T$ cost of phase correction. Therefore, with \textonehalf{} probability, gates 8--13 execute in full and cost as much as the unitary uncomputation.
\end{itemize}

\subsection{Interval MBU:\ Soundly and Efficiently Transforming Gate Sequences}
\label{sec:informal-interval}

\begin{figure}[t]
\centering
\includegraphics[width=\textwidth]{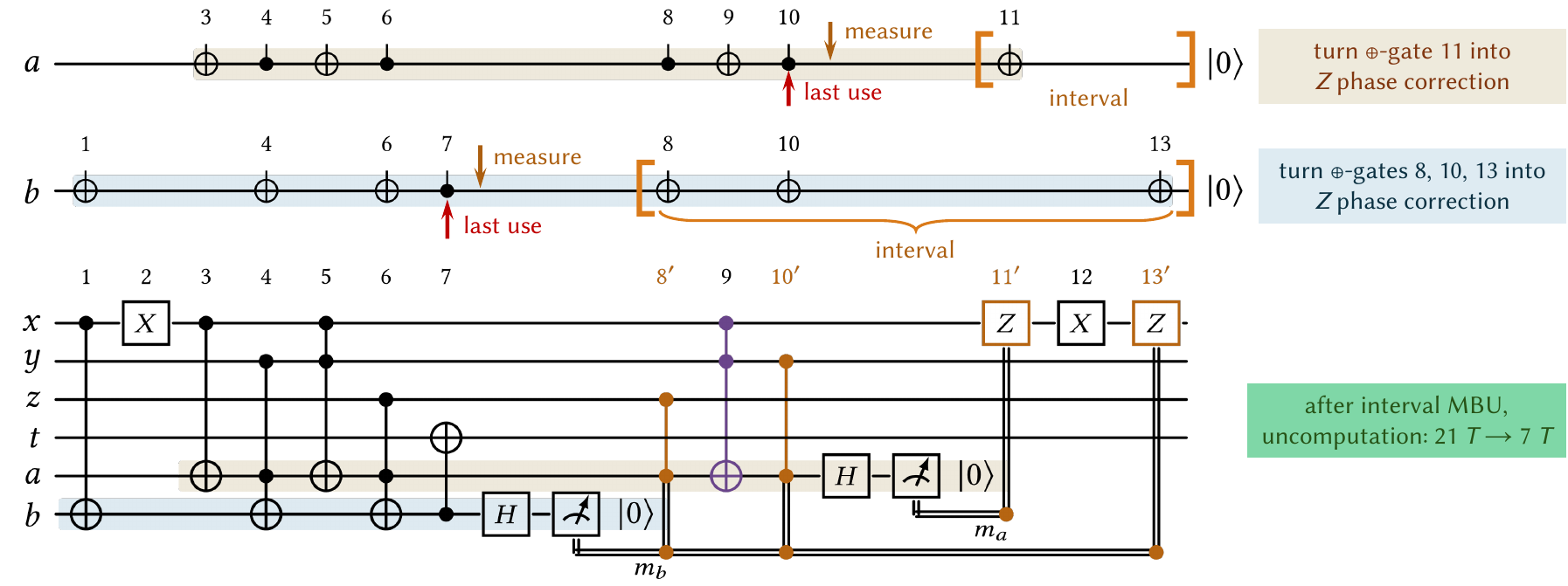}
\vspace{-4ex}
\caption{Interval MBU applied to the three-bit OR circuit of Fig.~\ref{fig:challenge-example}.
Top: the \emph{isolated lifetimes} of uses ($\bullet$) and writes ($\oplus$) of $a$ and $b$; each ancilla is measured right after its \emph{last use}, and the following writes form its \emph{interval}.
Bottom: the resulting circuit, in which each $\oplus$-gate of its interval is turned into a $Z$-rotation gate.}
\label{fig:interval-example}
\vspace{-2ex}
\end{figure}

In Section~\ref{sec:transform-interval}, we present \emph{interval MBU}, the first circuit transformation that soundly and efficiently converts general sequences of multiple unitary uncomputation gates on multiple ancillas into MBU\@.
We demonstrate its key insights in Figure~\ref{fig:interval-example} using the three-bit OR circuit from Figure~\ref{fig:challenge-example}.

Fundamentally, monolithic MBU fails because viewing unitary uncomputation as a contiguous block leads to that block entangling the measured ancilla with every other qubit it touches.
Instead of marking such blocks, interval MBU considers the lifetime of an ancilla in isolation from the rest of the circuit.
Figure~\ref{fig:interval-example} depicts these \emph{isolated lifetimes} of $a$ and $b$ derived from the input circuit.

Given the isolated lifetime of an ancilla, any dependencies on other ancillas and data qubits can be considered independently of the decisions on this ancilla.
Interval MBU is a policy for where to measure the ancilla and how to correct for the effect of each gate after the measurement:

\paragraph{Measure After Last Use}
To find the earliest point to measure an ancilla, one idea is to identify the start of unitary uncomputation for that ancilla, which in Figure~\ref{fig:challenge} is the first \textcolor{MBUCleanup}{purple $\oplus$} in its lifetime, located at gate~8 for $b$ and gate~9 for $a$.
But this choice is unsound: measuring $a$ at gate~9 destroys its value, and prevents gate~10 from continuing the uncomputation of $b$ using $a$ as a control.

Instead, the safe time to measure is right after the last gate that \emph{uses} the ancilla in its isolated lifetime in Fig.~\ref{fig:interval-example} --- reads it as a control $\bullet$ ($Z$) or applies a gate other than $\oplus$ ($X$) such as $H$ to it.

We call the suffix of an ancilla's lifetime after its last use, which only contains $\oplus$-gates, its \emph{interval}.
For $b$, the last use is gate~7 and the interval is the gates~8, 10, and~13; for $a$, the last use is gate~10, and the interval consists of gate~11.
After its last use, the only remaining operations on an ancilla are writes by $\oplus$-gates that restore it to zero, so measuring it cannot disturb any
other qubit.

\paragraph{Correct in Place}
Interval MBU measures $b$ at the start of its interval, after gate~7, and synthesizes the phase correction for each gate in a simple way.
Each $\oplus$ constituting the interval disappears from the circuit, leaving a (potentially controlled) $Z$ gate spanning its old control qubits and classically controlled by the measurement outcome $m_b$ --- like in endpoint MBU, but sound for a sequence of gates by the interval guarantee.
Nothing else in the circuit changes, including irrelevant intervening gates such as gates~9 and~12 that do not touch $b$.
The bottom of Figure~\ref{fig:interval-example} presents the result.

\paragraph{Result}
Interval MBU effectively reduces the $T$ cost of the three-bit OR circuit.
Whereas endpoint MBU replaces gates~11 and~13, interval MBU replaces gates~8, 10, 11, and~13 and retains only gate~9 for uncomputation.
Counting seven $T$ gates per Toffoli, unitary uncomputation (gates~8--13) costs $3 \times 7 = 21$ $T$ gates, endpoint MBU replaces only CNOTs and thus costs the same, and interval MBU leaves a single Toffoli, or $7$ $T$ gates, with all phase corrections being Clifford CZ or $Z$ gates.

\section{Static Analysis for Ancilla Lifetimes}
\label{sec:symbolic}

In this section, we formalize \textsf{MBUndo}'s static analysis for ancilla lifetimes --- the backward symbolic execution introduced informally in Section~\ref{sec:informal-analysis}.
The input of the analysis is a circuit $C=g_1\cdots g_n$ over unitary gates $g_i$ and a set $Q$ of qubits including the subset $A \,{\subsetneq}\, Q$ of \emph{ancillas}, qubits promised to be $\ket{0}$ at the end of $C$.
The output is, for every ancilla $a \in A$,
the circuit positions where $a$ is found to hold $\ket{0}$,
which we call its \emph{zero points}; they cut
the physical wire of the ancilla into \emph{lifetimes}.

\paragraph{Definitions}
The notations used in the analysis are defined below.
We refer to the point in the circuit after gate $g_i$ as position
$i$, so that position $0$ is the beginning of $C$ and position $n$ is
its end.
We write $\sigma_i$ for the \emph{symbolic state} at
position $i$, which maps each qubit to its \emph{symbolic value}.

\begingroup\medskip
\centering\small
\begin{tabular}{@{}rl@{\qquad}rl@{}}
qubits & $q,t,c\in Q$ & ancillas & $a\in A\subseteq Q$ \\
gates & $g_1\cdots g_n$ & positions & $0\le i\le n$ immediately after $g_i$ \\
$\qubits(g)$ & the qubits that $g$ acts on & $\written(g)$ & the target qubit(s) of $g$ \\
symbols & $s\in\Sym$ & values & $V,W\in\Val=\mathcal{P}(\Atom) \ni \emptyset$ \\
atoms & $\alpha\in\Atom=\{\one\}\cup\Sym\cup\{\prodatom{O}\}$ & product atoms & $\prodatom{O}$, $O\subseteq\Val$, $|O|\ge 2$ \\
symbolic states & $\sigma:Q\to\Val$, $\sigma_i$ at position $i$ & product table & $\Pi:\mathcal{P}(\Val)\rightharpoonup\Atom$ \\
\end{tabular}
\par\medskip
\endgroup

We first present the main algorithm of the backward symbolic execution
and its transfer rules, and then detail the symbolic values and how they
are updated during the execution.

\begin{wrapfigure}{r}{0.41\textwidth}
\vspace{-2.5ex}
\begin{algorithm}[H]
\caption{Backward symbolic execution.}
\label{alg:analysis}
\SetKwInOut{Input}{In}\SetKwInOut{Output}{Out}
\SetKw{KwDownTo}{downto}
\Input{circuit $C=g_1\cdots g_n$, ancillas $A\subseteq Q$}
\Output{$\mathcal{Z}(a)\subseteq\{0,\ldots,n\}$ for each $a\in A$}
$\sigma(a)\gets\emptyset$ for each $a\in A$\;
$\sigma(q)\gets\{\fresh\}$ for each $q\in Q\setminus A$\;
$\mathcal{Z}(a)\gets\{n\}$ for each $a\in A$\;
\For{$i\gets n$ \KwDownTo $1$}{
  $\sigma\gets\Step(\sigma,g_i)$\;
  \ForEach{$a\in A\cap\qubits(g_i)$}{\If{$\sigma(a)=\emptyset$}{$\mathcal{Z}(a)\gets \mathcal{Z}(a)\cup\{i-1\}$\;}}
}
\Return $\mathcal{Z}$\;
\end{algorithm}
\vspace{-2ex}
\end{wrapfigure}%
\paragraph{Main Algorithm}
Given input circuit $C$ and ancillas $A$, Algorithm~\ref{alg:analysis}
outputs the set $\mathcal{Z}(a)$ of zero points of every ancilla $a$,
the positions where $a$ is established to hold~$0$.
Conservatively, the gates between two consecutive zero points form a lifetime where $a$ holds a temporary value.
In Figure~\ref{fig:basic-example} from the overview, the zero points of $a$
are $0$, $3$, and $8$, which yield the two shaded lifetimes.

Alg.~\ref{alg:analysis} starts from the end of the circuit $C$.
By definition of $A$, every ancilla holds $0$ as denoted by an empty symbolic value $\emptyset$ (line~1), and every data qubit holds an unknown final value denoted by a fresh symbol (line~2).
Thus, $n$ is a zero point of every ancilla (line~3).
The main loop (lines~4--8) processes the gates from $g_n$ down to $g_1$, computing
$\sigma_{i-1}=\Step(\sigma_i,g_i)$ by the transfer rules that we define below. After each
gate $g_i$ that touches $a$, the algorithm examines the value of $a$: if $\sigma_{i-1}(a)=\emptyset$, then $a$ is established to
hold $0$ immediately before $g_i$, and position $i-1$ is recorded as a zero point (lines~6--8).

\paragraph{Transfer Rules}
The heart of the algorithm is the $\Step$ function, defined as follows:

\begingroup\medskip
\centering\small
\begin{tabular}{@{}l@{\quad}l@{\quad}l@{}}
\rulename{Xor} & $g=\XORg(t;\ell_1,\ldots,\ell_k)$ & $\Step(\sigma,g)=\sigma[t\mapsto\sigma(t)\symdiff\Product(\sigma(\ell_1),\ldots,\sigma(\ell_k))]$ \\[2pt]

\rulename{Diag} & $g$ is diagonal & $\Step(\sigma,g)=\sigma$ \\[2pt]
\rulename{Forget} & otherwise & $\Step(\sigma,g)=\sigma[q\mapsto\{\fresh\}\text{ for each }q\in\written(g)]$ \\
\end{tabular}
\par\medskip
\endgroup

Rule \rulename{Xor} handles a gate $\XORg(t;\ell_1,\ldots,\ell_k)$, where $k\ge 0$, that XORs the product of
its controls into its target $t$ and leaves controls unchanged.
Such gates include $X$ and $\mathrm{C}^k \mathrm{X}$ gates with arbitrary control polarities, which dominate the circuits we study.
Each control $\ell_j$ is either positive $c_j$ or negative $\bar{c}_j$ on a qubit $c_j\neq t$.
Here, $\symdiff$ is the symmetric difference of sets, $\sigma(q)$ is the symbolic value of qubit $q$, and $\sigma(\bar{c})=\sigma(c)\symdiff\{\one\}$ is that of a negative control.
$\Product$ computes the symbolic value of a product ($\land$) of the controls via several $\symdiff$ operations; it is defined below in Alg.~\ref{alg:product}.

Rule \rulename{Diag} handles a diagonal gate, e.g., $Z$, $S$, $T$, a phase rotation, or any controlled version of these.
This rule leaves the state unchanged because diagonal gates change no basis values.
Finally, rule \rulename{Forget} conservatively handles any other gate,
including those that exit the computational basis such as $H$ and $Y$
rotations, by replacing the value of every qubit it writes by a fresh
symbol recording that nothing is known about the value of that
qubit before the unknown gate.

The \rulename{Xor} rule recognizes gates with one distinguished target qubit and zero or more control qubits, as is the case for Toffoli.
Certain other well-known gates such as SWAP can decomposed into such gates in advance.
If such a decomposition is impossible, the \rulename{Forget} rule treats the gate conservatively as unknown.
For unknown gates, $\written(g)$ is defined to be equal to $\qubits(g)$.

\paragraph{Symbolic Values and Updates}
A symbolic \emph{value} $V \,{\in}\, \Val$ is a set of \emph{atoms} and
represents the XOR of its atoms; in particular, $\emptyset$ represents $0$.
An atom is the constant $\one$, a \emph{symbol} $s$, or a \emph{product atom}
$\prodatom{O}$, the product of a set $O$ of two or more
values built earlier. A symbol, created by $\fresh$, stands for a basis value that the
analysis does not know.
A \emph{valuation} function $\rho:\Sym\to\{0,1\}$ assigns values to symbols, and under $\rho$, atoms $\alpha$ and values $V$ denote basis
values as follows:
\[
\sem{\one}_\rho = 1,\qquad
\sem{s}_\rho=\rho(s),\qquad
\sem{\prodatom{O}}_\rho=\bigwedge\nolimits_{V\in O}\sem{V}_\rho,\qquad
\sem{V}_\rho=\bigoplus\nolimits_{\alpha\in V}\sem{\alpha}_\rho .
\]
\begin{wrapfigure}{r}{0.35\textwidth}
\vspace*{-2.75ex}%
\begin{algorithm}[H]
\caption{$\Product(V_1,\ldots,V_k)$.}
\label{alg:product}
\lIf{$V_j=\emptyset$ for some $j$}{\Return $\emptyset$}
$O\gets\{V_j \mid 1\le j\le k,\ V_j\neq\{\one\}\}$\;
\lIf{$O=\emptyset$}{\Return $\{\one\}$}
\lIf{$O=\{W\}$}{\Return $W$}
\lIf{$O=\{V,W\}$}{\Return $\PairProduct(V,W)$}
\Return $\{\Pi[O]\}$\;
\end{algorithm}
\vspace{-2.5ex}
\end{wrapfigure}%
\indent This definition is well-founded because the operands of a product atom
are values built earlier in the analysis.
When a valuation picks one basis state, a symbolic value gives the basis value of its qubit in that state as a function of the symbols.
In particular, $\emptyset$, the empty XOR, denotes $0$ under every
valuation, and so the constant $0$ needs no atom whereas the
constant $1$ is the atom $\one$.
This set representation makes the two updates needed by the rules, XOR and product, cheap.
It also enables a temporary written twice by the same gate to cancel.

Specifically, the XOR of two values is computed by their
symmetric difference,
$V\oplus W := V\symdiff W$, which realizes cancellation $s\oplus s=0$ by removing an atom on both sides.
It realizes commutativity and associativity for free, as values are sets.
The negation of a value follows as $V\symdiff\{\one\}$.

The product of $k$ values is computed by
Alg.~\ref{alg:product}, which first handles the two constant
values $\emptyset$ and $\{\one\}$, i.e., $0$ and $1$.
If some operand is $\emptyset$, the product is $\emptyset$ (line~1).
Operands equal to $\{\one\}$ do not change the product and are removed (line~2).
If no operand remains, the product is $\{\one\}$ (line~3), and if exactly one
operand remains, the product is that operand itself (line~4).
Duplicate operands collapse automatically because the operands are kept as a set, realizing the law $s \land s=s$.

For two operands, the algorithm calls the subprocedure
$\PairProduct$ (line~5), which is defined in Appendix~\ref{app:pair-product}.
$\PairProduct$ uses symmetric differences to apply a limited case of distributivity only to the constant $\one$ of a negative control, ensuring that the representation is small yet precise.

Given three or more operands, none of which is $\emptyset$ or $\{\one\}$ at this point, the algorithm performs hash-consing~\cite{ershov1958,modularHashConsing} using a global memoization table $\Pi$ that associates operand sets $O$ with product atoms $\prodatom{O}$.
On line~6, Algorithm~\ref{alg:product} looks up the operand set in the table $\Pi$.
If $O$ has not been seen, $\Pi$ maps a fresh product atom $O \mapsto \prodatom{O}$ and returns it.

\begin{wrapfigure}{r}{0.38\textwidth}
\vspace*{-3.25ex}%
\centering%
\scalebox{0.9}{%
\begin{quantikz}[row sep={0.435cm,between origins},column sep={0.75cm,between origins},/tikz/commutative diagrams/every matrix/.append style={/tikz/name=symex},/tikz/execute at end picture={\MBULifetimeX{symex}{3}{(symex-col2.west)}{($(symex-col4.center)!0.5!(symex-col5.center)+(-1pt,0)$)}{MBULifeOne}{}\MBULifetimeX{symex}{3}{($(symex-col5.center)!0.5!(symex-col6.center)+(1pt,0)$)}{(symex-col8.east)}{MBULifeTwo}{}\MBUNumberSteps{symex}{7}}]
\lstick{$x$} &[-0.45cm,between origins] \MBUCtrl{2} & & \MBUCtrl{2} & & & & &[-0.45cm,between origins] \rstick{$x'$} \\
\lstick{$y$} & \MBUCtrl{1} & \MBUDot & \MBUCtrl{1} & \MBUH & \MBUOCtrl{1} & & \MBUOCtrl{1} & \rstick{$y'$} \\
\lstick{$a$} & \MBUTarget{} & \MBUCtrl{-1} & \MBUTarget{} & & \MBUTarget{} & \MBUGate{T} & \MBUTarget{} & \rstick{$\ket{0}$}
\end{quantikz}}\\
{\footnotesize\setlength{\tabcolsep}{3.5pt}\renewcommand{\arraystretch}{1.02}%
\begin{tabular}[c]{@{}cl|ll@{}}
position & rule applied & $\sigma(y)$ & $\sigma(a)$ \\
\midrule
after $7$ & --- & $V_y$ & $\emptyset$ \\
after $6$ & \rulename{Xor} on $g_7$ & $V_y$ & $\{y',\one\}$ \\
after $5$ & \rulename{Diag} on $g_6$ & $V_y$ & $\{y',\one\}$ \\
after $4$ & \rulename{Xor} on $g_5$ & $V_y$ & $\emptyset$ \\
after $3$ & \rulename{Forget} on $g_4$ & $\{y''\}$ & $\emptyset$ \\
after $2$ & \rulename{Xor} on $g_3$ & $\{y''\}$ & $\{\patom{q}\}$ \\
after $1$ & \rulename{Diag} on $g_2$ & $\{y''\}$ & $\{\patom{q}\}$ \\
before $1$ & \rulename{Xor} on $g_1$ & $\{y''\}$ & $\emptyset$ \\
\end{tabular}}%
\vspace{-2ex}
\caption{Circuit with two lifetimes of $a$ and its symbolic states, where $\patom{q}=\prodatom{\{V_x,\{y''\}\}}$.}
\label{fig:sym-example}
\vspace{-3ex}
\end{wrapfigure}

\paragraph{Example}
The circuit in Fig.~\ref{fig:sym-example} has ancilla $a$.
The table lists its symbolic
states, where $V_x=\{x'\}$ and $V_y=\{y'\}$ are the final values of
$x$ and $y$. Gate~7 XORs the value of its negative control,
$V_y\symdiff\{\one\}=\{y',\one\}$, into $a$. Gate~6 preserves the
symbolic state, since $T$ is diagonal. Gate~5 XORs the same value into
$a$, which returns to $\emptyset$. Gate~4 replaces the value of $y$ by
the fresh symbol $y''$. Gate~3 XORs $\Product(V_x,\{y''\})$ into
$a$, which creates the product atom $\patom{q}$ for the operand set
$\{V_x,\{y''\}\}$ and adds it to $\Pi$.
Gate~2 preserves the symbolic state, and gate~1 XORs the same product into $a$, which is in $\Pi$, so $a$ becomes~$\emptyset$.
The zero points of $a$ are $0$, $4$, and~$7$, and its two
lifetimes are 1--3 and 5--7 as shaded.

\paragraph{Precision}
Values encode commutativity: when $V_x=\{x\}$ and $V_y=\{y\}$, $\Product(V_x,V_y)$ and $\Product(V_y,V_x)$ both yield the atom $\prodatom{\{V_x,V_y\}}$ and their XOR is $\emptyset$.
Meanwhile, $\Product(V_x,\{y,z\})$ remains as one atom not expanded into
$xy\oplus xz$, because we perform no distribution of $\wedge$ over $\oplus$ beyond the constant $\one$ and no flattening of nested products.
The values thus stay compact,
at the price of precision: with $V_z=\{z\}$, $\Product(V_x,\{y,z\})$ and
$\{\prodatom{\{V_x,V_y\}}\}\symdiff\{\prodatom{\{V_x,V_z\}}\}$ denote the
same function but differ as values.
Thus, the analysis may miss a zero point, but never reports a wrong one:

\begin{theorem}[Soundness]
\label{thm:soundness}
Let $\ket{\Psi_i}$ be the state of the circuit at position $i$ on any
input for which every ancilla holds $\ket{0}$ at the end of the circuit. For
every ancilla $a$ and every position $i\in \mathcal{Z}(a)$ output by
Alg.~\ref{alg:analysis}, $a$ holds $\ket{0}$ at position $i$, i.e.,
$\ket{\Psi_i}=\ket{\psi}\otimes\ket{0}_a$ for some state $\ket{\psi}$ of
the other qubits.
\end{theorem}
\begin{proof}[Proof Sketch]
Appendix~\ref{app:soundness} gives the full proof.
The key idea is that the symbolic state describes a set of possible
computational-basis bit strings as its symbols change.
As an invariant, this set contains at least every string with nonzero
amplitude in the actual quantum state.
If every string this set contains has a given ancilla equal to $0$, then that ancilla is also $\ket{0}$ in
the actual state.

Given a bit string $b$, we write $b(q)$ for its bit
at qubit $q$. We show by induction from position $n$ down to
$0$ that every basis state $\ket{b}$ with nonzero amplitude in $\ket{\Psi_i}$
is covered by $\sigma_i$: some valuation $\rho$ satisfies $b(q)=\sem{\sigma_i(q)}_\rho$
for all $q\in Q$. Then, $\sigma_i(a)=\emptyset$ forces $b(a)=0$ for every
such $b$.

At position $n$, zeroes for every $a \in A$ and final values for data qubits provide this valuation.
At each gate, rule \rulename{Xor} inverts the permutation of basis states by an XOR gate, rule \rulename{Diag} preserves the valuation, and rule \rulename{Forget} removes constraints on valuations on target qubits.
\end{proof}

\begin{theorem}[Complexity]
\label{thm:complexity}
Define $N=|Q|+\sum_{i=1}^{n}(1+|\qubits(g_i)|)$ as the size of $C$.
Algorithm~\ref{alg:analysis} runs in expected $O(N^2)$ time under standard
assumptions on hashing. This complexity is $O({(n+|Q|)}^2)$ when the arity of all
gates is bounded by a constant, i.e., $|\qubits(g_i)|=O(1)$ for all $i$.
\end{theorem}
\begin{proof}[Proof Sketch]
Appendix~\ref{app:complexity} gives the full proof.
Every atom comes from one qubit or by one gate, so there are $O(N)$ atoms, and every value (set of atoms) has size $O(N)$.
A gate on $k$ qubits uses $O(k+1)$ set operations on values, which costs $O((k+1)N)$ and sums to $O(N^2)$ over the circuit.
\end{proof}

\section{Automatic Transformations from Unitary to Measurement-Based Uncomputation}
\label{sec:transform}

In this section, we present the two circuit transformations of
\textsf{MBUndo}, endpoint MBU (Section~\ref{sec:transform-endpoint}) and
interval MBU (Section~\ref{sec:transform-interval}), which were introduced
informally in Sections~\ref{sec:informal-endpoint}
and~\ref{sec:informal-interval}.

\subsection{Endpoint MBU for Final Uncomputation Gates}
\label{sec:transform-endpoint}

\begin{wrapfigure}{r}{0.45\textwidth}
\vspace{-2.75ex}
\begin{algorithm}[H]
\caption{Endpoint MBU.}
\label{alg:endpoint}
\SetKwInOut{Input}{In}\SetKwInOut{Output}{Out}
\SetInd{0.4em}{0.8em}
\SetKwBlock{ReplaceBy}{\textnormal{replace $g$ in $C$ by}}{end}
\begingroup\color{gray}\hypersetup{linkcolor=gray}%
\Input{circuit $C=g_1\cdots g_n$, ancillas $A\subseteq Q$}
\Output{circuit $C$ after endpoint MBU}
compute the lifetimes of every $a\in A$ by Alg.~\ref{alg:analysis}\;
\endgroup
\ForEach{lifetime of an ancilla $a\in A$}{
  $g\gets$ the last gate of the lifetime\;
  \If{$g=\XORg(a;\ell_1,\ldots,\ell_k)$}{
    \ReplaceBy{
      $H(a)$; $m\gets\textsc{Measure}(a)$; $\textsc{Reset}(a)$\;
      \If{$m=1$}{
        $\mathrm{C}^{k-1}\mathrm{Z}(c_1,\ldots,c_k)$ conjugated by $X$ on the negative controls\;
      }
    }
  }
}
\Return $C$\;
\end{algorithm}
\vspace{2.75ex}
\centering\small
\begin{quantikz}[wire types={q,q,q,n},row sep={0.5cm,between origins},column sep=0.2cm,align equals at=2]
\lstick{$x$} & \MBUCleanupCtrl{1} & \\
\lstick{$y$} & \MBUOCtrl[MBUCleanup]{1} & \\
\lstick{$a$} & \MBUCleanupTarget & \rstick{$\ket{0}$} \\
 & &
\end{quantikz}
\hspace{0.05cm}$=$\hspace{0.05cm}
\begin{quantikz}[wire types={q,q,q,n},row sep={0.5cm,between origins},column sep=0.18cm,align equals at=2,classical gap=0.025cm,/tikz/commutative diagrams/every matrix/.append style={/tikz/name=negctrl},/tikz/execute at end picture={\MBUOutcomeLabel{negctrl}{4}{3}{5}{$m$}}]
 & & & & & \MBUPaymentCtrl{1} & & \\
 & & & & \MBUPaymentGate{X} & \MBUPaymentDot & \MBUPaymentGate{X} & \\
 & \MBUH & \MBUMeasure{1} & \MBUReset\setwiretype{n} & & & & \\
 & & & \MBUClassicalBegin & \MBUClassicalCtrl{-2} & \MBUClassicalCtrl{-2} & \MBUClassicalCtrl{-2} & \MBUClassicalEnd
\end{quantikz}%
\setlength{\abovecaptionskip}{1ex}%
\caption{Endpoint MBU on a Toffoli with a negative control on $y$, using $\mathrm{CZ}(x,y)$ conjugated by $X(y)$.}
\label{fig:endpoint-negctrl}
\vspace{-4ex}%
\end{wrapfigure}

Endpoint MBU (Algorithm~\ref{alg:endpoint}) automates the rule of \citet{gidney-toffli-mbu} and
\citet{amy-ross-rule} that applies to any gate $g=\XORg(a;\ell_1,\ldots,\ell_k)$,
i.e., a gate such as Toffoli that maps $\ket{c}\ket{a}$ to $\ket{c}\ket{a\oplus f(c)}$
for its controls $c_i$ or negations in the literals $\ell_i$.

Alg.~\ref{alg:endpoint} computes the lifetimes of every ancilla
$a$ by Alg.~\ref{alg:analysis} (line~1) and examines the last gate $g$ of each lifetime (lines~2--3).
If $g$ is an XOR gate with target $a$ (line~4), Alg.~\ref{alg:endpoint} replaces it by an $X$-basis measurement of $a$
with outcome $m$ and a reset of $a$ to $\ket{0}$ (line~6) and, when $m=1$, a
correction of the phase $(-1){}^{f(c)}$, which for $k\ge 1$ corresponds to a
$\mathrm{C}^{k-1}\mathrm{Z}$ gate on the qubits $c_1,\ldots,c_k$ of the
controls (lines~7--8).

For a negative control $\bar{c}_j$, the correction is a
$\mathrm{C}^{k-1}\mathrm{Z}$ gate conjugated by an $X$ gate on $c_j$ (line~8) to apply the phase when $\bar{c}_j=1$, so whether a gate is replaced does not depend on its control polarity.
Fig.~\ref{fig:endpoint-negctrl} gives an example, where a Toffoli with a negative control on $y$ is replaced by a $\mathrm{CZ}(x,y)$ gate between two $X$ gates on the negatively controlled $y$, which together apply the phase $(-1)^{x\bar{y}}$ when $m=1$.

\subsection{Interval MBU for Sequences of Multiple Uncomputation Gates}
\label{sec:transform-interval}

Instead of the last uncomputation gate,
interval MBU (Alg.~\ref{alg:interval}) finds in each lifetime of each ancilla~$a$ a sound
\emph{interval} of gates to replace.
Figure~\ref{fig:interval-lifetime} depicts interval MBU on the suffix
of a lifetime of $a$. After computing the lifetimes as Algorithm~\ref{alg:endpoint}
does (line~1), Algorithm~\ref{alg:interval} processes each lifetime in three
steps: it finds the last use of $a$ and its
interval (lines~3--4), measures $a$ right after the last use
(lines~5--6), and replaces every gate of the interval in place by its
phase correction (lines~7--9).

\paragraph{Last Use and Interval}
In its isolated lifetime, $a$ interacts with two kinds of gates: \emph{writes} by $\oplus$ ($X$) gates and \emph{uses} of $a$ as a control $\bullet$ ($Z$) or as a subject of non-$\oplus$ gates such as $Y$ or $H$.
In Fig.~\ref{fig:interval-lifetime}, the \emph{last use} $g^*$ of $a$ is as a control for $W$.
Its \emph{interval} consists of the $r$ writes after $g^*$, which restore $a$ back to $0$ (lines~3--4).
The intervening gates $U_1,\ldots,U_{r-1}$ do not touch $a$, and $g^*$ is its last use.

\paragraph{Measurement After Last Use}
Alg.~\ref{alg:interval} measures $a$ in the $X$ basis immediately after $g^*$,
with outcome $m$, and resets $a$ to $\ket{0}$ (lines~5--6).
The ancilla cannot be measured any earlier.
If $a$ were measured before $g^*$, then $W$ in Fig.~\ref{fig:interval-lifetime} would instead need to be controlled on all bits that computed $a$, which kills any cost savings; the same occurs if $g^*$ were a $Z$ or $H$ gate on $a$.
After $g^*$, the only gates on $a$ are $\oplus$-writes that turn into phase corrections. If $a$ has no use, its interval is its whole lifetime.

\paragraph{In-Place Phase Corrections}
Algorithm~\ref{alg:interval} replaces each $\oplus$-gate $g$ that has controls $c_1,\ldots,c_k$ in the interval of $a$ by the correction $\mathrm{C}^{k-1}\mathrm{Z}(c_1,\ldots,c_k)$ classically conditioned on $m$ (lines~7--9).
No gate is emitted if $k=0$, as the error is an undetectable global phase, and negative controls are conjugated by $X$ gates.
Nothing else changes, in particular
$U_1,\ldots,U_{r-1}$ and all gates before $g^*$ in
Fig.~\ref{fig:interval-lifetime}.

\begin{figure}[t]
\begin{minipage}[c]{0.5\textwidth}
\begin{algorithm}[H]
\caption{Interval MBU.}
\label{alg:interval}
\SetKwInOut{Input}{In}\SetKwInOut{Output}{Out}
\SetKwBlock{InsertAfter}{\textnormal{insert immediately after $g^*$}}{end}
\begingroup\color{gray}\hypersetup{linkcolor=gray}%
\Input{circuit $C=g_1\cdots g_n$, ancillas $A\subseteq Q$}
\Output{circuit $C$ after interval MBU}
compute the lifetimes of every $a\in A$ by Alg.~\ref{alg:analysis}\;
\endgroup
\ForEach{lifetime of an ancilla $a\in A$}{
  $g^*\gets$ the \textbf{\textit{last use}} of $a$ in the lifetime\;
  $\mathit{interval}\gets$ the XOR gates with target $a$ after $g^*$\;
  \InsertAfter{
    $H(a)$; $m\gets\textsc{Measure}(a)$; $\textsc{Reset}(a)$\;
  }
  \ForEach{$g=\XORg(a;\ell_1,\ldots,\ell_k)$ in $\mathit{interval}$}{
    replace $g$ in $C$ by: \If{$m=1$}{
      $\mathrm{C}^{k-1}\mathrm{Z}(c_1,\ldots,c_k)$ conjugated by $X$ on the negative controls\;
    }
  }
}
\Return $C$\;
\end{algorithm}
\end{minipage}\hfill
\begin{minipage}[c]{0.485\textwidth}
\centering\small
\begin{quantikz}[wire types={b,q,b,q,n},row sep={0.6cm,between origins},column sep={0.47cm,between origins},bundle gap=0.06cm,classical gap=0.025cm,/tikz/commutative diagrams/every matrix/.append style={/tikz/name=lifea},/tikz/execute at end picture={\tikzset{mbu number distance=8pt}\foreach \mbuCol/\mbuNum in {7/1,9/2,12/r}{\MBUNumberAt{lifea}{\mbuCol}{$\mbuNum$}}\foreach \mbuCol/\mbuNum in {7/1,9/2,12/r}{\node[above=8pt,text=MBUPayment,font=\scriptsize,inner sep=0pt] at (lifea-3-\mbuCol.center |- lifea-3-1.center) {$\mbuNum'$};}\MBULifetime{lifea}{2}{2}{12}{MBUShared}{}\MBULifetime{lifea}{4}{2}{6}{MBUShared}{}\draw[-{Stealth[length=4pt,width=3.5pt]},MBULastUse,semithick] ([yshift=-16pt]lifea-2-3.center) -- ([yshift=-6pt]lifea-2-3.center);\node[left=4pt,font=\scriptsize,text=MBULastUse,inner sep=0pt] at ([yshift=-12pt]lifea-2-3.center) {last use};\draw[-{Stealth[length=4pt,width=3.5pt]},MBUPayment,semithick] ([yshift=-16pt]lifea-2-5.center) -- ([yshift=-6pt]lifea-2-5.center);\node[right=4pt,font=\scriptsize,text=MBUPayment,inner sep=0pt] at ([yshift=-12pt]lifea-2-5.center) {measure};\MBUOutcomeLabel{lifea}{5}{5}{7}{$m$}}]
\lstick{$\ket{x}_c$} &[-0.1cm,between origins] \ \ldots\ &[0.05cm,between origins] \MBUGate{W} & &[0.06cm,between origins] & &[-0.16cm,between origins] \ctrl[style={mbu bundle dot,draw=MBUCleanup,fill=MBUCleanup},wire style={draw=MBUCleanup}]{1} & \MBUGate{U_1} & \ctrl[style={mbu bundle dot,draw=MBUCleanup,fill=MBUCleanup},wire style={draw=MBUCleanup}]{1} & \ \ldots\ &[0.06cm,between origins] \MBUWideGate{0.66cm}{U_{r-1}} & \ctrl[style={mbu bundle dot,draw=MBUCleanup,fill=MBUCleanup},wire style={draw=MBUCleanup}]{1} &[-0.3cm,between origins] \rstick{$\ket{x'}_c$} \\
\lstick{$\ket{f}_a$} & \ \ldots\ & \MBUCtrl{-1} & & & & \MBUCleanupTarget & & \MBUCleanupTarget & \ \ldots\ & & \MBUCleanupTarget & \rstick{$\ket{0}$} \\[0.55cm]
\lstick{$\ket{x}_c$} & \ \ldots\ & \MBUGate{W} & & & & \control[style={mbu bundle dot,draw=MBUPayment,fill=MBUPayment}]{} & \MBUGate{U_1} & \control[style={mbu bundle dot,draw=MBUPayment,fill=MBUPayment}]{} & \ \ldots\ & \MBUWideGate{0.66cm}{U_{r-1}} & \control[style={mbu bundle dot,draw=MBUPayment,fill=MBUPayment}]{} & \rstick{$\ket{x'}_c$} \\
\lstick{$\ket{f}_a$} & \ \ldots\ & \MBUCtrl{-1} & \MBUH & \MBUMeasure{1} & \MBUReset\setwiretype{n} & & & & & & & \\
 & & & & & \MBUClassicalBegin & \MBUClassicalCtrl{-2} & & \MBUClassicalCtrl{-2} & \ \ldots\ & & \MBUClassicalCtrl{-2} & \MBUClassicalEnd
\end{quantikz}
\caption{A lifetime of an ancilla $a$ (top) and its transformation by
Alg.~\ref{alg:interval} (bottom), with the other qubits represented as a bundle $c$.
The last use of $a$ controls $W$, and the $r$ writes of $a$ after it form
the interval.}
\label{fig:interval-lifetime}
\end{minipage}
\end{figure}

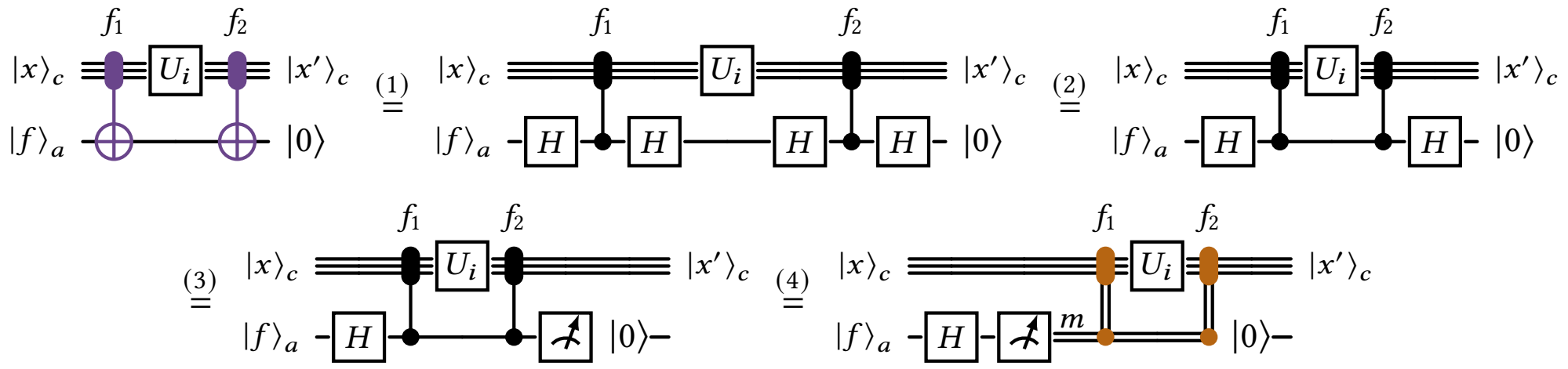
\begin{figure}[t]
\vspace*{-1.5ex}%
\centering\small
\begin{quantikz}[wire types={b,q},row sep={0.6cm,between origins},column sep=0.1cm,align equals at=1.5,bundle gap=0.06cm,/tikz/commutative diagrams/every matrix/.append style={/tikz/name=intva},/tikz/execute at end picture={\MBUNumberAt{intva}{2}{$f_1$}\MBUNumberAt{intva}{4}{$f_2$}}]
\lstick{\scriptsize$\ket{x}_c$} & \ctrl[style={mbu bundle dot,draw=MBUCleanup,fill=MBUCleanup},wire style={draw=MBUCleanup}]{1} & \MBUGate{U_i} & \ctrl[style={mbu bundle dot,draw=MBUCleanup,fill=MBUCleanup},wire style={draw=MBUCleanup}]{1} & \rstick{\scriptsize$\ket{x'}_c$} \\
\lstick{\scriptsize$\ket{f}_a$} & \MBUCleanupTarget & & \MBUCleanupTarget & \rstick{$\ket{0}$}
\end{quantikz}
$\overset{(1)}{=}$
\begin{quantikz}[wire types={b,q},row sep={0.6cm,between origins},column sep=0.1cm,align equals at=1.5,bundle gap=0.06cm,/tikz/commutative diagrams/every matrix/.append style={/tikz/name=intvb},/tikz/execute at end picture={\MBUNumberAt{intvb}{3}{$f_1$}\MBUNumberAt{intvb}{7}{$f_2$}}]
\lstick{\scriptsize$\ket{x}_c$} & & \ctrl[style={mbu bundle dot}]{1} & & \MBUGate{U_i} & & \ctrl[style={mbu bundle dot}]{1} & & \rstick{\scriptsize$\ket{x'}_c$} \\
\lstick{\scriptsize$\ket{f}_a$} & \MBUH & \control{} & \MBUH & & \MBUH & \control{} & \MBUH & \rstick{$\ket{0}$}
\end{quantikz}
$\overset{(2)}{=}$
\begin{quantikz}[wire types={b,q},row sep={0.6cm,between origins},column sep=0.1cm,align equals at=1.5,bundle gap=0.06cm,/tikz/commutative diagrams/every matrix/.append style={/tikz/name=intvc},/tikz/execute at end picture={\MBUNumberAt{intvc}{3}{$f_1$}\MBUNumberAt{intvc}{5}{$f_2$}}]
\lstick{\scriptsize$\ket{x}_c$} & & \ctrl[style={mbu bundle dot}]{1} & \MBUGate{U_i} & \ctrl[style={mbu bundle dot}]{1} & & \rstick{\scriptsize$\ket{x'}_c$} \\
\lstick{\scriptsize$\ket{f}_a$} & \MBUH & \control{} & & \control{} & \MBUH & \rstick{$\ket{0}$}
\end{quantikz}
\par\medskip
$\overset{(3)}{=}$
\begin{quantikz}[wire types={b,q},row sep={0.6cm,between origins},column sep=0.1cm,align equals at=1.5,bundle gap=0.06cm,/tikz/commutative diagrams/every matrix/.append style={/tikz/name=intvd},/tikz/execute at end picture={\MBUNumberAt{intvd}{3}{$f_1$}\MBUNumberAt{intvd}{5}{$f_2$}}]
\lstick{\scriptsize$\ket{x}_c$} & & \ctrl[style={mbu bundle dot}]{1} & \MBUGate{U_i} & \ctrl[style={mbu bundle dot}]{1} & & &[0.05cm] \rstick{\scriptsize$\ket{x'}_c$} \\
\lstick{\scriptsize$\ket{f}_a$} & \MBUH & \control{} & & \control{} & \MBUMeasureInline & \MBUReset\setwiretype{q} &
\end{quantikz}
$\overset{(4)}{=}$
\begin{quantikz}[wire types={b,q},row sep={0.6cm,between origins},column sep=0.1cm,align equals at=1.5,bundle gap=0.06cm,/tikz/commutative diagrams/every matrix/.append style={/tikz/name=intve},/tikz/execute at end picture={\MBUNumberAt{intve}{4}{$f_1$}\MBUNumberAt{intve}{6}{$f_2$}\MBUOutcomeLabelAbove{intve}{2}{3}{4}{\ \ $m$}}]
\lstick{\scriptsize$\ket{x}_c$} & & &[0.25cm] \control[style={mbu bundle dot,draw=MBUPayment,fill=MBUPayment}]{} & \MBUGate{U_i} & \control[style={mbu bundle dot,draw=MBUPayment,fill=MBUPayment}]{} & &[0.05cm] \rstick{\scriptsize$\ket{x'}_c$} \\
\lstick{\scriptsize$\ket{f}_a$} & \MBUH & \MBUMeasureInline & \MBUClassicalCtrl{-1} & & \MBUClassicalCtrl{-1} & \MBUReset\setwiretype{q} &
\end{quantikz}
\setlength{\abovecaptionskip}{0.25ex}%
\setlength{\belowcaptionskip}{-0.5ex}%
\caption{
Circuit equivalence rules that demonstrate the soundness of interval MBU\@.
Expressions $f_1$ and $f_2$ are the products of the controls (positive or negative) of two uncomputation gates, so $a$ holds $f=f_1\oplus f_2$ before the interval.
The gate $U_i$ stands for any intervening gates that do not touch the ancilla $a$.
}
\label{fig:interval-proof}
\end{figure}

\begin{theorem}[Soundness]
\label{thm:transform-soundness}
For every input where every ancilla ends in $\ket{0}$ and for every
measurement outcome, the circuits output by
Algs.~\ref{alg:endpoint} and~\ref{alg:interval} produce the same final
state as the original circuit.
\end{theorem}
\begin{proof}
Soundness of endpoint MBU follows directly from the rule of \citet{gidney-toffli-mbu} in Figure~\ref{fig:mbu-toffoli}, as generalized to multiply-controlled gates by \citet{amy-ross-rule}.
Algorithm~\ref{alg:endpoint} applies this rule precisely at the end of each lifetime, which satisfies the terminal $\ket{0}_a$ condition in Figure~\ref{fig:mbu-toffoli}.

Soundness of interval MBU follows from a sequence of circuit equivalence rules, illustrated in Figure~\ref{fig:interval-proof}.
Briefly, rule~(1) follows from $X=HZH$ and rule~(2) follows from $HH = I$.
Rule~(3) states that performing a unitary gate on a qubit and then (implicitly) measuring the qubit and discarding the outcome is equivalent to measuring the qubit immediately without performing the unitary gate.
Rule~(4) is the principle of deferred measurement~\citep{QCQI}.

Fix any given lifetime of ancilla $a$.
As shown in Figure~\ref{fig:interval-lifetime}, the interval of $a$ consists of $\oplus$-gates with controls from qubits $c$ and intervening gates $U_i$ on $c$ that do not interact with $a$.
By repeatedly applying rule (1) and cancelling adjacent $H$ gates via rule (2), every $\oplus$ ($X$) on $a$ is replaced with a multiply-controlled-$Z$ gate that has $\bullet$ ($Z$) in place of $\oplus$ on $a$.
This step corresponds to lines~7--9 in Algorithm~\ref{alg:interval}.
Now, each endpoint of the interval has an uncancelled Hadamard gate.
Theorem~\ref{thm:soundness} guarantees the terminal $\ket{0}$ condition of rule~(3),
so by applying rule~(3), the final Hadamard gate is replaced with a $Z$-basis measurement-and-reset gate.
By repeating rule~(4), this measurement commutes backwards through each $\bullet$ on $a$ by replacing quantum with classical control.
The end of this process leaves one Hadamard gate and measurement-and-reset gate at the beginning of the interval --- corresponding to lines~3--6 of Alg.~\ref{alg:interval}.
These rules do not change the basis state of any qubit other than $a$, and the gates they add do not touch any other ancilla between its last use and the end of its lifetime --- the same argument repeats for the remaining ancillas.
\end{proof}

\begin{theorem}[Cost]
\label{thm:transform-tcount}
Assuming that a $\mathrm{C}^{k-1}\mathrm{Z}$ gate costs no more $T$ gates than a
$\mathrm{C}^{k}\mathrm{X}$ gate, for every measurement outcome,
the circuits output by Algs.~\ref{alg:endpoint} and~\ref{alg:interval}
execute no more $T$ gates than the original.
\end{theorem}
\begin{proof}
Consider a replaced gate $g$, whose $T$ cost is $e$. The MBU transformation
removes $g$ and adds only its phase correction, an $H$ gate, and a measurement-and reset gate for the whole interval.
Only the correction can have nonzero $T$ cost.
It consists of $X$ gates and a $\mathrm{C}^{k-1}\mathrm{Z}$
gate, which has a $T$ cost of at most $e$ by the assumption -- which holds for standard decompositions~\cite{Selinger2013} --
and it executes only when $m=1$. Hence, no output executes more $T$ gates than the original.
\end{proof}

\paragraph{Complexity}
Neither transformation increases the asymptotic complexity of the static analysis
(Theorem~\ref{thm:complexity}). Once Alg.~\ref{alg:analysis} has computed
the zero points, one pass over the gates of each ancilla finds its
lifetimes, their last gates for endpoint MBU, and their last uses and
hence their intervals for interval MBU\@. This takes $O(N)$ time in total
for the size $N$ of Thm.~\ref{thm:complexity}. Each replaced gate on
$k+1$ qubits becomes $O(k)$ gates, namely one $\mathrm{C}^{k-1}\mathrm{Z}$
gate and at most $2k$ $X$ gates. Both transformations thus run in $O(N)$
time and output a circuit of size $O(N)$, so the total time of
\textsf{MBUndo} stays at the expected $O(N^2)$ of
Theorem~\ref{thm:complexity}, which remains dominated by the static analysis.

\section{Implementation}
\label{sec:impl}

We implement \textsf{MBUndo} as a compiler pass atop Qiskit~2.5.2~\cite{qiskit}.
\textsf{MBUndo} takes as input a circuit and a list of ancillas that the circuit uncomputes unitarily.
It performs the static analysis of Sec.~\ref{sec:symbolic}, followed by either the endpoint MBU (Sec.~\ref{sec:transform-endpoint}) or the interval MBU (Sec.~\ref{sec:transform-interval}) transformation.
It then outputs a circuit in which ancillas are uncomputed by measurement.

This design enables \textsf{MBUndo} to be applied as a compilation pass to the output circuits of existing synthesizers for unitary uncomputation, thereby further enhancing their value.
In this section, we detail several practical decisions made when implementing the techniques of Secs.~\ref{sec:symbolic} and~\ref{sec:transform}:

\paragraph{Multi-Target Gate Normalization}
Before the symbolic execution of Sec.~\ref{sec:symbolic}, \textsf{MBUndo} ensures
that every gate falls into one of the categories handled by the rules of
$\Step$. Specifically, a multi-target gate of Qiskit whose
definition consists of XOR gates only, e.g., SWAP, the Fredkin gate, and the
double-CNOT gate, is decomposed into those XOR gates via its
definition. Any other multi-target gate is conservatively treated as an unknown gate, which
is handled by the \rulename{Forget} rule.

\paragraph{Skipping Unprofitable Replacements}
In its implementation of the transformations of Section~\ref{sec:transform}, \textsf{MBUndo} applies a filter to skip unprofitable replacements under the cost model of this paper.
In particular, it leaves a lifetime unchanged if its last gate has fewer than two controls
(i.e., an $X$ or CNOT gate, which is Clifford) under endpoint
MBU, or if its interval is a single such gate under interval MBU, as such
a replacement is not profitable in either the $T$ or the Clifford count.

\paragraph{Relative-Phase Toffoli Gates}
It is common for uncomputation synthesizers such as Unqomp~\cite{unqomp} and Reqomp~\cite{reqomp} to postprocess their output circuits by replacing each compute--uncompute pair of Toffoli gates on an ancilla with a pair of \emph{relative-phase Toffoli gates}~\cite{maslov-rccx,barenco-elementary-gates}, named RCCX in Qiskit.
The RCCX gate $R$ controlled on $a$ and $b$ acts on target $t$ as $R\ket{a,b,t}=i^{ab}(-1){}^{at}\ket{a,b,t\oplus ab}$.
This substitution is sound because RCCX is equal to Toffoli except for a relative phase that a paired RCCX on the same qubits cancels.
It is profitable because an RCCX gate requires 4 $T$ gates, whereas Toffoli requires 7.

By nature, the RCCX substitution is orthogonal to our subject: turning unitary uncomputation into MBU\@.
The practical wrinkle is that MBU is only conditionally applicable to RCCX gates.
As we detail in Appendix~\ref{app:rccx}, the analogue of Gidney's rule (Fig.~\ref{fig:mbu-toffoli}) for the RCCX gate is sound under endpoint MBU for one RCCX, but is unsound under interval MBU for multiple RCCX gates.

To sidestep this unsoundness, \textsf{MBUndo} pre-processes its input circuit to restore RCCX pairs to Toffoli pairs, which does not change its semantics (Appendix~\ref{app:rccx}).
After synthesizing MBU, it post-processes its output circuit to re-introduce RCCX in place of Toffoli gates whenever Unqomp or Reqomp would do the same.
In brief, \textsf{MBUndo} first replaces every pair of Toffoli gates on an ancilla that the MBU transformation left intact with a pair of RCCX gates.
This step mirrors Unqomp and Reqomp, but unlike them it misses the Toffoli gates whose partners were eliminated by MBU\@.
To catch the benefit of RCCX for them as well, \textsf{MBUndo} replaces conservatively each such widowed Toffoli gate whose target is constant with an RCCX gate and a Clifford correction, which is sound by an analogue of Eq.~(12) of \citet{amy-ross-rule}.
Neither pre- nor post-processing affects the asymptotic complexity of \textsf{MBUndo}, and for fairness we apply the above post-processing to every baseline in our evaluation.
Overall, this approach matches existing practice, retains the $T$-count advantages of RCCX over Toffoli, and enables fair evaluation against existing systems.

\section{Evaluation}
\label{sec:eval}

In this section, we evaluate the performance of \textsf{MBUndo} via the following research questions:
\begin{enumerate}[leftmargin=3em]
  \item [\emph{RQ1.}] By how much do \textsf{MBUndo}'s automatic transformations, endpoint and interval MBU, reduce the gate costs of the circuits generated by synthesizers for unitary uncomputation?
  \item [\emph{RQ2.}] How do the results of \textsf{MBUndo}'s automatic transformations compare against baseline MBU constructions known in the literature for specific circuits?
  \item [\emph{RQ3.}] How well does \textsf{MBUndo} scale in compile time when given large circuits?
  \item [\emph{RQ4.}] How do gate savings from \textsf{MBUndo} compare against prior optimizers for quantum circuits?
\end{enumerate}

\subsection{RQ1: Enhancement Over Unitary Uncomputation Synthesizers}
\label{sec:eval-rq1}

\begin{enumerate}[leftmargin=3em]
  \item [\emph{RQ1.}]  By how much do \textsf{MBUndo}'s automatic transformations, endpoint and interval MBU, reduce the gate costs of the circuits generated by synthesizers for unitary uncomputation?
\end{enumerate}

\paragraph{Benchmarks}
We evaluate on 22 benchmark circuits, drawn from three sources:
\begin{enumerate}[topsep=1pt, leftmargin=15pt]
    \item[$a)$]\emph{Unitary Uncomputation Circuits.}
        All ten benchmarks evaluated by Reqomp~\cite{reqomp}, together with all three circuits evaluated by RwUn~\cite{dirty-and-clean-uncomputation} that assume clean\footnote{RwUn also studies circuits with dirty ancillas, i.e.\ those not initialized and returned to zero, which is a distinct model from standard uncomputation~\cite{Qurts,silq,unqomp,reqomp,modular-synthesis}.} ancillas, namely \texttt{Incrementer},
        \texttt{CondCleanMCS}, and \texttt{RiseCondCleanMCS}.
    \item[$b)$]\emph{MBU Circuits.}
        Six explicit circuits from the literature, to which MBU has been manually applied:
        \texttt{GidneyAdder}~\cite{gidney-toffli-mbu},
        \texttt{TableLookup}~\cite{windowed-quantum-arithmetic},
        \texttt{GF8Multiply}~\cite{MBU-GF2-multiplication},
        \texttt{TwoSidedComparator} and
        \texttt{VBEModAddition}~\cite{MBU-for-modular-arithmetic}, and
        \texttt{QLZOC}~\cite{MBU-QLZOC}.
        We force \textsf{MBUndo} to rediscover the MBU opportunities in these benchmarks by removing all uncomputation from them and passing them to synthesizers for unitary uncomputation.
    \item[$c)$]\emph{Additional Circuits.}
        Three other benchmarks for which MBU has not been previously explored, in which unitary uncomputation of an ancilla requires a sequence of gates:
        \texttt{KeccakChi}~\cite{keccak_chi},
        \texttt{SpeedySbox}~\cite{speedy-block-cipher}, and
        \texttt{MatrixVectorEq}~\cite{binary_matrix_vector_equality}.
\end{enumerate}

\paragraph{Methodology}
We evaluate against all existing circuit-level uncomputation synthesizers, namely
Unqomp~\cite{unqomp}, Reqomp\footnote{We fixed a number of implementation bugs in
Reqomp so that the tool behaves as formalized by \citet{reqomp}. The fixes enable Reqomp to
synthesize uncomputation for more benchmarks, and do not degrade its output
circuits.}~\cite{reqomp}, and RwUn~\cite{dirty-and-clean-uncomputation} with
its three clean-ancilla strategies \textsc{Sequential}, \textsc{Reverse}, and \textsc{Jointly}.%
\footnote{RwUn also provides a fourth strategy that rewrites all computation on ancillas onto data qubits so that no ancillas are used (at the cost of additional gates over data qubits), making it not suitable as an upstream of this evaluation.}
For each benchmark,
we run the synthesizer, obtain its output circuit and list of
ancillas, and apply \textsf{MBUndo}. Following the methodology
of \citet{unqomp-artifact,reqomp-artifact}, we decompose all $\mathrm{C}^k \mathrm{X}$ gates in
the circuits into Toffoli ladders before synthesizing uncomputation.
Uniquely, Reqomp can trade recomputation, i.e., more gates, for fewer ancillas on circuits that have specific dependency structures.
We specify the default ancilla budget so that Reqomp does not unnecessarily incur additional gates.

As discussed in Section~\ref{sec:impl}, \textsf{MBUndo} applies RCCX post-processing analogous to Unqomp and Reqomp.
For clarity, we report gate counts (i) \emph{without post-processing} after any tool and (ii)~\emph{with post-processing} after each tool.
For \textsf{MBUndo}, the \emph{worst-case} gate counts assume that every phase correction executes,
while \emph{expected} counts use the fact that each measurement has an independent $50\%$ probability of leaving no phase to correct and thus needing zero correction gates.
The two usually differ only in Clifford count, as Toffolis require zero $T$ phase corrections under MBU\@.

\paragraph{Results}
Table~\ref{tab:rq1} reports results against Reqomp, while Appendix~\ref{app:rq1-full} reports the remainder.
We focus on Reqomp as it is the only baseline that does not timeout or crash on all 22 benchmarks, and on all but one it matches or outperforms Unqomp and RwUn in $T$ count.
In the analysis below, we first compare endpoint MBU to Reqomp. Then, we compare interval to endpoint MBU on the benchmarks that compute ancillas using multiple gates, making endpoint MBU insufficient.

\begin{figure}[t]
\centering
\includegraphics[width=\textwidth]{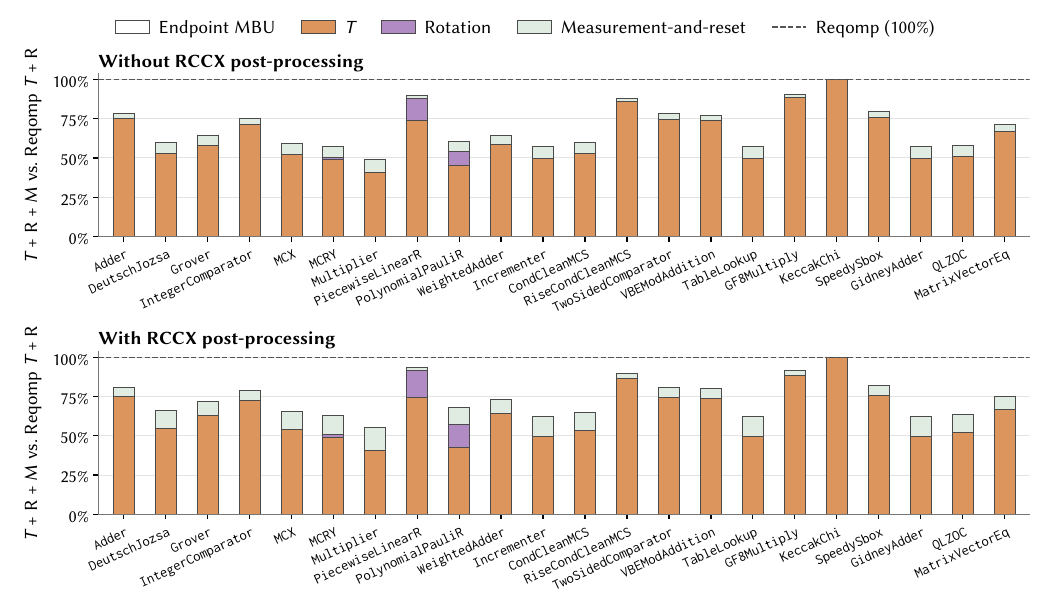}
\setlength{\abovecaptionskip}{-0.5ex}%
\caption{Non-Clifford costs, namely the \emph{T}, rotation (R), and measurement-and-reset (M) gates, of circuits produced by endpoint MBU given Reqomp's output, without (top) and with (bottom) RCCX post-processing.
All counts are normalized by the non-Clifford gate count of Reqomp's output (100\%). Lower is better.}
\label{fig:rq1-endpoint}
\end{figure}

\paragraph{Endpoint MBU vs.\ Reqomp}
Figure~\ref{fig:rq1-endpoint} compares the gate counts of Reqomp and endpoint MBU\@.
Without post-processing, endpoint MBU reduces $T$ gates over Reqomp by 37.5\% and
Clifford gates by 26.5\% (in expectation) in geomean.
After post-processing for both, endpoint MBU reduces $T$ gates by 36.5\% and Clifford by 15.2\% (in expectation).
The Clifford reduction shrinks because MBU retains Clifford gates in phase corrections even as Reqomp decreases them in RCCX gates.

The only benchmark on which endpoint MBU performs no transformation is \texttt{KeccakChi}.
Its last uncomputation gate in every lifetime is Clifford, which if replaced would save no $T$ gates but add Clifford corrections.
\textsf{MBUndo} filters out such unprofitable candidates (see Section~\ref{sec:impl}).

Assuming that uncomputation makes up roughly half of a circuit, MBU should not be expected to reduce gate counts by more than half in general.
Figure~\ref{fig:rq1-endpoint} depicts one exception, \texttt{Multiplier}, where \textsf{MBUndo}'s static analysis detects that certain Toffoli gates in the \emph{computation} half of the circuit always leave their target as zero, and can thus also be replaced with measurement.

\texttt{MCRY}, \texttt{PiecewiseLinearR}, and \texttt{PolynomialPauliR} use non-$T$ and non-Clifford rotation~gates not relevant to uncomputation.
\textsf{MBUndo} leaves them alone; we plot them for completeness.

\begin{figure}[t]
\centering
\includegraphics[width=\textwidth]{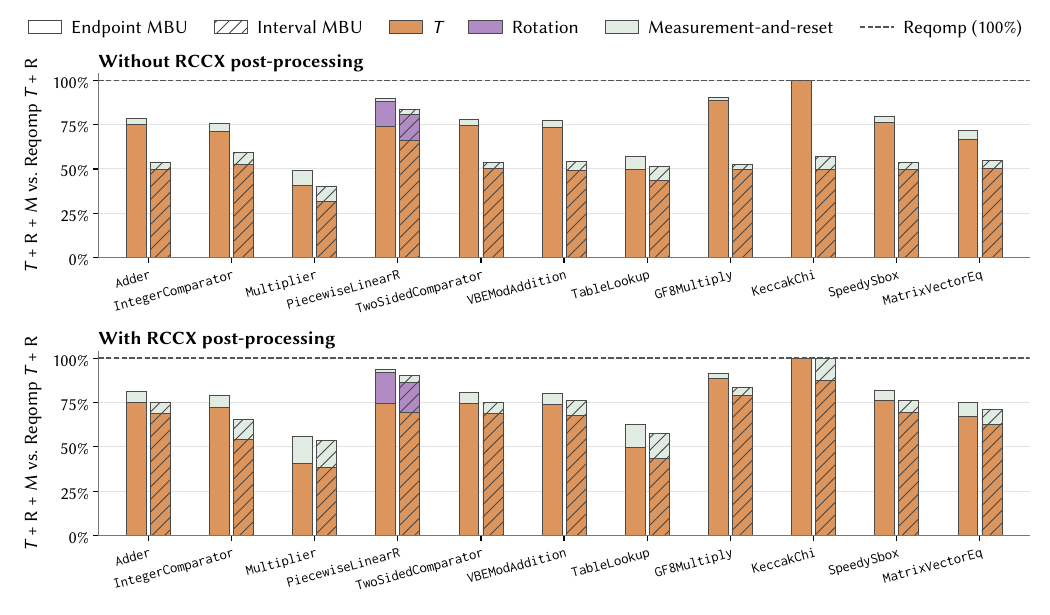}
\setlength{\abovecaptionskip}{-0.5ex}%
\setlength{\belowcaptionskip}{-1.4ex}%
\caption{\emph{T}, rotation, and measurement gate counts of the circuits produced by endpoint and interval MBU over Reqomp outputs, on the 11 benchmarks whose ancillas are uncomputed by several non-Clifford gates, without (top) and with (bottom) post-processing.
All counts are normalized as in Fig.~\ref{fig:rq1-endpoint}. Lower is better.}
\label{fig:rq1-interval}
\end{figure}

\paragraph{Interval vs.\ Endpoint}
Figure~\ref{fig:rq1-interval} compares the gate counts of interval and endpoint MBU for the 11 benchmarks where they differ, which contain ancillas that are uncomputed by multiple non-Clifford gates.
Without post-processing, endpoint MBU reduces the $T$ and Clifford counts for these circuits by 28.9\% and 19.8\% over Reqomp, whereas interval MBU reduces them by 50.5\% and 38.2\%.

The results of interval and endpoint MBU are close for \texttt{PiecewiseLinearR} because the input to Reqomp accounts for 80.6\%, not the typical ${\sim}50\%$, of the non-Clifford gates of Reqomp's output. Excluding these, interval MBU eliminates all $T$ gates in the uncomputation of this circuit.

With RCCX post-processing, endpoint MBU reduces gates by 28.5\% and 11.7\%, as compared to 36.0\% and 18.4\% for interval MBU\@.
The gap narrows after post-processing, as RCCX substitution (Section~\ref{sec:impl}) is disproportionately effective for Toffoli gates not at the end of a lifetime.
Even when not eliminated, Toffoli gates only cost 4 rather than 7 $T$ gates when replaced with RCCX\@.
The remaining gap is largest for \texttt{IntegerComparator} (18.4\%) and smallest for \texttt{Multiplier} (2.3\%).

Over the full set of 22 benchmarks, interval MBU reduces $T$ gates by 47.9\% and Clifford by 35.6\% without post-processing, and by 40.0\% and 18.7\% with post-processing.
Both interval and endpoint MBU reduce $T$ gates by 43.7\% on the benchmarks not shown in Fig.~\ref{fig:rq1-interval} with post-processing.

\paragraph{Other Synthesizers.}
The outputs of Unqomp and RwUn exhibit similar behavior (Appendix~\ref{app:rq1-full}).
Over Unqomp, without post-processing,
endpoint and interval MBU reduce $T$ gates by 38.6\% and 48.4\% and Clifford by 26.9\% and 35.2\%, respectively.
The three strategies of RwUn succeed on 9, 13, and 11 benchmarks, respectively,
where endpoint MBU reduces $T$ gates by 18.4\%, 17.8\%, and 17.9\%, and interval MBU by 59.0\%, 55.5\%, and 47.8\%.
The same pattern appears with post-processing.

These results are distinctive for RwUn, which rewrites the entire circuit instead of replaying the computation gates --- leading to fewer opportunities for endpoint MBU but more for interval MBU\@.
Uniquely, RwUn uncomputes some ancillas via multi-controlled gates whose MBU phase corrections use $T$ gates.
Its expected and worst-case $T$ counts differ: 216 vs.\ 282 for \texttt{MCRY}.

\begin{table}[t]
\caption{Gate counts of the circuits produced by Reqomp and by \textsf{MBUndo} when given Reqomp's output, without and with the RCCX post-processing of Section~\ref{sec:impl}\@. \emph{T} and C are the counts of \emph{T} and Clifford gates on expectation, rounded to integers. Q and A are the numbers of qubits and of ancillas of Reqomp's output, which \textsf{MBUndo} does not change. LT is the number of lifetimes identified by our analysis, and M the number of measurement-and-reset gates introduced by endpoint (End.) and interval (Int.) MBU.}%
\vspace*{-1ex}%
\label{tab:rq1}
\centering\scriptsize
\setlength{\tabcolsep}{2.4pt}
\setlength{\cmidrulekern}{\tabcolsep}
\newcolumntype{N}{>{\raggedleft\arraybackslash}p{5mm}}
\begin{tabular}{@{}l rr rrr NN NN NN NN NN NN@{}}
\toprule
 & & & & & & \multicolumn{6}{c}{Without RCCX post-processing} & \multicolumn{6}{c@{}}{With RCCX post-processing} \\
\cmidrule(lr){7-12} \cmidrule(l){13-18}
 & & & & \multicolumn{2}{c}{M} & \multicolumn{2}{c}{Reqomp} & \multicolumn{2}{c}{End.\ MBU} & \multicolumn{2}{c}{Int.\ MBU} & \multicolumn{2}{c}{Reqomp} & \multicolumn{2}{c}{End.\ MBU} & \multicolumn{2}{c@{}}{Int.\ MBU} \\
\cmidrule(lr){5-6} \cmidrule(lr){7-8} \cmidrule(lr){9-10} \cmidrule(lr){11-12} \cmidrule(lr){13-14} \cmidrule(lr){15-16} \cmidrule(l){17-18}
Benchmark & Q & A & LT & End. & Int. & \multicolumn{1}{c}{$T$} & \multicolumn{1}{c}{C} & \multicolumn{1}{c}{$T$} & \multicolumn{1}{c}{C} & \multicolumn{1}{c}{$T$} & \multicolumn{1}{c}{C} & \multicolumn{1}{c}{$T$} & \multicolumn{1}{c}{C} & \multicolumn{1}{c}{$T$} & \multicolumn{1}{c}{C} & \multicolumn{1}{c}{$T$} & \multicolumn{1}{c@{}}{C} \\
\midrule
\texttt{Adder} & 36 & 12 & 11 & 11 & 11 & 308 & 420 & 231 & 349 & 154 & 266 & 176 & 288 & 132 & 261 & 121 & 244 \\
\texttt{DeutschJozsa} & 19 & 8 & 8 & 8 & 8 & 119 & 158 & 63 & 106 & 63 & 106 & 71 & 110 & 39 & 90 & 39 & 90 \\
\texttt{Grover} & 9 & 3 & 20 & 20 & 20 & 336 & 479 & 196 & 349 & 196 & 349 & 216 & 359 & 136 & 309 & 136 & 309 \\
\texttt{IntegerComparator} & 24 & 11 & 11 & 6 & 10 & 147 & 199 & 105 & 160 & 77 & 125 & 87 & 139 & 63 & 124 & 47 & 109 \\
\texttt{MCX} & 23 & 10 & 10 & 10 & 10 & 147 & 168 & 77 & 103 & 77 & 103 & 87 & 108 & 47 & 83 & 47 & 83 \\
\texttt{MCRY}$^\dagger$ & 24 & 11 & 11 & 11 & 11 & 154 & 178 & 77 & 107 & 77 & 107 & 88 & 112 & 44 & 85 & 44 & 85 \\
\texttt{Multiplier} & 24 & 9 & 65 & 65 & 65 & 770 & 970 & 315 & 548 & 245 & 473 & 440 & 640 & 180 & 438 & 170 & 423 \\
\texttt{PiecewiseLinearR}$^\dagger$ & 13 & 6 & 42 & 21 & 36 & 1078 & 1530 & 931 & 1394 & 833 & 1277 & 868 & 1320 & 784 & 1268 & 728 & 1221 \\
\texttt{PolynomialPauliR}$^\dagger$ & 9 & 3 & 45 & 45 & 45 & 630 & 780 & 315 & 488 & 315 & 488 & 360 & 510 & 180 & 398 & 180 & 398 \\
\texttt{WeightedAdder} & 25 & 9 & 77 & 74 & 74 & 1246 & 1499 & 728 & 1018 & 728 & 1013 & 832 & 1085 & 536 & 890 & 536 & 885 \\
\texttt{Incrementer} & 12 & 5 & 5 & 5 & 5 & 70 & 87 & 35 & 55 & 35 & 55 & 40 & 57 & 20 & 45 & 20 & 45 \\
\texttt{CondCleanMCS} & 121 & 68 & 93 & 93 & 93 & 1389 & 1602 & 738 & 998 & 738 & 998 & 795 & 1008 & 423 & 776 & 423 & 776 \\
\texttt{RiseCondCleanMCS} & 11 & 2 & 2 & 2 & 2 & 101 & 128 & 87 & 115 & 87 & 115 & 59 & 86 & 51 & 81 & 51 & 81 \\
\texttt{TwoSidedComparator} & 241 & 96 & 97 & 97 & 97 & 2674 & 3530 & 1995 & 2900 & 1344 & 2202 & 1534 & 2390 & 1146 & 2148 & 1053 & 2008 \\
\texttt{VBEModAddition} & 264 & 132 & 135 & 105 & 134 & 2800 & 3972 & 2065 & 3290 & 1379 & 2540 & 1600 & 2772 & 1180 & 2504 & 1082 & 2342 \\
\texttt{TableLookup} & 29 & 14 & 71 & 60 & 69 & 840 & 1224 & 420 & 834 & 364 & 763 & 480 & 864 & 240 & 714 & 208 & 659 \\
\texttt{GF8Multiply} & 12 & 3 & 3 & 2 & 3 & 126 & 153 & 112 & 140 & 63 & 87 & 72 & 99 & 64 & 94 & 57 & 83 \\
\texttt{KeccakChi} & 401 & 1 & 200 & 0 & 200 & 2800 & 4200 & 2800 & 4200 & 1400 & 2800 & 1600 & 3000 & 1600 & 3000 & 1400 & 2800 \\
\texttt{SpeedySbox} & 43 & 19 & 116 & 106 & 116 & 3108 & 3604 & 2366 & 2915 & 1554 & 2044 & 1776 & 2272 & 1352 & 2007 & 1236 & 1832 \\
\texttt{GidneyAdder} & 389 & 129 & 129 & 129 & 129 & 1806 & 2835 & 903 & 1997 & 903 & 1933 & 1032 & 2061 & 516 & 1739 & 516 & 1675 \\
\texttt{QLZOC} & 231 & 63 & 152 & 152 & 152 & 2177 & 3119 & 1113 & 2131 & 1113 & 2131 & 1265 & 2207 & 657 & 1827 & 657 & 1827 \\
\texttt{MatrixVectorEq} & 321 & 126 & 126 & 126 & 126 & 2667 & 3304 & 1785 & 2485 & 1337 & 1909 & 1527 & 2164 & 1023 & 1849 & 959 & 1657 \\
\bottomrule
\end{tabular}
\par\vspace{2pt}\begin{minipage}{\textwidth}\footnotesize $^\dagger$\,\texttt{MCRY}, \texttt{PiecewiseLinearR}, and \texttt{PolynomialPauliR} also contain 2, 182, and 61 rotation gates, which no tool changes.\end{minipage}
\vspace*{-4ex}%
\end{table}

\subsection{RQ2: Comparison with Existing MBU Constructions}
\label{sec:eval-rq2}

\begin{enumerate}[leftmargin=3em]
  \item [\emph{RQ2.}] How do the results of \textsf{MBUndo}'s automatic transformations compare against baseline MBU constructions known in the literature for specific circuits?
\end{enumerate}

\paragraph{Benchmarks}
We group the known applications of MBU to individual circuits, to the best of
our knowledge, into three categories: (i) circuits that apply Gidney's rule to
uncompute each ancilla computed by a Toffoli gate, including
\texttt{GidneyAdder}~\cite{gidney-toffli-mbu} and \texttt{QLZOC}~\cite{MBU-QLZOC};
(ii) specialized constructions that exploit the structure of a particular
computation, including the MBU of \texttt{TableLookup}~\cite[Sec.~2]{windowed-quantum-arithmetic} and of
\texttt{GF8Multiply}~\cite[Appendix~C.3]{MBU-GF2-multiplication}; and (iii)
the monolithic MBU of \citet{MBU-for-modular-arithmetic},
which applies to an uncomputation block that operates on a single ancilla and
neither modifies data qubits nor depends on other ancillas.

\paragraph{Methodology}
For categories (i) and (ii), we compare three ways to construct each benchmark: via the unitary uncomputation synthesized by Reqomp, via the output of interval MBU in \textsf{MBUndo}, and via the published circuit.
For category (iii), there currently exists no software tool to automatically transform a circuit by monolithic MBU\@.
Instead, we optimistically calculate its best possible performance on the 22 benchmarks from RQ1.
Specifically, we compute the number of unitary uncomputation gates for each benchmark by subtracting the number of gates input to Reqomp from the number of gates it outputs.
We divide this number by two, as monolithic MBU executes each uncomputation block half the time, and add it to the number of input gates.
Throughout this question, to isolate the effect of MBU itself, we apply no RCCX post-processing to any tool.

\DeclareRobustCommand{\legendsym}[3]{\tikz[baseline=-0.5ex]{\draw[#1,line width=0.8pt,#2] (0,0)--(1.6em,0);\node[#1,draw,fill=white,line width=0.8pt,inner sep=0pt,minimum size=3.2pt,#3] at (0.8em,0){};}}
\begin{table}[t]
\centering
\begin{minipage}[c]{0.49\textwidth}
\centering\scriptsize
\caption{Cost of the uncomputation in the circuits of categories (i, above) and (ii, below) through unitary uncomputation, \textsf{MBUndo}, and prior constructions.}%
\vspace*{-1ex}%
\label{tab:rq2}
\setlength{\tabcolsep}{4pt}
\begin{tabular}{@{}l l r r r@{}}
\toprule
Benchmark & Uncomputation & \emph{T} & C & M \\
\midrule
\texttt{GidneyAdder} & Reqomp (unitary) & 903 & 1160 & 0 \\
 & \textsf{MBUndo} (interval) & 0 & 257.5\,(386) & 129 \\
 & \citet{gidney-toffli-mbu} & 0 & 321.5\,(386) & 129 \\
\addlinespace[2pt]
\texttt{QLZOC} & Reqomp (unitary) & 1064 & 1216 & 0 \\
 & \textsf{MBUndo} (interval) & 0 & 228\,(304) & 152 \\
 & \citet{MBU-QLZOC} & 0 & 228\,(304) & 152 \\
\midrule
\texttt{GF8Multiply} & Reqomp (unitary) & 63 & 75 & 0 \\
 & \textsf{MBUndo} (interval) & 0 & 9\,(15) & 3 \\
 & \citet{MBU-GF2-multiplication} & 0 & 7.5\,(10) & 3 \\
\addlinespace[2pt]
\texttt{TableLookup} & Reqomp (unitary) & 448 & 636 & 0 \\
 & \textsf{MBUndo} (interval) & 182 & 372\,(405) & 43 \\
 & \citet{windowed-quantum-arithmetic} & 49 & 127\,(151) & 19 \\
\bottomrule
\end{tabular}
\end{minipage}\hfill
\begin{minipage}[c]{0.50\textwidth}
\centering%
\includegraphics[width=\textwidth]{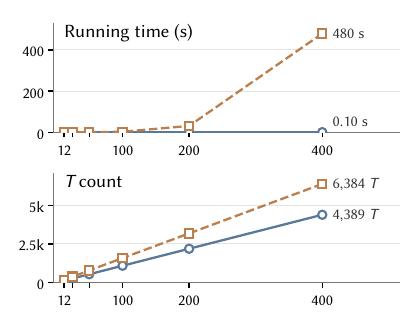}%
\setlength{\abovecaptionskip}{-1.5ex}%
\captionof{figure}{Comparison of \textsf{MBUndo} (\legendsym{color={rgb,255:red,88;green,121;blue,152}}{solid}{circle}) and Reqomp (\legendsym{color={rgb,255:red,187;green,127;blue,78}}{dashed}{rectangle}) on \texttt{Adder} with varying bit width \emph{n} on \emph{x}-axis.}
\label{fig:rq3-adder}
\end{minipage}%
\vspace*{-1.5ex}%
\end{table}

\paragraph{Results}
In Table~\ref{tab:rq2}, we present the results for categories (i) and (ii).
This table reports only the costs of uncomputation rather than the whole circuit.
For category (i), \textsf{MBUndo} reproduces the manual constructions: it
removes every $T$ uncomputation gate, introduces the same
measurements, and matches the worst-case Clifford counts, with equal or
lower expected Clifford count.

For category (ii), on \texttt{GF8Multiply}, \textsf{MBUndo} similarly eliminates all $T$ gates.
On \texttt{TableLookup}, \textsf{MBUndo} obtains 67\% of the $T$ reduction of the specialized design, leaving 182 of 448 gates where the prior work leaves 49.
This result is expected, as~\citet{windowed-quantum-arithmetic}
performs a structural rewrite beyond a general-purpose transformation like \textsf{MBUndo} --- measuring the lookup register and correcting the remaining phases with another lookup over a table of only $\sqrt{L}$ entries for $L$ addresses.

We present the results for (iii) in Appendix~\ref{app:rq2-mono}.
\textsf{MBUndo} surpasses the best possible performance of monolithic MBU on all 22 benchmarks, in $T$ and expected and worst-case Clifford counts.
Relative to Reqomp, monolithic MBU reduces $T$ gates by at most 23.1\% (geomean) and Clifford gates by at most 21.4\%, whereas interval MBU reduces them by 47.9\% and 35.6\% (in expectation).

\subsection{RQ3: Scalability of \textsf{MBUndo}}
\label{sec:eval-rq3}

\begin{enumerate}[leftmargin=3em]
  \item [\emph{RQ3.}]  How well does \textsf{MBUndo} scale in compile time when given large circuits?
\end{enumerate}

\paragraph{Methodology}
The 22 benchmarks of RQ1 span a variety of sizes, and \textsf{MBUndo} completes each within 43 ms on a laptop.
To cover a wider range of sizes, we use the ten benchmarks of Reqomp, each of which is built from a scaling parameter $n$, e.g., the number of bits of the operands of \texttt{Adder}.
RQ1 evaluates them at the smallest size used by~\citet[Sec.~6.1]{reqomp}; here we evaluate six sizes per benchmark.
We report the running time and $T$ count of \textsf{MBUndo} applying interval MBU to the output of Reqomp, in the same setting of RQ1 with post-processing.
We also report running times of Reqomp itself, to compare the time that \textsf{MBUndo} adds on top of its upstream tool.

As another baseline specific to the static analysis of Section~\ref{sec:symbolic}, we compare against the symbolic executor of \citet{CaretteANF} for simulating quantum circuits over basis states.
Like \textsf{MBUndo}, it tracks every qubit as a Boolean function of symbolic variables.
However, it represents the function in algebraic normal form (ANF), which applies the full distributivity of $\land$ over $\oplus$ and can blow up in the size of a symbolic value.
To compare the two representations, we run the tool of \citet{CaretteANF} on the outputs of Reqomp for the 15 of the 22 benchmarks of RQ1 that consist of XOR-based gates only (the tool does not support other gates).
We obtain the ANF after each write to an ancilla to compare the zero points (see Section~\ref{sec:symbolic}) identified by that tool and by \textsf{MBUndo}.

\paragraph{Results}
Figure~\ref{fig:rq3-adder} shows representative timing results on \texttt{Adder};
the other nine circuits (Appendix~\ref{app:rq3-charts}) exhibit a similar pattern.
\textsf{MBUndo} stays efficient even on large circuits: on \texttt{Adder},
its running time grows from 4 milliseconds at $n=12$ to 0.10 seconds at $n=400$.
Over all benchmarks, it takes at most 1.0 second, on the largest instance of \texttt{PolynomialPauliR}.
For reference, the upstream tool Reqomp takes 480 seconds to synthesize the uncomputation of \texttt{Adder} with $n=400$.

When run on the 15 RQ1 benchmarks, the symbolic executor of \citet{CaretteANF} times out on five circuits after 1,000 seconds.
On the ten benchmarks it completes, it identifies the same zero points as those found by \textsf{MBUndo}'s static analysis.
This comparison indicates that \textsf{MBUndo} can precisely identify ancilla lifetimes on the evaluated circuits in time less than prior tools.

\subsection{RQ4: Comparison with Circuit Optimizers}
\label{sec:eval-rq4}

\begin{enumerate}[leftmargin=3em]
  \item [\emph{RQ4.}] How do gate savings from \textsf{MBUndo} compare against prior optimizers for quantum circuits?
\end{enumerate}

\paragraph{Methodology}
\textsf{MBUndo} is designed to reduce the gate cost of circuits that rely on unitary uncomputation.
Among existing circuit optimizers, SPARE~\cite{spare} is most similar in purpose: it targets the compute--use--uncompute structure of ancillas to optimize circuits.

Also relevant are optimizers such as Tzap~\citep{tzap} and Feynman~\citep{feynman} that perform rewrites unrelated to uncomputation such as phase folding.
However, these tools currently handle circuits with measurements only by treating every measurement as opaque and ignoring the relationship between measured outcomes and classically controlled gates.
This behavior significantly reduces their optimization opportunities on any circuit that uses MBU, so applying them to the outputs of both Reqomp and \textsf{MBUndo} would not compare the two on equal footing.

Instead, we take a conservative approach: give Reqomp the benefit of the optimizer and deny it to \textsf{MBUndo}.
We obtain the gate counts of Reqomp followed by the Tzap optimizer with its recommended setting \texttt{-O3} that includes phase folding to combine and cancel rotation gates.
Feynman, which also uses phase folding, never leaves fewer $T$ gates than Tzap on our benchmarks.

We provide to each optimizer the output circuit of Reqomp on the 22 benchmarks of RQ1.
We apply the RCCX post-processing of RQ1 to the outputs of Reqomp and \textsf{MBUndo}; Tzap does not output any Toffoli gates to post-process, and SPARE performs its own RCCX post-processing that covers no fewer gates than ours.
We give SPARE a time limit of 1,000 seconds per circuit.

\paragraph{Results}
Appendix~\ref{app:rq4-full} lists the gate counts and running times of the evaluated optimizers on all benchmarks.
SPARE completes 18 of the 22 benchmarks without timing out or crashing.
On those 18 cases, SPARE reduces the $T$ count of Reqomp's output by 13.8\% in geomean.
\textsf{MBUndo} using interval MBU reduces the $T$ count by 41.4\% on these 18 benchmarks, and uses fewer $T$ gates than SPARE on all but \texttt{PolynomialPauliR}, where SPARE trades $T$ gates for rotation gates.
SPARE takes between 2.6 and 879 seconds, whereas \textsf{MBUndo} takes at most 43 milliseconds per circuit.

\textsf{MBUndo} produces fewer $T$ gates than Reqomp optimized by Tzap on 13 of the 22 benchmarks, even in this disadvantaged setting as explained above.
On these 13 benchmarks, \textsf{MBUndo} reduces the $T$ count of Reqomp's output by 44.5\% in geomean, whereas Tzap reduces it by 10.2\%; on the other nine, \textsf{MBUndo} reduces it by 32.8\% and Tzap reduces it by 49.2\%.
These results suggest that transforming unitary uncomputation into MBU can by itself trade blows with a state-of-the-art circuit optimizer.
We hope they motivate greater support for dynamic circuits in optimizers, which if combined with \textsf{MBUndo} can yield improvements beyond what either tool achieves alone.

\section{Related Work}

\paragraph{Uncomputation}
The study of uncomputation originates in reversible computing~\cite{reversibility}.
For quantum programs, uncomputation has received architectural support~\cite{architecture-support-uncomputation}, automation via type systems in languages~\cite{silq,Qurts}, and modular synthesis by compilers for languages~\cite{modular-synthesis,reoc} and circuits~\cite{unqomp,reqomp,safe-dirty-uncomputation}.
For example, \citet{tiwari2025uncomputing} uncompute ancillas by partial uncomputation when dependencies are cyclic,
and RwUn~\cite{dirty-and-clean-uncomputation} synthesizes uncomputation by rewrite rules that extend to dirty ancillas.
All of these prior approaches synthesize circuits through unitary rather than measurement-based uncomputation.

Since its introduction, MBU has been applied by hand to numerous circuit designs~\cite{windowed-quantum-arithmetic,MBU-QLZOC,MBU-GF2-multiplication,MBU-for-modular-arithmetic}.
Programming frameworks such as Q\#~\cite{qsharp}, Qualtran~\cite{qualtran}, and Qrisp~\cite{qrisp} incorporate MBU as primitives taken from published circuit designs, and recent work furthers understanding of MBU from an error-correction perspective~\cite{mbu-QEC-recast}.
Our contribution of \textsf{MBUndo} complements these specific MBU constructions by automatically converting unitary uncomputation in arbitrary circuits to MBU, without requiring a specific circuit structure or uncomputed function.

Our work focuses on saving $T$ gates and not qubits.
A rich body of work studies the time--space trade-offs of re-materializing temporary values, as shown in reversible pebble games~\cite{reversible-pebble-game,reversible-line-DP}.
Spooky pebble games~\cite{gidney-spooky-pebble} similarly arise from MBU, with advantages in minimizing qubits at the cost of more gates~\cite{spooky-pebble-game-optimize-space,spooky-line-pebbling-complexity}.

\paragraph{Symbolic Execution}
Symbolic representations of quantum states have been applied to debugging~\cite{fang2024qse}, verification~\cite{symqv}, and simulation~\cite{chen2024loop} of quantum programs.
Closest to our work, \citet{CaretteANF} simulate Hadamard--Toffoli circuits by tracking the \emph{basis value} of each qubit as a Boolean formula in algebraic normal form.
Our analysis instead tracks basis values conservatively as sets of XOR terms, which keeps them linear in the circuit size (Theorem~\ref{thm:complexity}) for scalability (Section~\ref{sec:eval-rq3}).
Our symbolic values share the spirit of XOR-AND graphs~\cite{meuli-xaig}, which represent a Boolean function as a graph of XOR and AND nodes from which a compiler emits a circuit;
our values differ in that they are updated gate by gate during symbolic execution, using cheap set operations and hash lookups.

\paragraph{Quantum Languages and Compilation}
Other work in the programming languages community has furthered understanding of the implementation~\citep{fang2026hadamard,buckley2025language,wright2024quff}, compilation~\citep{peduri2022qssa,tarabkah2025synthesis}, optimization~\citep{xu2022quartz,shi2019optimized,zhang2022uniq,xu2023synthesizing,chen2026dead}, and verification~\citep{paykin2017qwire,abdulla2025verifying,tao2022giallar,yu2021abstract,liu2019formal,lee2025exact} of quantum programs.
\textsf{MBUndo} acts as an additional compilation pass that complements and enhances this work.

\section{Limitations and Future Directions}

\paragraph{Language-Level}
\textsf{MBUndo} works at the circuit level, where it transforms arbitrary circuits with unitary uncomputation.
A natural alternative direction is to automate MBU at the language level.
For instance, a language in the spirit of Silq~\cite{silq} and Qurts~\cite{Qurts} could decide by its type system when a program variable admits MBU, and then synthesize the measurement and phase corrections from a language-level specification of the function.

\paragraph{Architecture-Level}
\textsf{MBUndo} measures cost mainly by $T$ count, the standard metric for fault-tolerant circuits, because $T$ gates are the major bottleneck on fault-tolerant hardware~\cite{beverland-scaling,gidney2024} whereas Clifford gates and measurements are far cheaper~\cite{litinski-gosc}.
In a concrete setting, the actual cost of each operation may depend on the architecture and its error-correction scheme, both of which are evolving rapidly.
Future work could incorporate into the transformations a finer-grained cost model of a concrete architecture, which accounts for the cost of $T$ gates in terms of the production of magic states, as well as the cost of measurement and of the classical processing it requires such as decoding and feed-forward, to decide where MBU pays off for circuits on a specific architecture and error-correction scheme.

\section{Conclusion}

This work initiates the study of automatically and systematically transforming quantum circuits to use measurement-based uncomputation.
Our efficient static analysis discovers opportunities for MBU in an input circuit, while our circuit transformations soundly replace unitary uncomputation gates by measurements and phase corrections that can be efficient on fault-tolerant hardware.

In the near term, our work brings the benefit of MBU to a broad class of quantum programs; in particular, it cuts the $T$-gate cost of the circuits output by existing uncomputation synthesizers and complements their methods.
In the long run, we hope this work motivates compiler techniques for dynamic circuits, which combine unitary gates with measurement and classical feed-forward.
Such directions reveal new ways to reduce cost that unitary gate optimization alone cannot reach.

\medskip
\noindent
\textbf{\textsf{AI Use}}\quad
The ideas of this work and the design of all of its analyses and transformations are due to the authors.
The authors used Claude Fable 5.1 by Anthropic to implement the algorithms, to write tests for the implementation and evaluation, and to write scripts for running the experiments.

\bibliographystyle{ACM-Reference-Format}
\bibliography{ref/ref}

\appendix
\clearpage
\section{Algorithm for Product of Two Symbolic Values}
\label{app:pair-product}

\begin{wrapfigure}{r}{0.36\textwidth}
\vspace{-2ex}
\begin{algorithm}[H]
\caption{$\PairProduct(V,W)$.}
\label{alg:pair-product}
$V'\gets V\setminus\{\one\}$,\ $W'\gets W\setminus\{\one\}$\;
$P\gets V'$ \textbf{if} $V'=W'$ \textbf{else} $P\gets\{\Pi[\{V',W'\}]\}$\;
\lIf{$\one\in V$}{$P\gets P\symdiff W'$}
\lIf{$\one\in W$}{$P\gets P\symdiff V'$}
\lIf{$\one\in V\cap W$}{$P\gets P\symdiff\{\one\}$}
\Return $P$\;
\end{algorithm}
\vspace{-2ex}
\end{wrapfigure}
Alg.~\ref{alg:product} delegates the product of exactly two operands $V$ and $W$ to $\PairProduct$, which is defined in Alg.~\ref{alg:pair-product}.
This algorithm distributes the constant $\one$ out of $V$ and $W$ for precision.

To see its correctness, let $V'=V\setminus\{\one\}$ and $W'=W\setminus\{\one\}$ be the operands without the constant.
Since $\bar{x}y=xy\oplus y$
and $\bar{x}\bar{y}=xy\oplus x\oplus y\oplus 1$, the product of $V$ and $W$ is the product of $V'$ and $W'$, XORed with $W'$ if $V$ contains $\one$, with $V'$ if $W$ does, and with $\{\one\}$ if both do (lines~3--5). The product of $V'$ and $W'$ is looked up in $\Pi$ under $\{V',W'\}$ just like any other product in Alg.~\ref{alg:product}, or is $V'$ itself when $V'=W'$ (line~2).

The result of this method is that all four polarity combinations of two controls share one product atom, e.g.,
$\Product(V_x\symdiff\{\one\},V_y)=\{\prodatom{\{V_x,V_y\}}\}\symdiff V_y$.
This property makes the static analysis more precise on temporaries computed and uncomputed by Toffoli gates, the most common XOR gates, without incurring additional asymptotic cost.
For instance, a circuit may compute a temporary by a Toffoli with a negative control $\bar{x}$ and uncompute it by a Toffoli with the positive control $x$ followed by a CNOT\@.
Algorithm~\ref{alg:pair-product} ensures that these equivalent writes yield the same value.

\section{Proof of Theorem~\ref{thm:soundness}}
\label{app:soundness}

\begin{proof}[Proof of Theorem~\ref{thm:soundness}]
The key idea is that the symbolic state may represent extra basis
bit strings, but must include every string with nonzero amplitude in
the actual state. If an ancilla has the empty symbolic value, all
represented strings have that bit equal to $0$, so the actual state
must also have that ancilla in $\ket{0}$ and hence unentangled from
the other qubits.
Fix an input satisfying the final-zero promise on $A$. A basis bit
string is a map $b:Q\to\{0,1\}$, with $b(q)$ its bit at qubit $q$;
$b\in\mathrm{supp}\,\Psi_i$ means that $\ket{b}$ has nonzero amplitude
in $\ket{\Psi_i}$. We prove the following invariant by backward
induction on the position $i$:
\[
\text{for every }b\in\mathrm{supp}\,\Psi_i,\quad
\text{there exists }\rho\text{ such that }\quad
b(q)=\sem{\sigma_i(q)}_\rho\quad\text{for all }q\in Q.
\]
Such a valuation \emph{covers} $b$ in $\sigma_i$: it matches all qubits
of $b$, although different strings may use different valuations, since
the invariant chooses the valuation $\rho$ after the string $b$.

\paragraph{Base case.}
At position $n$, every supported string $b$ has $b(a)=0$ for all
$a\in A$ by the promise on $A$. This matches $\sigma_n(a)=\emptyset$, which
denotes $0$ under every valuation.
Each data qubit $q\notin A$ has its own fresh symbol $s_q$, so setting
$\rho(s_q)=b(q)$ gives a valuation covering $b$.

\paragraph{Inductive step.}
Assume the invariant at position $i$, and choose any
$b\in\mathrm{supp}\,\Psi_{i-1}$. We construct a valuation covering $b$
in $\sigma_{i-1}=\Step(\sigma_i,g_i)$ for each transfer rule.
\medskip

\noindent
\begingroup
\setlength{\abovedisplayskip}{3pt plus 1pt minus 1pt}%
\setlength{\belowdisplayskip}{3pt plus 1pt minus 1pt}%
\setlength{\abovedisplayshortskip}{0pt plus 1pt}%
\setlength{\belowdisplayshortskip}{3pt plus 1pt minus 1pt}%
\rulename{Xor}~~
An XOR gate permutes basis states, so $b'=g_i(b)$ belongs to
$\mathrm{supp}\,\Psi_i$. By the induction hypothesis, some valuation
$\rho$ covers $b'$. We first show the symbolic operations realize XOR and product
under every valuation:
\[
\sem{V\symdiff W}_\rho
  =\sem{V}_\rho\oplus\sem{W}_\rho, \qquad
\sem{\Product(V_1,\ldots,V_k)}_\rho
  =\bigwedge\nolimits_{j=1}^{k}\sem{V_j}_\rho.
\]
The first identity holds because $\symdiff$ drops exactly the atoms
common to $V$ and $W$, which cancel in the XOR. The second holds
because every value that Alg.~\ref{alg:product} returns denotes the
product of its operands: the constants of lines~1 and~3, the sole
operand of line~4, the atom $\prodatom{O}$ of line~6, which denotes the
product of $O$ by definition, and the value of $\PairProduct$ (Appendix~\ref{app:pair-product});
dropping operands equal to $\{\one\}$ and duplicates (line~2) uses
$1x=x$ and $xx=x$.
For $g_i=\XORg(t;\ell_1,\ldots,\ell_k)$, the gate leaves the controls
unchanged, and XORing into the target is self-inverse. Thus
$\sem{\sigma_{i-1}(t)}_\rho
=b'(t)\oplus \textstyle \bigwedge\nolimits_{j=1}^{k}b'(\ell_j)
=b(t)$ by the two identities, with $b'(\bar c)=b'(c)\oplus1$ for a
negative control, as in the symbolic rule. All other bits and symbolic
values are unchanged, so $\rho$ covers $b$.
\endgroup
\medskip

\noindent
\rulename{Diag}~~
A diagonal gate multiplies each amplitude by a nonzero phase, so the
support is unchanged. As the rule also leaves $\sigma$ unchanged,
coverage follows from the induction hypothesis.
\medskip

\noindent
\rulename{Forget}~~
The rule keeps $\sigma_i(q)$ for $q\notin\written(g_i)$ and gives each
$q\in\written(g_i)$ a fresh symbol $s_q$, which can be assigned any
bit. It thus suffices to find some $b'\in\mathrm{supp}\,\Psi_i$ that
agrees with $b$ outside $\written(g_i)$. Consider the group of all
basis states that agree with $b$ outside $\written(g_i)$. By the
definition of $\written(g_i)$, the gate maps this group into itself,
and being unitary, it preserves the total probability of the group.
That probability is positive before the gate, as it includes $b$, and
hence after it, so some member $b'$ of the group is in
$\mathrm{supp}\,\Psi_i$. Choose a valuation covering $b'$ by the
induction hypothesis and extend it by $s_q\mapsto b(q)$ for each
$q\in\written(g_i)$. The retained values contain no fresh symbol, so
they still denote $b'(q)=b(q)$ for $q\notin\written(g_i)$.
The resulting valuation covers $b$ in
$\sigma_{i-1}$, which completes the induction.
\medskip

\noindent
Finally, Alg.~\ref{alg:analysis} records $i\in \mathcal{Z}(a)$ only when
$\sigma_i(a)=\emptyset$, including its initialization at $i=n$.
By the invariant, every $b\in\mathrm{supp}\,\Psi_i$ then satisfies
$b(a)=\sem{\emptyset}_\rho=0$. Hence all amplitudes with $a=1$ vanish,
and $\ket{\Psi_i}=\ket{\psi}\otimes\ket{0}_a$ for some state
$\ket{\psi}$ of the other qubits, as required.
\end{proof}

\section{Proof of Theorem~\ref{thm:complexity}}
\label{app:complexity}

\begin{proof}[Proof of Theorem~\ref{thm:complexity}]
Let $n_{\mathrm{XOR}}$ be the number of XOR gates and $F$ the set of
gates handled by rule \rulename{Forget}. The symbolic execution creates
one symbol per data qubit, one symbol per written qubit of each gate in $F$, and
at most one product atom per XOR gate, whether by the general lookup
(Alg.~\ref{alg:product}, line~6) or by $\PairProduct$
(Alg.~\ref{alg:pair-product}, line~2), so the number of atoms is
\[
B\le 1+|Q\setminus A|+n_{\mathrm{XOR}}+\textstyle\sum\nolimits_{g\in F}|\written(g)|
= O(N),\qquad\text{where } N=|Q|+\textstyle\sum\nolimits_{i=1}^{n}(1+|\qubits(g_i)|).
\]
Every value is a set of at most $B$ atoms.
Processing an XOR gate with $k$ controls builds at most $k$ literal values
(a negation is a symmetric difference with $\{\one\}$),
builds and hashes the operand set (which contains at most $k$ values of size at most $B$),
performs at most one table lookup,
and performs one symmetric difference on the target,
which costs $O((k+1)B)$ in total.

With exactly two operands, $\PairProduct$ instead removes the constant $\one$ from both operands,
compares the two results, hashes and looks up the constant-free pair,
and performs at most three further symmetric differences, all on values
of size at most $B$, which costs $O(B)$, so the bound per XOR gate is
unchanged. The other rules cost $O(k)$ for a gate on $k$ qubits.

A gate $g_i$ has at most $|\qubits(g_i)|$ controls, so its cost is $O((1+|\qubits(g_i)|)\cdot B)$,
and summing over the gates gives $O(B\cdot\sum_i(1+|\qubits(g_i)|))$.
We have $B=O(N)$ from above, and the sum is at most $N$ by the definition of $N$,
so the total is $O(N\cdot B)=O(N^2)$ expected time under the usual assumptions on hashing.
When the arity of the gates is bounded by a constant $c$, i.e., $|\qubits(g_i)|\le c$ for all $i$,
we have $N\le|Q|+(1+c)\cdot n=O(n+|Q|)$, and the bound becomes $O({(n+|Q|)}^2)$.
Finally, Alg.~\ref{alg:analysis} adds only constant work per touched ancilla and gate,
namely the emptiness test of lines~6--8, i.e., $O(\sum_i|\qubits(g_i)|)=O(N)$ in total,
which does not affect the bound.
\end{proof}

\section{Handling Relative-Phase Toffoli Gates in MBUndo}
\label{app:rccx}

This appendix details how \textsf{MBUndo} treats relative-phase Toffoli
gates~\cite{maslov-rccx,barenco-elementary-gates}, named RCCX in Qiskit
(Sec.~\ref{sec:impl}). Recall that the RCCX gate $R$ with controls $a$
and $b$ acts on its target $t$ as
$R\ket{a,b,t}=i^{ab}(-1)^{at}\ket{a,b,t\oplus ab}$, i.e., as a Toffoli
gate up to the relative phase $i^{ab}(-1)^{at}$. The gate is
self-inverse, so a pair of RCCX gates on the same qubits cancels like a
pair of Toffoli gates when the controls hold the same values at both
gates and the target holds the same value before the second gate as
after the first. Uncomputation synthesizers such as Unqomp~\cite{unqomp}
and Reqomp~\cite{reqomp} guarantee this for every Toffoli gate on an
ancilla by construction, as they pair each gate that computes an ancilla
with one that uncomputes it on the same controls with the same values,
and thus substitute RCCX for all of them.

\paragraph{The Obstacle to MBU on a Sequence of RCCX Gates}
Consider an ancilla $t$ that is computed by a sequence of RCCX gates,
used, and uncomputed by the same sequence in reverse order. If $t$ is
measured after its use as in interval MBU, the correction that replaces
one uncomputing gate $R(a,b;t)$ must apply
$(-1)^{m\,ab}\cdot i^{ab}\cdot(-1)^{at}$, where $m$ is the measurement
outcome, the first factor is the correction for a Toffoli gate
(Sec.~\ref{sec:transform-interval}), and the other two factors reproduce
the relative phase of $R$. The quantum value of $t$, however, is no
longer available after the measurement. Only when $t$ is uncomputed by
a single gate does $t$ hold $ab$ before it, so that the remaining phase
reduces to $(-i)^{ab}$, which can be corrected without the value of
$t$. MBU on one RCCX gate is thus sound, whereas it is unsound for a
sequence of RCCX uncomputation gates.

\paragraph{Pre- and Post-Processing}
To sidestep this obstacle while keeping the benefit of RCCX,
\textsf{MBUndo} applies MBU to Toffoli gates only, with the following
pre- and post-processing. The pre-processing restores every pair of
RCCX gates in the input circuit to a pair of Toffoli gates. Unlike
Unqomp and Reqomp, which can pair every Toffoli gate on an ancilla
because their algorithms guarantee its partner by construction,
\textsf{MBUndo} must handle input circuits in general and thus pairs
gates by the static analysis of Sec.~\ref{sec:symbolic}: two gates form
a pair when the symbolic execution gives the same symbolic values to
their controls and target after the earlier gate and before the later
gate, which proves that their relative phases cancel. 
The pairing is thus sound for every input circuit,
not only for the outputs of the existing uncomputation synthesizers.

After the transformation of Sec.~\ref{sec:transform}, the
post-processing rewrites the output circuit in two steps. First, every
pair of Toffoli gates on an ancilla that the transformation left intact,
paired as above, is replaced by a pair of RCCX gates. The symbolic
execution handles the measurements, resets, and classically controlled
gates that the transformation introduces conservatively: a measurement
and a classically controlled gate are treated as unknown gates by the
\rulename{Forget} rule, and a reset sets its qubit to $\emptyset$.
This step alone, however, leaves every Toffoli gate whose partner was
replaced by MBU as it is, whereas Unqomp and Reqomp substitute RCCX for
both gates of a pair. To retain the benefit of RCCX for such gates as
much as possible, the second step rewrites them individually: every
remaining Toffoli gate on an ancilla whose partner was replaced by MBU
is replaced by an RCCX gate and a Clifford correction, provided that
its target has the symbolic value $\emptyset$ ($\ket{0}$) or $\{\one\}$
($\ket{1}$), as in Eq.~(12) of \citet{amy-ross-rule}.
For $\ket{0}$, the RCCX gate yields $i^{ab}\ket{a,b,ab}$, and an
$S^\dagger$ gate on the target cancels the phase, as it contributes
$(-i)^{t}=(-i)^{ab}$:
\[
S_t^\dagger R\ket{a,b,0}=(-i)^{ab}\,i^{ab}\ket{a,b,ab}=\ket{a,b,ab}.
\]
For $\ket{1}$, we have $R\ket{a,b,1}=i^{ab}(-1)^{a}\ket{a,b,1\oplus ab}$,
and the phase is cancelled by an $S$ gate on the target, a $Z$ gate on
$a$, and a global phase of $-\pi/2$:
\[
e^{-i\pi/2}\,Z_a S_t R\ket{a,b,1}
=e^{-i\pi/2}\,(-1)^{a}\,i^{1\oplus ab}\,i^{ab}(-1)^{a}\ket{a,b,1\oplus ab}
=\ket{a,b,1\oplus ab},
\]
as $i^{ab}\,i^{1\oplus ab}=i$ for either value of $ab$. In both cases,
the $T$ count of the gate drops from 7 to 4, so the $T$-count reduction
of the RCCX substitution is conservatively retained for the Toffoli gates that
lose their partners to MBU\@. Neither the pre-processing nor the
post-processing changes the complexity bound of
Theorem~\ref{thm:complexity}, as both rely on the same symbolic
execution.

\section{Complete Results of RQ1}
\label{app:rq1-full}

Table~\ref{tab:rq1-full} lists the complete results of RQ1
(Sec.~\ref{sec:eval-rq1}) for all uncomputation synthesizers, without and with
the RCCX post-processing of Sec.~\ref{sec:impl}, including the worst-case
counts of \textsf{MBUndo}.

{\scriptsize\setlength{\tabcolsep}{1.6pt}
\begin{longtable}{@{}l l c r rr rrr rrr@{}}
\caption{Complete results of RQ1.}
\label{tab:rq1-full} \\
\toprule
Benchmark & Synthesizer & Post. & LT & \multicolumn{2}{c}{Baseline} & \multicolumn{3}{c}{Endpoint MBU} & \multicolumn{3}{c}{Interval MBU} \\
\cmidrule(lr){5-6} \cmidrule(lr){7-9} \cmidrule(l){10-12}
 & & & & \emph{T} & C & \emph{T} & C & M & \emph{T} & C & M \\
\midrule
\endfirsthead
\multicolumn{12}{@{}l}{\emph{Table~\ref{tab:rq1-full} (continued)}} \\
\toprule
Benchmark & Synthesizer & Post. & LT & \multicolumn{2}{c}{Baseline} & \multicolumn{3}{c}{Endpoint MBU} & \multicolumn{3}{c}{Interval MBU} \\
\cmidrule(lr){5-6} \cmidrule(lr){7-9} \cmidrule(l){10-12}
 & & & & \emph{T} & C & \emph{T} & C & M & \emph{T} & C & M \\
\midrule
\endhead
\midrule \multicolumn{12}{r@{}}{\emph{continued on the next page}} \\
\endfoot
\bottomrule
\endlastfoot
\texttt{Adder} & Unqomp & no & 11 & 308 & 420 & 231\,(231) & 348.5\,(354) & 11 & 154\,(154) & 266\,(277) & 11 \\
 &  & yes &  & 176 & 288 & 132\,(132) & 260.5\,(266) & 11 & 121\,(121) & 244\,(255) & 11 \\
 & Reqomp & no & 11 & 308 & 420 & 231\,(231) & 348.5\,(354) & 11 & 154\,(154) & 266\,(277) & 11 \\
 &  & yes &  & 176 & 288 & 132\,(132) & 260.5\,(266) & 11 & 121\,(121) & 244\,(255) & 11 \\
 & RwUn-Seq & & & \multicolumn{8}{c}{Timeout (${>}1000$ s)} \\
 & RwUn-Rev & no & 11 & 385 & 486 & 385\,(385) & 486\,(486) & 0 & 154\,(154) & 238.5\,(266) & 11 \\
 &  & yes &  & 346 & 458 & 346\,(346) & 458\,(458) & 0 & 121\,(121) & 216.5\,(244) & 11 \\
 & RwUn-Join & no & 11 & 385 & 486 & 385\,(385) & 486\,(486) & 0 & 154\,(154) & 238.5\,(266) & 11 \\
 &  & yes &  & 346 & 458 & 346\,(346) & 458\,(458) & 0 & 121\,(121) & 216.5\,(244) & 11 \\
\texttt{DeutschJozsa} & Unqomp & no & 8 & 119 & 158 & 63\,(63) & 106\,(110) & 8 & 63\,(63) & 106\,(110) & 8 \\
 &  & yes &  & 71 & 110 & 39\,(39) & 90\,(94) & 8 & 39\,(39) & 90\,(94) & 8 \\
 & Reqomp & no & 8 & 119 & 158 & 63\,(63) & 106\,(110) & 8 & 63\,(63) & 106\,(110) & 8 \\
 &  & yes &  & 71 & 110 & 39\,(39) & 90\,(94) & 8 & 39\,(39) & 90\,(94) & 8 \\
 & RwUn-Seq & & & \multicolumn{8}{c}{Error} \\
 & RwUn-Rev & & & \multicolumn{8}{c}{Error} \\
 & RwUn-Join & & & \multicolumn{8}{c}{Error} \\
\texttt{Grover} & Unqomp & no & 20 & 336 & 479 & 196\,(196) & 349\,(359) & 20 & 196\,(196) & 349\,(359) & 20 \\
 &  & yes &  & 216 & 359 & 136\,(136) & 309\,(319) & 20 & 136\,(136) & 309\,(319) & 20 \\
 & Reqomp & no & 20 & 336 & 479 & 196\,(196) & 349\,(359) & 20 & 196\,(196) & 349\,(359) & 20 \\
 &  & yes &  & 216 & 359 & 136\,(136) & 309\,(319) & 20 & 136\,(136) & 309\,(319) & 20 \\
 & RwUn-Seq & & & \multicolumn{8}{c}{Error} \\
 & RwUn-Rev & & & \multicolumn{8}{c}{Error} \\
 & RwUn-Join & & & \multicolumn{8}{c}{Error} \\
\texttt{IntegerComparator} & Unqomp & no & 11 & 147 & 215 & 105\,(105) & 176\,(179) & 6 & 77\,(77) & 141\,(146) & 10 \\
 &  & yes &  & 87 & 155 & 63\,(63) & 140\,(143) & 6 & 47\,(47) & 125\,(130) & 10 \\
 & Reqomp & no & 11 & 147 & 199 & 105\,(105) & 160\,(163) & 6 & 77\,(77) & 125\,(130) & 10 \\
 &  & yes &  & 87 & 139 & 63\,(63) & 124\,(127) & 6 & 47\,(47) & 109\,(114) & 10 \\
 & RwUn-Seq & no & 36 & 17787 & 20361 & 16751\,(17332) & 19244.5\,(19941) & 35 & 6961.5\,(13846) & 8294.5\,(16450) & 35 \\
 &  & yes &  & 17757 & 20342 & 16721\,(17302) & 19225.5\,(19922) & 35 & 6931.5\,(13816) & 8275.5\,(16431) & 35 \\
 & RwUn-Rev & no & 11 & 147 & 197 & 105\,(105) & 158\,(161) & 6 & 77\,(77) & 127\,(136) & 10 \\
 &  & yes &  & 99 & 154 & 75\,(75) & 139\,(142) & 6 & 47\,(47) & 108\,(117) & 10 \\
 & RwUn-Join & no & 11 & 147 & 191 & 105\,(105) & 152\,(155) & 6 & 77\,(77) & 121\,(128) & 10 \\
 &  & yes &  & 99 & 148 & 75\,(75) & 133\,(136) & 6 & 47\,(47) & 102\,(109) & 10 \\
\texttt{MCX} & Unqomp & no & 10 & 147 & 168 & 77\,(77) & 103\,(108) & 10 & 77\,(77) & 103\,(108) & 10 \\
 &  & yes &  & 87 & 108 & 47\,(47) & 83\,(88) & 10 & 47\,(47) & 83\,(88) & 10 \\
 & Reqomp & no & 10 & 147 & 168 & 77\,(77) & 103\,(108) & 10 & 77\,(77) & 103\,(108) & 10 \\
 &  & yes &  & 87 & 108 & 47\,(47) & 83\,(88) & 10 & 47\,(47) & 83\,(88) & 10 \\
 & RwUn-Seq & no & 30 & 2772 & 3168 & 1725.5\,(2359) & 2031.5\,(2785) & 30 & 1074.5\,(2072) & 1308\,(2498) & 30 \\
 &  & yes &  & 2739 & 3144 & 1695.5\,(2329) & 2011.5\,(2765) & 30 & 1044.5\,(2042) & 1288\,(2478) & 30 \\
 & RwUn-Rev & no & 10 & 147 & 168 & 77\,(77) & 103\,(108) & 10 & 77\,(77) & 103\,(108) & 10 \\
 &  & yes &  & 87 & 108 & 47\,(47) & 83\,(88) & 10 & 47\,(47) & 83\,(88) & 10 \\
 & RwUn-Join & no & 10 & 147 & 168 & 77\,(77) & 103\,(108) & 10 & 77\,(77) & 103\,(108) & 10 \\
 &  & yes &  & 87 & 108 & 47\,(47) & 83\,(88) & 10 & 47\,(47) & 83\,(88) & 10 \\
\texttt{MCRY}$^\dagger$ & Unqomp & no & 11 & 154 & 178 & 77\,(77) & 106.5\,(112) & 11 & 77\,(77) & 106.5\,(112) & 11 \\
 &  & yes &  & 88 & 112 & 44\,(44) & 84.5\,(90) & 11 & 44\,(44) & 84.5\,(90) & 11 \\
 & Reqomp & no & 11 & 154 & 178 & 77\,(77) & 106.5\,(112) & 11 & 77\,(77) & 106.5\,(112) & 11 \\
 &  & yes &  & 88 & 112 & 44\,(44) & 84.5\,(90) & 11 & 44\,(44) & 84.5\,(90) & 11 \\
 & RwUn-Seq & no & 1 & 296 & 342 & 215.5\,(282) & 252\,(329) & 1 & 215.5\,(282) & 252\,(329) & 1 \\
 &  & yes &  & 296 & 342 & 215.5\,(282) & 252\,(329) & 1 & 215.5\,(282) & 252\,(329) & 1 \\
 & RwUn-Rev & no & 1 & 296 & 342 & 215.5\,(282) & 252\,(329) & 1 & 215.5\,(282) & 252\,(329) & 1 \\
 &  & yes &  & 296 & 342 & 215.5\,(282) & 252\,(329) & 1 & 215.5\,(282) & 252\,(329) & 1 \\
 & RwUn-Join & no & 1 & 296 & 342 & 215.5\,(282) & 252\,(329) & 1 & 215.5\,(282) & 252\,(329) & 1 \\
 &  & yes &  & 296 & 342 & 215.5\,(282) & 252\,(329) & 1 & 215.5\,(282) & 252\,(329) & 1 \\
\texttt{Multiplier} & Unqomp & no & 65 & 770 & 1010 & 315\,(315) & 587.5\,(620) & 65 & 245\,(245) & 512.5\,(550) & 65 \\
 &  & yes &  & 440 & 680 & 180\,(180) & 477.5\,(510) & 65 & 170\,(170) & 462.5\,(500) & 65 \\
 & Reqomp & no & 65 & 770 & 970 & 315\,(315) & 547.5\,(580) & 65 & 245\,(245) & 472.5\,(510) & 65 \\
 &  & yes &  & 440 & 640 & 180\,(180) & 437.5\,(470) & 65 & 170\,(170) & 422.5\,(460) & 65 \\
 & RwUn-Seq & & & \multicolumn{8}{c}{Timeout (${>}1000$ s)} \\
 & RwUn-Rev & & & \multicolumn{8}{c}{Timeout (${>}1000$ s)} \\
 & RwUn-Join & & & \multicolumn{8}{c}{Timeout (${>}1000$ s)} \\
\texttt{PiecewiseLinearR}$^\dagger$ & Unqomp & no & 44 & 1078 & 1564 & 931\,(931) & 1427.5\,(1438) & 21 & 833\,(833) & 1311\,(1329) & 36 \\
 &  & yes &  & 868 & 1354 & 784\,(784) & 1301.5\,(1312) & 21 & 728\,(728) & 1255\,(1273) & 36 \\
 & Reqomp & no & 42 & 1078 & 1530 & 931\,(931) & 1393.5\,(1404) & 21 & 833\,(833) & 1277\,(1295) & 36 \\
 &  & yes &  & 868 & 1320 & 784\,(784) & 1267.5\,(1278) & 21 & 728\,(728) & 1221\,(1239) & 36 \\
 & RwUn-Seq & & & \multicolumn{8}{c}{Error} \\
 & RwUn-Rev & & & \multicolumn{8}{c}{Error} \\
 & RwUn-Join & & & \multicolumn{8}{c}{Error} \\
\texttt{PolynomialPauliR}$^\dagger$ & Unqomp & no & 45 & 630 & 780 & 315\,(315) & 487.5\,(510) & 45 & 315\,(315) & 487.5\,(510) & 45 \\
 &  & yes &  & 360 & 510 & 180\,(180) & 397.5\,(420) & 45 & 180\,(180) & 397.5\,(420) & 45 \\
 & Reqomp & no & 45 & 630 & 780 & 315\,(315) & 487.5\,(510) & 45 & 315\,(315) & 487.5\,(510) & 45 \\
 &  & yes &  & 360 & 510 & 180\,(180) & 397.5\,(420) & 45 & 180\,(180) & 397.5\,(420) & 45 \\
 & RwUn-Seq & & & \multicolumn{8}{c}{Error} \\
 & RwUn-Rev & & & \multicolumn{8}{c}{Error} \\
 & RwUn-Join & & & \multicolumn{8}{c}{Error} \\
\texttt{WeightedAdder} & Unqomp & no & 90 & 1386 & 1699 & 798\,(798) & 1153\,(1195) & 84 & 798\,(798) & 1148\,(1195) & 84 \\
 &  & yes &  & 912 & 1225 & 576\,(576) & 1005\,(1047) & 84 & 576\,(576) & 1000\,(1047) & 84 \\
 & Reqomp & no & 77 & 1246 & 1499 & 728\,(728) & 1018\,(1055) & 74 & 728\,(728) & 1013\,(1055) & 74 \\
 &  & yes &  & 832 & 1085 & 536\,(536) & 890\,(927) & 74 & 536\,(536) & 885\,(927) & 74 \\
 & RwUn-Seq & & & \multicolumn{8}{c}{Timeout (${>}1000$ s)} \\
 & RwUn-Rev & & & \multicolumn{8}{c}{Timeout (${>}1000$ s)} \\
 & RwUn-Join & & & \multicolumn{8}{c}{Timeout (${>}1000$ s)} \\
\texttt{Incrementer} & Unqomp & no & 5 & 70 & 87 & 35\,(35) & 54.5\,(57) & 5 & 35\,(35) & 54.5\,(57) & 5 \\
 &  & yes &  & 40 & 57 & 20\,(20) & 44.5\,(47) & 5 & 20\,(20) & 44.5\,(47) & 5 \\
 & Reqomp & no & 5 & 70 & 87 & 35\,(35) & 54.5\,(57) & 5 & 35\,(35) & 54.5\,(57) & 5 \\
 &  & yes &  & 40 & 57 & 20\,(20) & 44.5\,(47) & 5 & 20\,(20) & 44.5\,(47) & 5 \\
 & RwUn-Seq & no & 9 & 2870 & 3322 & 2800\,(2807) & 3253.5\,(3266) & 7 & 658\,(1281) & 896.5\,(1737) & 9 \\
 &  & yes &  & 2855 & 3312 & 2785\,(2792) & 3243.5\,(3256) & 7 & 643\,(1266) & 886.5\,(1727) & 9 \\
 & RwUn-Rev & no & 5 & 70 & 94 & 70\,(70) & 94\,(94) & 0 & 35\,(35) & 57.5\,(63) & 5 \\
 &  & yes &  & 55 & 84 & 55\,(55) & 84\,(84) & 0 & 20\,(20) & 47.5\,(53) & 5 \\
 & RwUn-Join & no & 5 & 70 & 94 & 70\,(70) & 94\,(94) & 0 & 35\,(35) & 57.5\,(63) & 5 \\
 &  & yes &  & 55 & 84 & 55\,(55) & 84\,(84) & 0 & 20\,(20) & 47.5\,(53) & 5 \\
\texttt{CondCleanMCS} & Unqomp & & & \multicolumn{8}{c}{Error} \\
 & Reqomp & no & 93 & 1389 & 1602 & 738\,(738) & 997.5\,(1044) & 93 & 738\,(738) & 997.5\,(1044) & 93 \\
 &  & yes &  & 795 & 1008 & 423\,(423) & 775.5\,(822) & 93 & 423\,(423) & 775.5\,(822) & 93 \\
 & RwUn-Seq & no & 2 & 4105 & 4695 & 3720\,(4077) & 4259\,(4669) & 2 & 2649\,(3979) & 3041.5\,(4571) & 2 \\
 &  & yes &  & 4105 & 4695 & 3720\,(4077) & 4259\,(4669) & 2 & 2649\,(3979) & 3041.5\,(4571) & 2 \\
 & RwUn-Rev & no & 2 & 3398 & 3887 & 3034\,(3370) & 3475\,(3861) & 2 & 2302.5\,(3286) & 2644.5\,(3777) & 2 \\
 &  & yes &  & 3398 & 3887 & 3034\,(3370) & 3475\,(3861) & 2 & 2302.5\,(3286) & 2644.5\,(3777) & 2 \\
 & RwUn-Join & no & 2 & 3398 & 3887 & 3034\,(3370) & 3475\,(3861) & 2 & 2302.5\,(3286) & 2644.5\,(3777) & 2 \\
 &  & yes &  & 3398 & 3887 & 3034\,(3370) & 3475\,(3861) & 2 & 2302.5\,(3286) & 2644.5\,(3777) & 2 \\
\texttt{RiseCondCleanMCS} & Unqomp & & & \multicolumn{8}{c}{Error} \\
 & Reqomp & no & 2 & 101 & 128 & 87\,(87) & 115\,(116) & 2 & 87\,(87) & 115\,(116) & 2 \\
 &  & yes &  & 59 & 86 & 51\,(51) & 81\,(82) & 2 & 51\,(51) & 81\,(82) & 2 \\
 & RwUn-Seq & no & 10 & 766 & 888 & 559.5\,(647) & 670.5\,(779) & 10 & 286.5\,(486) & 370\,(618) & 10 \\
 &  & yes &  & 727 & 850 & 523.5\,(611) & 636.5\,(745) & 10 & 250.5\,(450) & 336\,(584) & 10 \\
 & RwUn-Rev & no & 2 & 633 & 736 & 608.5\,(612) & 711.5\,(717) & 2 & 241\,(395) & 307\,(500) & 2 \\
 &  & yes &  & 594 & 698 & 572.5\,(576) & 677.5\,(683) & 2 & 205\,(359) & 273\,(466) & 2 \\
 & RwUn-Join & no & 2 & 633 & 736 & 608.5\,(612) & 711.5\,(717) & 2 & 241\,(395) & 307\,(500) & 2 \\
 &  & yes &  & 594 & 698 & 572.5\,(576) & 677.5\,(683) & 2 & 205\,(359) & 273\,(466) & 2 \\
\texttt{TwoSidedComparator} & Unqomp & no & 97 & 2674 & 3814 & 1995\,(1995) & 3183.5\,(3232) & 97 & 1344\,(1344) & 2485.5\,(2581) & 97 \\
 &  & yes &  & 1534 & 2674 & 1146\,(1146) & 2431.5\,(2480) & 97 & 1053\,(1053) & 2291.5\,(2387) & 97 \\
 & Reqomp & no & 97 & 2674 & 3530 & 1995\,(1995) & 2899.5\,(2948) & 97 & 1344\,(1344) & 2201.5\,(2297) & 97 \\
 &  & yes &  & 1534 & 2390 & 1146\,(1146) & 2147.5\,(2196) & 97 & 1053\,(1053) & 2007.5\,(2103) & 97 \\
 & RwUn-Seq & & & \multicolumn{8}{c}{Timeout (${>}1000$ s)} \\
 & RwUn-Rev & & & \multicolumn{8}{c}{Timeout (${>}1000$ s)} \\
 & RwUn-Join & & & \multicolumn{8}{c}{Timeout (${>}1000$ s)} \\
\texttt{VBEModAddition} & Unqomp & & & \multicolumn{8}{c}{Error} \\
 & Reqomp & no & 135 & 2800 & 3972 & 2065\,(2065) & 3289.5\,(3342) & 105 & 1379\,(1379) & 2540\,(2656) & 134 \\
 &  & yes &  & 1600 & 2772 & 1180\,(1180) & 2503.5\,(2556) & 105 & 1082\,(1082) & 2342\,(2458) & 134 \\
 & RwUn-Seq & & & \multicolumn{8}{c}{Timeout (${>}1000$ s)} \\
 & RwUn-Rev & no & 136 & 3290 & 4387 & 2961\,(2961) & 4081.5\,(4105) & 47 & 1379\,(1379) & 2344.5\,(2580) & 133 \\
 &  & yes &  & 2804 & 4001 & 2490\,(2490) & 3709.5\,(3733) & 47 & 1082\,(1082) & 2146.5\,(2382) & 133 \\
 & RwUn-Join & & & \multicolumn{8}{c}{Timeout (${>}1000$ s)} \\
\texttt{TableLookup} & Unqomp & no & 71 & 840 & 1224 & 420\,(420) & 834\,(864) & 60 & 364\,(364) & 763\,(817) & 69 \\
 &  & yes &  & 480 & 864 & 240\,(240) & 714\,(744) & 60 & 208\,(208) & 659\,(713) & 69 \\
 & Reqomp & no & 71 & 840 & 1224 & 420\,(420) & 834\,(864) & 60 & 364\,(364) & 763\,(817) & 69 \\
 &  & yes &  & 480 & 864 & 240\,(240) & 714\,(744) & 60 & 208\,(208) & 659\,(713) & 69 \\
 & RwUn-Seq & no & 42 & 2478 & 2972 & 2110.5\,(2149) & 2615.5\,(2683) & 40 & 616\,(1050) & 983.5\,(1585) & 41 \\
 &  & yes &  & 2388 & 2912 & 2020.5\,(2059) & 2555.5\,(2623) & 40 & 538\,(972) & 931.5\,(1533) & 41 \\
 & RwUn-Rev & no & 40 & 2254 & 2716 & 1921.5\,(1953) & 2395.5\,(2453) & 38 & 588\,(994) & 937.5\,(1495) & 39 \\
 &  & yes &  & 2164 & 2656 & 1831.5\,(1863) & 2335.5\,(2393) & 38 & 510\,(916) & 885.5\,(1443) & 39 \\
 & RwUn-Join & no & 59 & 686 & 999 & 497\,(497) & 823.5\,(837) & 27 & 315\,(315) & 630\,(686) & 59 \\
 &  & yes &  & 482 & 825 & 350\,(350) & 725.5\,(739) & 27 & 180\,(180) & 540\,(596) & 59 \\
\texttt{GF8Multiply} & Unqomp & no & 3 & 126 & 153 & 112\,(112) & 140\,(141) & 2 & 84\,(84) & 110\,(114) & 3 \\
 &  & yes &  & 72 & 99 & 64\,(64) & 94\,(95) & 2 & 60\,(60) & 88\,(92) & 3 \\
 & Reqomp & no & 3 & 126 & 153 & 112\,(112) & 140\,(141) & 2 & 63\,(63) & 87\,(93) & 3 \\
 &  & yes &  & 72 & 99 & 64\,(64) & 94\,(95) & 2 & 57\,(57) & 83\,(89) & 3 \\
 & RwUn-Seq & & & \multicolumn{8}{c}{Error} \\
 & RwUn-Rev & & & \multicolumn{8}{c}{Error} \\
 & RwUn-Join & no & 3 & 126 & 153 & 112\,(112) & 140\,(141) & 2 & 84\,(84) & 110\,(114) & 3 \\
 &  & yes &  & 72 & 99 & 64\,(64) & 94\,(95) & 2 & 60\,(60) & 88\,(92) & 3 \\
\texttt{KeccakChi} & Unqomp & no & 200 & 2800 & 4600 & 2800\,(2800) & 4600\,(4600) & 0 & 1400\,(1400) & 3200\,(3400) & 200 \\
 &  & yes &  & 1600 & 3400 & 1600\,(1600) & 3400\,(3400) & 0 & 1400\,(1400) & 3200\,(3400) & 200 \\
 & Reqomp & no & 200 & 2800 & 4200 & 2800\,(2800) & 4200\,(4200) & 0 & 1400\,(1400) & 2800\,(3000) & 200 \\
 &  & yes &  & 1600 & 3000 & 1600\,(1600) & 3000\,(3000) & 0 & 1400\,(1400) & 2800\,(3000) & 200 \\
 & RwUn-Seq & no & 200 & 2800 & 4400 & 2800\,(2800) & 4400\,(4400) & 0 & 1400\,(1400) & 2900\,(3200) & 200 \\
 &  & yes &  & 2800 & 4400 & 2800\,(2800) & 4400\,(4400) & 0 & 1400\,(1400) & 2900\,(3200) & 200 \\
 & RwUn-Rev & no & 200 & 2800 & 4400 & 2800\,(2800) & 4400\,(4400) & 0 & 1400\,(1400) & 2900\,(3200) & 200 \\
 &  & yes &  & 2800 & 4400 & 2800\,(2800) & 4400\,(4400) & 0 & 1400\,(1400) & 2900\,(3200) & 200 \\
 & RwUn-Join & no & 200 & 2800 & 4400 & 2800\,(2800) & 4400\,(4400) & 0 & 1400\,(1400) & 2900\,(3200) & 200 \\
 &  & yes &  & 2800 & 4400 & 2800\,(2800) & 4400\,(4400) & 0 & 1400\,(1400) & 2900\,(3200) & 200 \\
\texttt{SpeedySbox} & Unqomp & no & 116 & 3108 & 3604 & 2366\,(2366) & 2915\,(2968) & 106 & 1554\,(1554) & 2044\,(2164) & 116 \\
 &  & yes &  & 1776 & 2272 & 1352\,(1352) & 2007\,(2060) & 106 & 1236\,(1236) & 1832\,(1952) & 116 \\
 & Reqomp & no & 116 & 3108 & 3604 & 2366\,(2366) & 2915\,(2968) & 106 & 1554\,(1554) & 2044\,(2164) & 116 \\
 &  & yes &  & 1776 & 2272 & 1352\,(1352) & 2007\,(2060) & 106 & 1236\,(1236) & 1832\,(1952) & 116 \\
 & RwUn-Seq & no & 124 & 4060 & 4692 & 2786\,(2982) & 3427\,(3728) & 114 & 1750\,(1946) & 2316\,(2700) & 124 \\
 &  & yes &  & 2788 & 3452 & 1736\,(1932) & 2483\,(2784) & 114 & 1432\,(1628) & 2104\,(2488) & 124 \\
 & RwUn-Rev & no & 116 & 5852 & 6740 & 5110\,(5110) & 6051\,(6104) & 106 & 2198\,(2842) & 2884\,(3844) & 116 \\
 &  & yes &  & 5222 & 6112 & 4792\,(4792) & 5839\,(5892) & 106 & 1880\,(2524) & 2672\,(3632) & 116 \\
 & RwUn-Join & no & 116 & 3108 & 3604 & 2366\,(2366) & 2915\,(2968) & 106 & 1554\,(1554) & 2044\,(2164) & 116 \\
 &  & yes &  & 1776 & 2272 & 1352\,(1352) & 2007\,(2060) & 106 & 1236\,(1236) & 1832\,(1952) & 116 \\
\texttt{GidneyAdder} & Unqomp & no & 129 & 1806 & 2835 & 903\,(903) & 1996.5\,(2061) & 129 & 903\,(903) & 1932.5\,(2061) & 129 \\
 &  & yes &  & 1032 & 2061 & 516\,(516) & 1738.5\,(1803) & 129 & 516\,(516) & 1674.5\,(1803) & 129 \\
 & Reqomp & no & 129 & 1806 & 2835 & 903\,(903) & 1996.5\,(2061) & 129 & 903\,(903) & 1932.5\,(2061) & 129 \\
 &  & yes &  & 1032 & 2061 & 516\,(516) & 1738.5\,(1803) & 129 & 516\,(516) & 1674.5\,(1803) & 129 \\
 & RwUn-Seq & & & \multicolumn{8}{c}{Timeout (${>}1000$ s)} \\
 & RwUn-Rev & no & 129 & 3598 & 5012 & 2702\,(2702) & 4180\,(4244) & 128 & 903\,(903) & 2125\,(2446) & 129 \\
 &  & yes &  & 3211 & 4754 & 2315\,(2315) & 3922\,(3986) & 128 & 516\,(516) & 1867\,(2188) & 129 \\
 & RwUn-Join & & & \multicolumn{8}{c}{Timeout (${>}1000$ s)} \\
\texttt{QLZOC} & Unqomp & no & 152 & 2177 & 3119 & 1113\,(1113) & 2131\,(2207) & 152 & 1113\,(1113) & 2131\,(2207) & 152 \\
 &  & yes &  & 1265 & 2207 & 657\,(657) & 1827\,(1903) & 152 & 657\,(657) & 1827\,(1903) & 152 \\
 & Reqomp & no & 152 & 2177 & 3119 & 1113\,(1113) & 2131\,(2207) & 152 & 1113\,(1113) & 2131\,(2207) & 152 \\
 &  & yes &  & 1265 & 2207 & 657\,(657) & 1827\,(1903) & 152 & 657\,(657) & 1827\,(1903) & 152 \\
 & RwUn-Seq & & & \multicolumn{8}{c}{Timeout (${>}1000$ s)} \\
 & RwUn-Rev & & & \multicolumn{8}{c}{Timeout (${>}1000$ s)} \\
 & RwUn-Join & & & \multicolumn{8}{c}{Timeout (${>}1000$ s)} \\
\texttt{MatrixVectorEq} & Unqomp & no & 126 & 2667 & 3304 & 1785\,(1785) & 2485\,(2548) & 126 & 1337\,(1337) & 1909\,(2036) & 126 \\
 &  & yes &  & 1527 & 2164 & 1023\,(1023) & 1849\,(1912) & 126 & 959\,(959) & 1657\,(1784) & 126 \\
 & Reqomp & no & 126 & 2667 & 3304 & 1785\,(1785) & 2485\,(2548) & 126 & 1337\,(1337) & 1909\,(2036) & 126 \\
 &  & yes &  & 1527 & 2164 & 1023\,(1023) & 1849\,(1912) & 126 & 959\,(959) & 1657\,(1784) & 126 \\
 & RwUn-Seq & & & \multicolumn{8}{c}{Timeout (${>}1000$ s)} \\
 & RwUn-Rev & no & 126 & 2667 & 3304 & 1785\,(1785) & 2485\,(2548) & 126 & 1337\,(1337) & 1909\,(2036) & 126 \\
 &  & yes &  & 1527 & 2164 & 1023\,(1023) & 1849\,(1912) & 126 & 959\,(959) & 1657\,(1784) & 126 \\
 & RwUn-Join & & & \multicolumn{8}{c}{Timeout (${>}1000$ s)} \\
\end{longtable}\par\vspace{-4pt}\noindent\footnotesize $^\dagger$\,\texttt{MCRY}, \texttt{PiecewiseLinearR}, and \texttt{PolynomialPauliR} also contain 2, 182, and 61 rotation gates, which no tool changes.\par}

\section{Complete Results of RQ2}
\label{app:rq2-mono}

Table~\ref{tab:rq2-mono} lists, for every benchmark of RQ1, the $T$ and Clifford
counts of the circuit of Reqomp, the lowest counts that monolithic
MBU~\cite{MBU-for-modular-arithmetic} can reach as explained in
Sec.~\ref{sec:eval-rq2}, and the counts of the circuit that \textsf{MBUndo}
produces from the output of Reqomp with interval MBU\@.

\begin{table}[ht]
\caption{Comparison with monolithic MBU~\cite{MBU-for-modular-arithmetic} on the output of Reqomp. For each benchmark, the table lists the \emph{T} and Clifford counts of the input circuit of Reqomp without uncomputation, of the output circuit of Reqomp, of the best case of monolithic MBU, i.e., with half of the uncomputation gates saved in expectation, and of the circuit of \textsf{MBUndo} with interval MBU (in expectation, with worst case in parentheses where it differs), and the reduction (-$\Delta$) of the latter two relative to the output of Reqomp.}
\label{tab:rq2-mono}
\centering\footnotesize
\setlength{\tabcolsep}{2.6pt}
\begin{tabular}{@{}l rr rr rrrr rrrr@{}}
\toprule
 & \multicolumn{2}{c}{Reqomp input} & \multicolumn{2}{c}{Reqomp} & \multicolumn{4}{c}{Monolithic MBU (best case)} & \multicolumn{4}{c}{\textsf{MBUndo} (interval MBU)} \\
\cmidrule(lr){2-3} \cmidrule(lr){4-5} \cmidrule(lr){6-9} \cmidrule(l){10-13}
Benchmark & \emph{T} & C & \emph{T} & C & \emph{T} & -$\Delta$ & C & -$\Delta$ & \emph{T} & -$\Delta$ & C & -$\Delta$ \\
\midrule
\texttt{Adder} & 154 & 222 & 308 & 420 & 231 & 25.0\% & 321 & 23.6\% & 154 & 50.0\% & 266\,(277) & 36.7\% \\
\texttt{DeutschJozsa} & 63 & 94 & 119 & 158 & 91 & 23.5\% & 126 & 20.3\% & 63 & 47.1\% & 106\,(110) & 32.9\% \\
\texttt{Grover} & 196 & 319 & 336 & 479 & 266 & 20.8\% & 399 & 16.7\% & 196 & 41.7\% & 349\,(359) & 27.1\% \\
\texttt{IntegerComparator} & 77 & 114 & 147 & 199 & 112 & 23.8\% & 156.5 & 21.4\% & 77 & 47.6\% & 125\,(130) & 37.2\% \\
\texttt{MCX} & 77 & 88 & 147 & 168 & 112 & 23.8\% & 128 & 23.8\% & 77 & 47.6\% & 103\,(108) & 38.7\% \\
\texttt{MCRY} & 77 & 90 & 154 & 178 & 115.5 & 25.0\% & 134 & 24.7\% & 77 & 50.0\% & 106.5\,(112) & 40.2\% \\
\texttt{Multiplier} & 385 & 530 & 770 & 970 & 577.5 & 25.0\% & 750 & 22.7\% & 245 & 68.2\% & 472.5\,(510) & 51.3\% \\
\texttt{PiecewiseLinearR} & 833 & 1209 & 1078 & 1530 & 955.5 & 11.4\% & 1369.5 & 10.5\% & 833 & 22.7\% & 1277\,(1295) & 16.5\% \\
\texttt{PolynomialPauliR} & 315 & 420 & 630 & 780 & 472.5 & 25.0\% & 600 & 23.1\% & 315 & 50.0\% & 487.5\,(510) & 37.5\% \\
\texttt{WeightedAdder} & 763 & 937 & 1246 & 1499 & 1004.5 & 19.4\% & 1218 & 18.7\% & 728 & 41.6\% & 1013\,(1055) & 32.4\% \\
\texttt{Incrementer} & 35 & 47 & 70 & 87 & 52.5 & 25.0\% & 67 & 23.0\% & 35 & 50.0\% & 54.5\,(57) & 37.4\% \\
\texttt{CondCleanMCS} & 682 & 786 & 1389 & 1602 & 1035.5 & 25.4\% & 1194 & 25.5\% & 738 & 46.9\% & 997.5\,(1044) & 37.7\% \\
\texttt{RiseCondCleanMCS} & 87 & 112 & 101 & 128 & 94 & 6.9\% & 120 & 6.2\% & 87 & 13.9\% & 115\,(116) & 10.2\% \\
\texttt{TwoSidedComparator} & 1344 & 1916 & 2674 & 3530 & 2009 & 24.9\% & 2723 & 22.9\% & 1344 & 49.7\% & 2201.5\,(2297) & 37.6\% \\
\texttt{VBEModAddition} & 1400 & 2114 & 2800 & 3972 & 2100 & 25.0\% & 3043 & 23.4\% & 1379 & 50.8\% & 2540\,(2656) & 36.1\% \\
\texttt{TableLookup} & 420 & 617 & 840 & 1224 & 630 & 25.0\% & 920.5 & 24.8\% & 364 & 56.7\% & 763\,(817) & 37.7\% \\
\texttt{GF8Multiply} & 63 & 78 & 126 & 153 & 94.5 & 25.0\% & 115.5 & 24.5\% & 63 & 50.0\% & 87\,(93) & 43.1\% \\
\texttt{KeccakChi} & 1400 & 2400 & 2800 & 4200 & 2100 & 25.0\% & 3300 & 21.4\% & 1400 & 50.0\% & 2800\,(3000) & 33.3\% \\
\texttt{SpeedySbox} & 1554 & 1808 & 3108 & 3604 & 2331 & 25.0\% & 2706 & 24.9\% & 1554 & 50.0\% & 2044\,(2164) & 43.3\% \\
\texttt{GidneyAdder} & 903 & 1675 & 1806 & 2835 & 1354.5 & 25.0\% & 2255 & 20.5\% & 903 & 50.0\% & 1932.5\,(2061) & 31.8\% \\
\texttt{QLZOC} & 1113 & 1903 & 2177 & 3119 & 1645 & 24.4\% & 2511 & 19.5\% & 1113 & 48.9\% & 2131\,(2207) & 31.7\% \\
\texttt{MatrixVectorEq} & 1337 & 1656 & 2667 & 3304 & 2002 & 24.9\% & 2480 & 24.9\% & 1337 & 49.9\% & 1909\,(2036) & 42.2\% \\
\bottomrule
\end{tabular}
\end{table}

\section{Complete Results of RQ3}
\label{app:rq3-charts}

\begin{figure}[p]
\centering
\includegraphics[width=\textwidth]{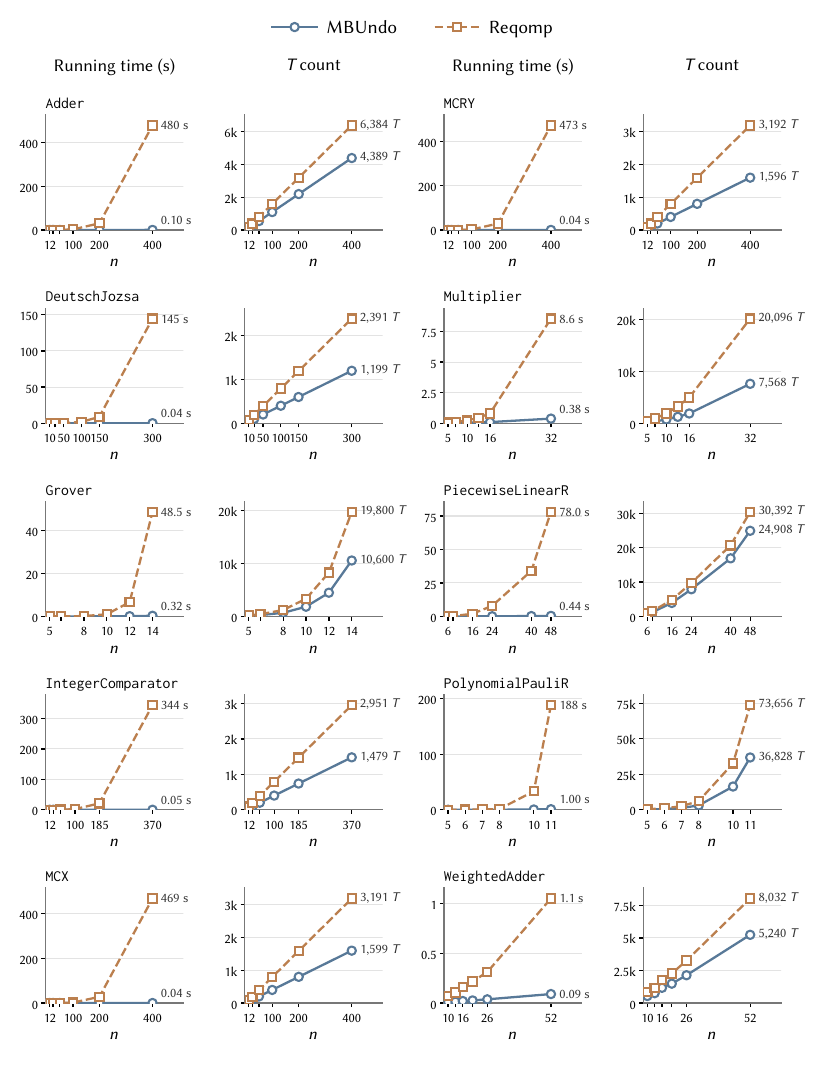}
\caption{Running time and \emph{T} count of \textsf{MBUndo} and Reqomp on the ten benchmarks of Reqomp at six sizes \emph{n} each, where \textsf{MBUndo} is run on top of the output circuit of Reqomp.}
\label{fig:rq3-all}
\end{figure}

Figure~\ref{fig:rq3-all} shows the results of RQ3 (Sec.~\ref{sec:eval-rq3}) on all ten benchmarks of Reqomp, in the format of Figure~\ref{fig:rq3-adder}: for each benchmark, the running time and $T$ count of \textsf{MBUndo} applying interval MBU to the output of Reqomp, and those of Reqomp itself, at six sizes $n$.

\section{Complete Results of RQ4}
\label{app:rq4-full}

Table~\ref{tab:rq4-full} lists, for every benchmark of RQ4
(Sec.~\ref{sec:eval-rq4}), the gate counts of the output of Reqomp and of the
circuits that SPARE, Feynman, and Tzap produce from it, together with the
running time of each tool. The corresponding counts of \textsf{MBUndo} are
those of interval MBU with RCCX post-processing.

\begin{table}[t]
\caption{Results of the circuit optimizers of RQ4 on the output of Reqomp: running time and the numbers of $T$ gates, rotation gates (R), and Clifford gates after RCCX post-processing (for SPARE, after its own post-processing). The Reqomp rows characterize the input circuits for the optimizers.}
\label{tab:rq4-full}
\centering\footnotesize
\setlength{\tabcolsep}{2pt}
\begin{tabular}{@{}l l r r r r @{\hspace{8pt}} l l r r r r@{}}
\toprule
Benchmark & Tool & Time (s) & $T$ & R & Clifford & Benchmark & Tool & Time (s) & $T$ & R & Clifford \\
\midrule
\texttt{Adder} & Reqomp & -- & 176 & 0 & 288 & \texttt{CondCleanMCS} & Reqomp & -- & 795 & 0 & 1008 \\
 & SPARE & 9.0 & 180 & 0 & 264 &  & SPARE & \multicolumn{4}{c}{timeout ($>$1000\,s)} \\
 & Feynman & 0.006 & 88 & 0 & 242 &  & Feynman & 0.026 & 721 & 0 & 926 \\
 & Tzap & 0.027 & 88 & 0 & 176 &  & Tzap & 0.027 & 721 & 0 & 930 \\
\addlinespace[2.5pt]
\texttt{DeutschJozsa} & Reqomp & -- & 71 & 0 & 110 & \texttt{RiseCondCleanMCS} & Reqomp & -- & 59 & 0 & 86 \\
 & SPARE & 3.0 & 71 & 0 & 111 &  & SPARE & \multicolumn{4}{c}{Error} \\
 & Feynman & 0.001 & 71 & 0 & 108 &  & Feynman & 0.001 & 59 & 0 & 86 \\
 & Tzap & 0.024 & 71 & 0 & 108 &  & Tzap & 0.025 & 59 & 0 & 85 \\
\addlinespace[2.5pt]
\texttt{Grover} & Reqomp & -- & 216 & 0 & 359 & \texttt{TwoSidedComparator} & Reqomp & -- & 1534 & 0 & 2390 \\
 & SPARE & 8.1 & 216 & 0 & 314 &  & SPARE & 305.8 & 1534 & 0 & 2488 \\
 & Feynman & 0.005 & 178 & 0 & 302 &  & Feynman & 0.085 & 784 & 0 & 2107 \\
 & Tzap & 0.026 & 178 & 0 & 292 &  & Tzap & 0.029 & 784 & 0 & 1873 \\
\addlinespace[2.5pt]
\texttt{IntegerComparator} & Reqomp & -- & 87 & 0 & 139 & \texttt{VBEModAddition} & Reqomp & -- & 1600 & 0 & 2772 \\
 & SPARE & 3.1 & 79 & 0 & 137 &  & SPARE & 838.3 & 1567 & 0 & 3042 \\
 & Feynman & 0.001 & 87 & 0 & 139 &  & Feynman & 0.105 & 1024 & 0 & 2516 \\
 & Tzap & 0.025 & 87 & 0 & 139 &  & Tzap & 0.030 & 1016 & 0 & 2128 \\
\addlinespace[2.5pt]
\texttt{MCX} & Reqomp & -- & 87 & 0 & 108 & \texttt{TableLookup} & Reqomp & -- & 480 & 0 & 864 \\
 & SPARE & 2.7 & 87 & 0 & 108 &  & SPARE & 85.7 & 320 & 0 & 589 \\
 & Feynman & 0.001 & 87 & 0 & 108 &  & Feynman & 0.015 & 232 & 0 & 766 \\
 & Tzap & 0.025 & 87 & 0 & 108 &  & Tzap & 0.029 & 172 & 0 & 500 \\
\addlinespace[2.5pt]
\texttt{MCRY} & Reqomp & -- & 88 & 2 & 112 & \texttt{GF8Multiply} & Reqomp & -- & 72 & 0 & 99 \\
 & SPARE & 2.7 & 80 & 6 & 108 &  & SPARE & 2.6 & 84 & 0 & 96 \\
 & Feynman & 0.001 & 88 & 2 & 112 &  & Feynman & 0.001 & 56 & 0 & 83 \\
 & Tzap & 0.024 & 88 & 2 & 116 &  & Tzap & 0.025 & 52 & 0 & 83 \\
\addlinespace[2.5pt]
\texttt{Multiplier} & Reqomp & -- & 440 & 0 & 640 & \texttt{KeccakChi} & Reqomp & -- & 1600 & 0 & 3000 \\
 & SPARE & 16.7 & 236 & 0 & 326 &  & SPARE & \multicolumn{4}{c}{Timeout ($>$1000\,s)} \\
 & Feynman & 0.009 & 248 & 0 & 568 &  & Feynman & 0.113 & 1600 & 0 & 3000 \\
 & Tzap & 0.026 & 248 & 0 & 448 &  & Tzap & 0.028 & 882 & 0 & 3160 \\
\addlinespace[2.5pt]
\texttt{PiecewiseLinearR} & Reqomp & -- & 868 & 182 & 1320 & \texttt{SpeedySbox} & Reqomp & -- & 1776 & 0 & 2272 \\
 & SPARE & 173.3 & 830 & 175 & 1341 &  & SPARE & 168.3 & 1726 & 0 & 2208 \\
 & Feynman & 0.036 & 548 & 182 & 1255 &  & Feynman & 0.090 & 1008 & 0 & 1640 \\
 & Tzap & 0.031 & 476 & 168 & 1216 &  & Tzap & 0.029 & 950 & 0 & 1584 \\
\addlinespace[2.5pt]
\texttt{PolynomialPauliR} & Reqomp & -- & 360 & 61 & 510 & \texttt{GidneyAdder} & Reqomp & -- & 1032 & 0 & 2061 \\
 & SPARE & 13.3 & 80 & 131 & 260 &  & SPARE & \multicolumn{4}{c}{Timeout ($>$1000\,s)} \\
 & Feynman & 0.008 & 120 & 61 & 254 &  & Feynman & 0.057 & 1032 & 0 & 2061 \\
 & Tzap & 0.025 & 120 & 31 & 358 &  & Tzap & 0.026 & 1032 & 0 & 2060 \\
\addlinespace[2.5pt]
\texttt{WeightedAdder} & Reqomp & -- & 832 & 0 & 1085 & \texttt{QLZOC} & Reqomp & -- & 1265 & 0 & 2207 \\
 & SPARE & 35.7 & 831 & 0 & 1123 &  & SPARE & 878.7 & 1265 & 0 & 2511 \\
 & Feynman & 0.022 & 678 & 0 & 945 &  & Feynman & 0.070 & 1087 & 0 & 2029 \\
 & Tzap & 0.027 & 678 & 0 & 945 &  & Tzap & 0.027 & 1087 & 0 & 2028 \\
\addlinespace[2.5pt]
\texttt{Incrementer} & Reqomp & -- & 40 & 0 & 57 & \texttt{MatrixVectorEq} & Reqomp & -- & 1527 & 0 & 2164 \\
 & SPARE & 3.6 & 39 & 0 & 54 &  & SPARE & 249.9 & 1527 & 0 & 2290 \\
 & Feynman & 0.001 & 40 & 0 & 57 &  & Feynman & 0.077 & 1271 & 0 & 1908 \\
 & Tzap & 0.024 & 40 & 0 & 57 &  & Tzap & 0.028 & 1271 & 0 & 1908 \\
\bottomrule
\end{tabular}
\end{table}

\end{document}